\pdfoutput=1
\documentclass[a4paper,fleqn,usenatbib]{mnras}

\usepackage{mathptmx}

\usepackage[T1]{fontenc}
\usepackage{ae,aecompl}

\usepackage{graphicx}	
\usepackage{amsmath}	
\usepackage{amssymb}	
\usepackage{natbib}
\usepackage{float}
\usepackage{placeins}
\usepackage{tabularx}

\title[AzTEC 1.1 mm observations of high-z protoclusters]{AzTEC 1.1 mm observations of high-z protocluster environments: SMG overdensities and misalignment between AGN jets and SMG distribution}

\author[Zeballos et al.]{M. Zeballos$^{1,2}$\thanks{E-mail:zeballos@inaoep.mx}, I. Aretxaga$^{2}$, D.H. Hughes$^{2}$, A. Humphrey$^{3}$, G. W. Wilson$^{4}$, 
\newauthor J. Austermann$^{5}$, J. S. Dunlop$^{6}$, H. Ezawa$^{7}$, D. Ferrusca$^{2}$, B. Hatsukade$^{8}$,
\newauthor R. J. Ivison$^{6,9}$, R. Kawabe$^{10,11,12}$, S. Kim$^{13}$, T. Kodama$^{10,12}$, K. Kohno$^{8,14}$ ,
\newauthor A. Monta\~{n}a$^{2}$, K. Nakanishi$^{10,12}$, M. Plionis$^{15,16}$, D. S\'{a}nchez-Arg\"{u}elles$^{2}$, 
\newauthor J. A. Stevens$^{17}$, Y. Tamura$^{8}$, M. Velazquez$^{2}$, and M. S. Yun$^{4}$\\
$^{1}$Universidad de las Am\'{e}ricas Puebla, Sta. Catarina M\'{a}rtir, San Andr\'{e}s Cholula, 72810 Puebla, Mexico\\
$^{2}$Instituto Nacional de Astrof\'{i}sica, \'{O}ptica y Electr\'{o}nica (INAOE), Aptdo. Postal 51 y 216, 72000 Puebla, Mexico\\
$^{3}$Instituto de Astrof\'isica e Ci\^encias do Espa\c co, Universidade do Porto, CAUP, Rua das Estrelas, P-4150-762 Porto, Portugal\\
$^{4}$Department of Astronomy, University of Massachusetts, Amherst, MA 01003, USA\\
$^{5}$Center for Astrophysics and Space Astronomy, University of Colorado, Boulder, CO 80309, USA\\
$^{6}$Institute for Astronomy, University of Edinburgh, Royal Observatory, Blackford Hill, Edinburgh EH9 3HJ, UK\\
$^{7}$ALMA Project Office, National Astronomical Observatory of Japan, 2-21-1 Osawa, Mitaka, Tokyo 181-0015, Japan\\
$^{8}$Institute of Astronomy, University of Tokyo, 2-21-1 Osawa, Mitaka, Tokyo 181-0015, Japan\\
$^{9}$European Southern Observatory, Karl Schwarzschild Strasse 2, Garching, Germany\\
$^{10}$National Astronomical Observatory of Japan, 2-21-1 Osawa, Mitaka, Tokyo 181-8588, Japan\\
$^{11}$Department of Astronomy, School of Science, The University of Tokyo, 7-3-1 Hongo, Bunkyo-ku, Tokyo 113-0033, Japan\\
$^{12}$The Graduate University for Advanced Studies (SOKENDAI), 2-21-1 Osawa, Mitaka, Tokyo 181-8588, Japan\\
$^{13}$Astronomy and Space Science Department, Sejong University, Seoul, South Korea\\
$^{14}$Research Center for the Early Universe, University of Tokyo, 7-3-1 Hongo, Bunkyo-ku, Tokyo 113-0033, Japan\\
$^{15}$National Observatory of Athens, Lofos Nymfon, Athens 11810, Greece\\
$^{16}$Physics Department, Aristotle University of Thessaloniki, 54124 Greece\\
$^{17}$Centre for Astrophysics, Science and Technology Research Institute, University of Hertfordshire, Hatfield AL10 9AB, UK}

\date{Accepted XXX. Received YYY; in original form ZZZ}

\pubyear{2018}

\begin{document}
\label{firstpage}
\pagerange{\pageref{firstpage}--\pageref{lastpage}}
\maketitle

\begin{abstract}
We present observations at 1.1 mm towards 16 powerful radio galaxies and a radio-quiet quasar at $0.5<z<6.3$ acquired with the AzTEC camera mounted at the JCMT and ASTE to study the spatial distribution of submillimeter galaxies towards possible protocluster regions. The survey covers a total area of 1.01 square degrees with rms depths of 0.52 - 1.44 mJy and detects 728 sources above 3$\sigma$. We find overdensities of a factor of $\sim2$ in the source counts of 3 individual fields (4C+23.56, PKS1138-262 and MRC0355-037) over areas of $\sim$200 sq deg. When combining all fields, the source-count analysis finds an overdensity that reaches a factor $\gtrsim 3$ at S$_{\rm{1.1mm}} \ge 4$ mJy covering a 1.5-arcmin-radius area centred on the AGN. The large size of our maps allows us to establish that beyond a radius of 1.5 arcmin, the radial surface density of SMGs falls to that of a blank field. In addition, we find a trend for SMGs to align closely to a perpendicular direction with respect to the radio jets of the powerful central radio galaxies (73$_{+13}^{-14}$ degrees). This misalignment is found over projected co-moving scales of 4-20 Mpc, departs from perfect alignment (0 deg) by $\sim 5\sigma$, and apparently has no dependence on SMG luminosity. Under the assumption that the AzTEC sources are at the redshift of the central radio galaxy, the misalignment reported here can be interpreted as SMGs preferentially inhabiting mass-dominant filaments funneling material towards the protoclusters, which are also the parent structures of the radio galaxies.
\end{abstract}




\section{Introduction}

Clusters of galaxies are the largest virialized structures in the Universe, with masses that range $10^{14}-10^{16}$ M$_\odot$. They originate from the gravitational collapse of matter and represent extreme potential wells developed from the initial conditions in the density field of the Universe. Thus, they are natural laboratories for studying evolutionary scenarios for the formation of large-scale structure (e.g.~\citealp{Jimenez09,Harrison12}), and specifically, for the formation and evolution of galaxies (e.g.~\citealp{Demarco10a}).

Despite extensive multi-wavelength studies towards clusters in the nearby Universe, their progenitors remain relatively unexplored. Protocluster identification is difficult because of the small difference in density between the forming cluster and its surroundings. In addition, classical cluster detection techniques such as searching for extended X-ray emission (e.g.~\citealp{Rosati98,Pierre04}) and identifying red galaxy overdensities (e.g.~\citealp{Andreon09,Demarco10b}) fail to work for $z > 2.5$. Both the X-ray emission produced by the hot intracluster medium (ICM) and the optical light emitted by red galaxies suffer from cosmological dimming due to the expansion of the Universe. Furthermore, these identification techniques, as well as the Sunyaev-Zeld\'ovich effect (e.g.~\citealp{Menanteau09,PlanckC13}), usually detect signals of an already evolved cluster, with an old galaxy population and a virialized environment heating the ICM.

One the most popular techniques to search for protoclusters has been targeting the fields of the most powerful high redshift radio galaxies and quasars (AGN; e.g.~\citealp{Kurk00,McLure01}). Since these AGN are hosted by the most massive galaxies in the Universe, they pinpoint the location of the highest density regions under the currently most accepted paradigm of structure formation, the $\Lambda$CDM model.

Evidence for the large masses of powerful AGN host galaxies comes from the radio/optical luminosities of the AGN itself, which indicate the presence of a supermassive black hole (SMBH). In the nearby Universe, there is a well-studied correlation between SMBH mass and the bulge luminosity of the host galaxy (e.g.~\citealp{McLure02}), and although at higher redshifts this correlation is still under scrutiny (e.g.~\citealp{Wang13,Willott15}), powerful AGN are expected to inhabit massive galaxies. In addition, radio galaxies at high redshifts were found to have the largest \emph{K}-band luminosities in the early Universe, which indicates stellar masses of up to $10^{12}$ M$_{\odot}$ (e.g.~\citealp{Rocca-Volmerange04,Seymour07,Targett11}). Another piece of evidence comes from the existence of giant nebulae of ionized gas surrounding them, with sizes of up to $\sim 200$ kpc, which contain enough gas to produce systems as large as cD-like galaxies (e.g.~\citealp{Reuland03}).

Previous narrow-band filter observations towards the environments of high-redshift AGN found overdensities of a variety of star-forming galaxies such as Ly$\alpha$ emitters (LAE; e.g.~\citealp{Venemans07}), H$\alpha$ emitters (HAE; e.g.~\citealp{Kurk04,Hatch11,Hayashi12}), and Lyman break galaxies (LBG; e.g.~\citealp{Miley04,Overzier06,Intema06}). These type of observations, however, trace relatively low mass galaxies with unobscured star formation. Since about 50\% of the cosmic star formation is obscured by dust (e.g.~\citealp{Dole06}), far-infrared (far-IR), submillimetre (submm) or millimetre (mm) observations are required to fully understand the formation history of stellar mass in galaxy clusters. Besides recent systematic searches in Planck all sky maps (e.g.~\citealp{Planck15,Clements14,Flores-Cacho16}), the primary protocluster identification technique used at these wavelengths has been targeting the fields of powerful AGN (e.g.~\citealp{Stevens03,DeBreuck04,Priddey08,Stevens10,Wylezalek13b,Dannerbauer14,Rigby14}) with the highest radio, optical, or X-ray luminosities and at the highest redshifts ($z \gtrsim 2$). In most cases, these studies found a number density of sources $\geq2$ larger than blank-field estimates, consistent with these regions being extreme density peaks in the Universe.

Far-IR/submm/mm studies towards these biased regions are sensitive to a heavily obscured star-forming galaxy population characterized by extreme far-infrared luminosities (L$_{\rm{FIR}} > 10^{12}$ L$_\odot$), large star formation rates (SFR $> 100 - 1000$ M$_\odot$ yr$^{-1}$), and a redshift distribution with $\sim50$\% of bright sources at $2< z < 3$ (e.g.~\citealp{Chapman03,Chapman05,Aretxaga03,Aretxaga07,Pope05,Wardlow11,Smolcic12,Simpson14}) with a possible tail towards higher values (e.g.~\citealp{Coppin09,Riechers10,Cox11,Yun12,Walter12}). This implies that they are a young population capable of building large stellar masses in $< 1$ Gyr (provided the star-forming activity is sustained for the whole period of time; see \citealp{Casey14} for a review). Therefore, they are very good candidates to be the progenitors of the massive galaxies that we see today as the dominant population in the centre of rich galaxy clusters. In addition, recent studies of SMGs at $z = 1-3$ estimate stellar masses of $>10^{11}$ M$_{\odot}$ (e.g.~\citealp{Dye08,Targett12}), molecular gas masses of $\sim 5 \times 10^{10}$ M$_{\odot}$ (e.g.~\citealp{Greve05,Bothwell13}), and dust masses of $\sim 10^{8}$ M$_{\odot}$ (e.g.~\citealp{Chapman05,Magnelli12}), values that also support the theory that SMGs are destined to evolve into massive ellipticals in the low-redshift Universe. Studying the properties of SMGs towards protocluster regions will improve our understanding of the still elusive formation history of these massive galaxies, and possibly give us an insight into the formation of the stellar population of the richest galaxy clusters.

In this paper we present 1.1 mm continuum imaging observations towards the environments of 16 powerful high-redshift radio galaxies and a quasar acquired with the AzTEC camera \citep{Wilson08a}. This sample is a subset of the AzTEC Cluster Environment Survey (ACES), which observed 40 fields towards powerful AGN and massive galaxy clusters. We introduce the sample in section \ref{sec:sample}. Section \ref{sec:obs} describes the details of the observations, and section \ref{sec:datarec} explains the data reduction process in order to obtain clean and optimally filtered sky maps. Section \ref{sec:nc} estimates the number density of sources as a function of flux density for individual and combined fields, and section \ref{sec:alignment} analyzes a possible relation between the spatial distribution of SMGs and the radio jet directions of the radio galaxies in the sample. In section \ref{sec:discussion} we discuss our results and in section \ref{sec:conclusions} we summarize the conclusions of the paper.

Throughout this paper, we adopt a flat cosmology with $\Omega_M = 0.27$, $\Omega_\Lambda = 0.73$ and $H_0 =71$ km s$^{-1}$ Mpc$^{-1}$.

\section{Sample selection}
\label{sec:sample}

\subsection{Target fields}

\begin{figure}
\includegraphics[width=0.5\textwidth, trim=5cm 9cm 4cm 9cm, clip]{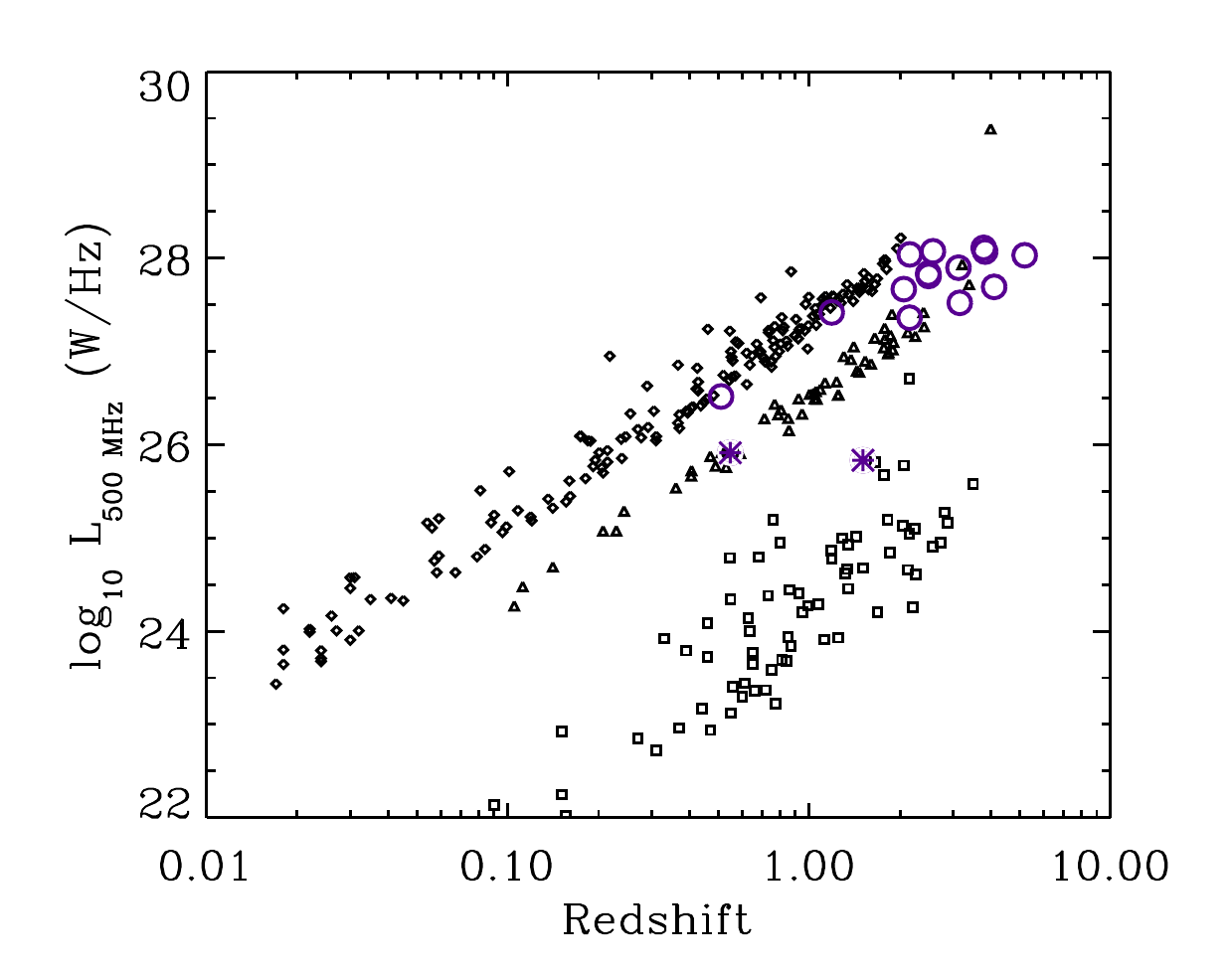}
\caption{Radio luminosity-redshift ($L_{\rm{500\,MHz}} - z$) plane for radio surveys 3CRR (diamonds; \citealp{3CRR}), 6CE (triangles; \citealp{6CE1,6CE2}) and LBDS (squares, Lynx and Hercules fields only; \citealp{Rigby07}), which have successively deeper flux-density limits. These radio luminosities were calculated using a typical spectral index of 0.8 \citep{DeZotti10}. Open circles denote ACES radio galaxies whose synchrotron self-absorption frequency appears to be at $< 500$ MHz (rest frame). Asterisks denote the 2 Gigahertz Peaked Spectrum (GPS) sources we include in our sample. ACES AGN luminosities were estimated by fitting a straight line or a parabola to their spectra, depending on the number of data points available.}
\label{PZplane}
\end{figure}

We targeted fields centred on luminous AGN at 0.5 < z < 6.3 which were known or suspected to be hosted by massive galaxies, as potential signposts of high redshift overdensities. Our powerful high-redshift radio galaxies are among the prime AGN to pinpoint protoclusters due to their location at the high-end of the luminosity-redshift ($L_{\rm{500\,MHz}}-z$) plane, with luminosities in the range $26.5 <$ log$_{10}$(L$_{\rm{500\,MHz}}$) $< 28.1$ W Hz$^{-1}$ (Figure~\ref{PZplane}) and their high-mass end SMBHs. We also include two Gigahertz Peaked Spectrum (GPS) sources, TXS2322-040 and MRC2008-068, that show lower 500-MHz luminosities due to synchrotron self-absorption, but some theories place them as young counterparts of classical extended radio sources (e.g.~\citealp{ODea98}). More importantly, \emph{R} and \emph{K} band images show that these type of sources are also massive galaxies in their own right \citep{Snellen96a,Snellen96b}.

All fields are located far from the galactic plane and, in general, far from where there is large contamination by galactic cirrus. On average, these regions show background dust emission variations at 100 $\mu$m that are $< 0.3$ MJy sr$^{-1}$. This yields a cirrus confusion noise $< 0.6$ mJy/beam at the wavelength (1.1 mm) and beam size (30 \& 18 arcsec) of our observations under the assumption of an isothermal dust spectrum with $T = 19$ K and emissivity index $\beta = 1.4$ (\citealp{Bracco11}).

Although seven of these selected targets were already confirmed as rich protoclusters via narrow line emission observations (TNJ0924-2201, TNJ1338-1942, MRC0316-257, PKS0529-549, MRC2104-242, 4C+23.56 and PKS1138-262), the selection criteria of the fields were unbiased regarding previous overdensity detections.

We complete the ACES protocluster sample with one extra field centred at the optically luminous (M$_B = -27.7$) radio quiet quasar SDSSJ1030+0524 ($z = 6.28$), which at the time of our observations was the highest-redshift known quasar. The list of selected targets is shown in Table~\ref{ACESRGtable_radio}.

\begin{table*}
\begin{center}
\caption{\label{ACESRGtable_radio} General properties of the AGN targeted in the AzTEC observations. There are 16 radio galaxies (RG) and a radio quiet quasar (RQ-QSO). The columns show: 1) AGN name; 2) AGN type; 3) \& 4) their most accurate coordinates as determined from radio, optical or mid(near)-infrared data; 5) redshift; 6) radio luminosity estimated at a rest-frame frequency of 500 MHz; 7) position angle (PA) of the radio emission measured North to East; and 8) reference for PA.}
\begin{tabular}{lcccccccl}
\hline
Central AGN 	& Type	& R.A. 				& Dec. 				& z	& log$_{10}$(L)   		& PA 		&References	\\
			& 		& \small{(J2000)}		& \small{(J2000)}		& 	& 					& 			& 			\\
     			& 		& \small{(hr:m:s)}		& \small{(d:am:as)}		& 	& \small{(W Hz$^{-1}$)}  	& \small{(deg)} 	&    			\\
\hline
SDSSJ1030+0524	 & \small{RQ-QSO}    &	 10:30:27.10       &	 +05:24:55.0	  &	 6.311    	  &	         	   &	               	 &\\
TNJ0924-2201 	& \small{ 	 RG	 } 	&  	 09:24:19.91 	  &  	 -22:01:41.5 	  &  	 5.190 	  &  	 28.03 &  	 74 		 &\cite{DeBreuck00} \\
TNJ1338-1942 	& \small{ 	 RG 	 } 	&  	 13:38:26.23 	  &  	 -19:42:33.6 	  &  	 4.110 	  &  	 27.69 &  	 152		 &\cite{DeBreuck00}\\
TNJ2007-1316 	& \small{ 	 RG 	 } 	&  	 20:07:53.22 	  &  	 -13:16:43.4 	  &  	 3.837 	  &  	 28.07 &  	 27 		 &C. De Breuck, priv. comm.\\
4C+41.17 	  	& \small{ 	 RG 	 } 	&  	 06:50:52.35 	  &  	 +41:30:31.4 	  &  	 3.792 	  &  	 28.11 &  	 48 	 	 &\cite{Chambers90}\\
TNJ2009-3040 	& \small{ 	 RG 	 } 	&  	 20:09:48.08 	  &  	 -30:40:07.4 	  &  	 3.160 	  &  	 27.52 &  	 144		 &\cite{DeBreuck00}\\
MRC0316-257 	  	& \small{ 	 RG 	 } 	&  	 03:18:12.14 	  &  	 -25:35:10.2 	  &  	 3.130 	  &  	 27.90 &  	 51 		 &\cite{McCarthy90}\\ 
PKS0529-549 	  	& \small{ 	 RG 	 } 	&  	 05:30:25.43 	  &  	 -54:54:23.3 	  &  	 2.575 	  &  	 28.07 &  	 104		 &\cite{Broderick07}\\
MRC2104-242 	  	& \small{ 	 RG 	 } 	&  	 21:06:58:27 	  &  	 -24:05:09.1 	  &  	 2.491 	  &  	 27.81 &  	 12 		 &\cite{Pentericci01}\\
4C+23.56 	  	& \small{ 	 RG 	 } 	&  	 21:07:14.82 	  &  	 +23:31:45.1 	  &  	 2.483 	  &  	 27.84 &  	 52 		 &\cite{Chambers96}\\
PKS1138-262 	  	& \small{ 	 RG 	 } 	&  	 11:40:48.35 	  &  	 -26:29:08.6 	  &  	 2.156 	  &  	 28.04 &  	 90 		 &\cite{Pentericci97}\\
MRC0355-037 	  	& \small{ 	 RG 	 } 	&  	 03:57:48.06 	  &  	 -03:34:09.5 	  &  	 2.153 	  &  	 27.36 &  	 120		 &\cite{Gopal-Krishna05}\\
MRC2048-272 	  	& \small{ 	 RG 	 } 	&  	 20:51:03.49 	  &  	 -27:03:03.7 	  &  	 2.060 	  &  	 27.67 &  	 45 		 &\cite{Kapahi98}\\
TXS2322-040 	  	& \small{ 	 RG 	 } 	&  	 23:25:10.23 	  &  	 -03:44:46.7 	  &  	 1.509 	  &  	 25.83 &  	 -4 		 &\cite{Xiang06}\\
MRC2322-052 	  	& \small{ 	 RG 	 } 	&  	 23:25:19.62 	  &  	 -04:57:36.6 	  &  	 1.188 	  &  	 27.42 &  	 107		 &\cite{Best99}\\
MRC2008-068 	  	& \small{ 	 RG 	 } 	&  	 20:11:14.22 	  &  	 -06:44:03.6 	  &  	 0.547 	  &  	 25.92 &  	 -28		 &\cite{Morganti93}\\
MRC2201-555 	  	& \small{ 	 RG 	 } 	&  	 22:05:04.83 	  &  	 -55:17:44.0 	  &  	 0.510 	  &  	 26.52 &  	 -85		 &\cite{Burgess06}\\
\hline
\end{tabular}
\end{center}
\end{table*}

\subsection{Control field}

AzTEC observed, together with the ACES sample, a set of well-known blank fields with available multi-wavelength data (\citealp{ScottKS12} and references therein). This set covers a total area of 1.6 square degrees with a depth range between 0.4 and 1.7 mJy beam$^{-1}$, and provides the best estimation to-date of the surface density of SMGs towards blank fields at flux densities between S$_{\rm{1.1mm}} = 1 - 12$ mJy. We use these data to compare the properties of the SMGs in our sample against those of SMGs in unbiased environments.

\section{Observations}
\label{sec:obs}

The ACES protocluster sample was surveyed at a wavelength of 1.1 mm with the AzTEC camera as a visiting instrument at the 15-m James Clerk Maxwell Telescope (JCMT; FWHM = 18 arcsec) in Hawaii and at the 10-m Atacama Submillimeter Telescope Experiment (ASTE; FWHM = 30 arcsec) in the northern part of Chile. The field towards 4C+41.17 was observed at the JCMT in December 2005, and the other 16 fields were observed at the ASTE from May to October 2007 and July to December 2008. The sample was observed under very good weather conditions: for 95\% of the observing time the zenith atmospheric opacity was $\tau_{\rm{220\,GHz}} < 0.095$ at the ASTE and $\tau_{\rm{225\,GHz}} < 0.115$ at the JCMT. The zenith opacity mean and standard deviation values for each field are given in Table~\ref{ACESRGtable_mm}.

The 4C+41.17 field was mapped using a raster scanning technique, while the ASTE fields were mapped using a Lissajous pattern centred on the AGN. Integration times varied between 16 and 35 hours per field (excluding calibration and pointing observations), and the resulting maps cover uniform noise areas ranging from 170 to 300 sq arcmin.

AzTEC maps were calibrated using planets as primary calibrators. Each night Uranus or Neptune was imaged to derive the flux conversion factor for each detector. In a single observation of a field the typical statistical calibration error was 6-13\% \citep{Wilson08a}. When all observations in each field are considered, the calibration errors on our measured source flux densities integrate down to 1.7 - 2.5\%, which need to be combined in quadrature with the 5\% absolute uncertainty on the flux densities of the planets.

In order to correct the observations for small pointing offsets between the centre of the AzTEC array and the telescope boresight, well-known bright point sources ($>1$ Jy) a few degrees away from the science targets were periodically observed. They were taken every 1 or 2 hours, always bracketing the ACES protocluster observations. The resulting absolute pointing uncertainty of the AzTEC maps is $< 2$ arcsec, much smaller than our beam sizes (18 and 30 arcsec).

\begin{table*}
\begin{center}
\caption{\label{ACESRGtable_mm} Main properties of the AGN fields targeted in the AzTEC observations. The columns are: 1) radio galaxy or quasar at which the map is centred; 2) telescope used; 3) uniform noise area determined by a 50\% coverage cut; 4) range of noise rms values inside this area; 5) integration time (excluding calibration and pointing observations); 6) zenith opacity mean and standard deviation values measured at 220 GHz for the ASTE fields and at 225 GHz for the JCMT field; and 7) scanning pattern used to observe the target.}
\begin{tabular}{lccccccl}
\hline
Central AGN 	&  Telescope     & Area          		& Noise rms range 	& Int. time	&$\tau$   		& Scan type\\
     			&                   	&(arcmin$^{2}$) 	& (mJy/beam)      	&   (hr)	&		 	&   \\
\hline
SDSSJ1030+0524 	& ASTE 	&	212.6	&	  0.52-0.77 	&	31.82	&	$0.035 \pm 0.020$ & Lissajous\\
TNJ0924-2201 	& ASTE 	&	210.6	&	  0.85-1.26 	&	30.14	&	$0.041 \pm 0.023$ & Lissajous\\
TNJ1338-1942 	& ASTE 	&	209.5	&	  0.96-1.44 	&	16.48	&	$ 0.064 \pm 0.023$ & Lissajous\\
TNJ2007-1316 	& ASTE 	&	211.3	&	  0.90-1.34 	&	19.50	&	$0.051 \pm 0.015$ & Lissajous\\
4C+41.17 		& JCMT 	&	303.3	&	  0.97-1.42 	&	35.44	&	$0.074 \pm 0.027$  & Raster\\
TNJ2009-3040 	& ASTE 	&	211.9	&	  0.88-1.32 	&	19.49	&	$0.063 \pm 0.048$ & Lissajous\\
MRC0316-257 		& ASTE 	&	211.7	&	  0.65-0.96 	&	22.28	&	$0.045 \pm 0.016$ & Lissajous\\
PKS0529-549 		& ASTE 	&	213.4	&	  0.62-0.92 	&	28.33	&	$0.038 \pm 0.016$ & Lissajous\\
MRC2104-242 		& ASTE 	&	209.1	&	  0.83-1.24 	&	24.87	&	$0.046 \pm 0.015$ & Lissajous\\
4C+23.56 		& ASTE 	&	166.0	&	  0.55-0.85 	&	34.43	&	$0.057 \pm 0.030$ & Lissajous\\
PKS1138-262 		& ASTE 	&	211.6	&	  0.70-1.04 	&	42.14	&	$0.042 \pm 0.025 $ & Lissajous\\
MRC0355-037 		& ASTE 	&	212.4	&	  0.78-1.15 	&	25.72	&	$0.043 \pm 0.016$ & Lissajous\\
MRC2048-272 		& ASTE 	&	211.7	&	  0.73-1.09 	&	19.88	&	$0.054 \pm 0.017$ & Lissajous\\
TXS2322-040 		& ASTE 	&	212.1	&	  0.61-0.91 	&	23.67	&	$0.037 \pm 0.014$ & Lissajous\\
MRC2322-052 		& ASTE 	&	212.4	&	  0.68-1.01 	&	23.68	&	$0.051\pm 0.019$ & Lissajous\\
MRC2008-068 		& ASTE 	&	211.9	&	  0.83-1.24 	&	20.70	&	$0.045 \pm 0.020$ & Lissajous\\
MRC2201-555 		& ASTE 	&	213.4	&	  0.74-1.11 	&	26.45	&	$0.089 \pm 0.062$ & Lissajous\\
\hline
\end{tabular}
\end{center}
\end{table*}

\section{Data reduction and analysis}
\label{sec:datarec}

We reduced the AzTEC data in a manner similar to that described in detail by \cite{ScottKS08,ScottKS12}.

The raw timestream data from the instrument, which include both bolometer and pointing data, are despiked and then cleaned of atmospheric contamination using the standard principal component analysis (PCA) technique. An astrometric correction is made to all pointing signals in the timestream based on a linear interpolation of the pointing offsets measured by the bracketing pointing observations. With this correction in place, the bolometer signals are flux-calibrated and binned into $3 \times 3$ ($2\times2$) sq arcsec pixels for ASTE (JCMT) observations. Performing this process for observations of each field results in independent maps which are then co-added to make a preliminary image of the sky around the AGN of choice. As in previous AzTEC analyses, we also produce a weight map and 100 noise-only realisations of each field. The weight map is built by adding in quadrature the inverse of the variance of all bolometer samples that contribute to a pixel, i.e.~it represents the inverse of the squared noise level per pixel (pixel weight). The noise maps, on the other hand, are produced by jackknifing the timestream data and used in the characterisation of the map properties.

In order to remove pixel-to-pixel variations, each preliminary AzTEC map needs to be convolved with the point-source response of the instrument. This point-source kernel is obtained by inserting 3 fake sources in the timestreams that made the original map and tracing them through the entire reduction process. A detailed description can be found in \cite{Downes12}. We use the mean power spectral density of the noise maps and the estimated point-source kernel to construct an optimal filter for point source detection. The final set of filtered maps for each ACES protocluster field is composed of: a filtered signal map, a filtered weight map, the corresponding S/N map, and a set of 100 filtered noise realisations.

\subsection{The maps}
\label{subsec:maps}

\begin{figure}
\centering
\includegraphics[width=0.5\textwidth, trim=1.5cm 6cm 1cm 5cm, clip]{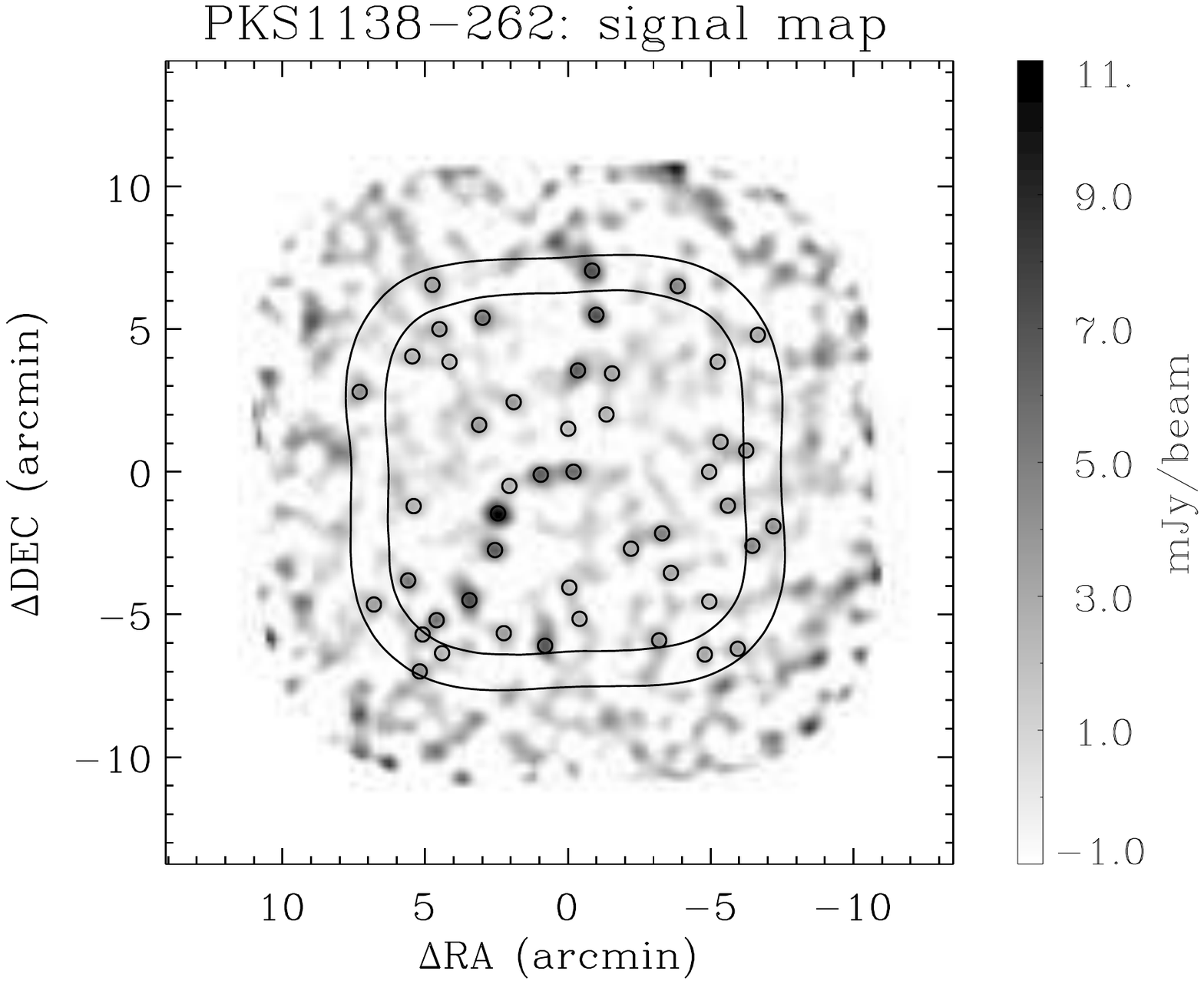}
\includegraphics[width=0.5\textwidth, trim=1.5cm 6cm 1cm 5cm, clip]{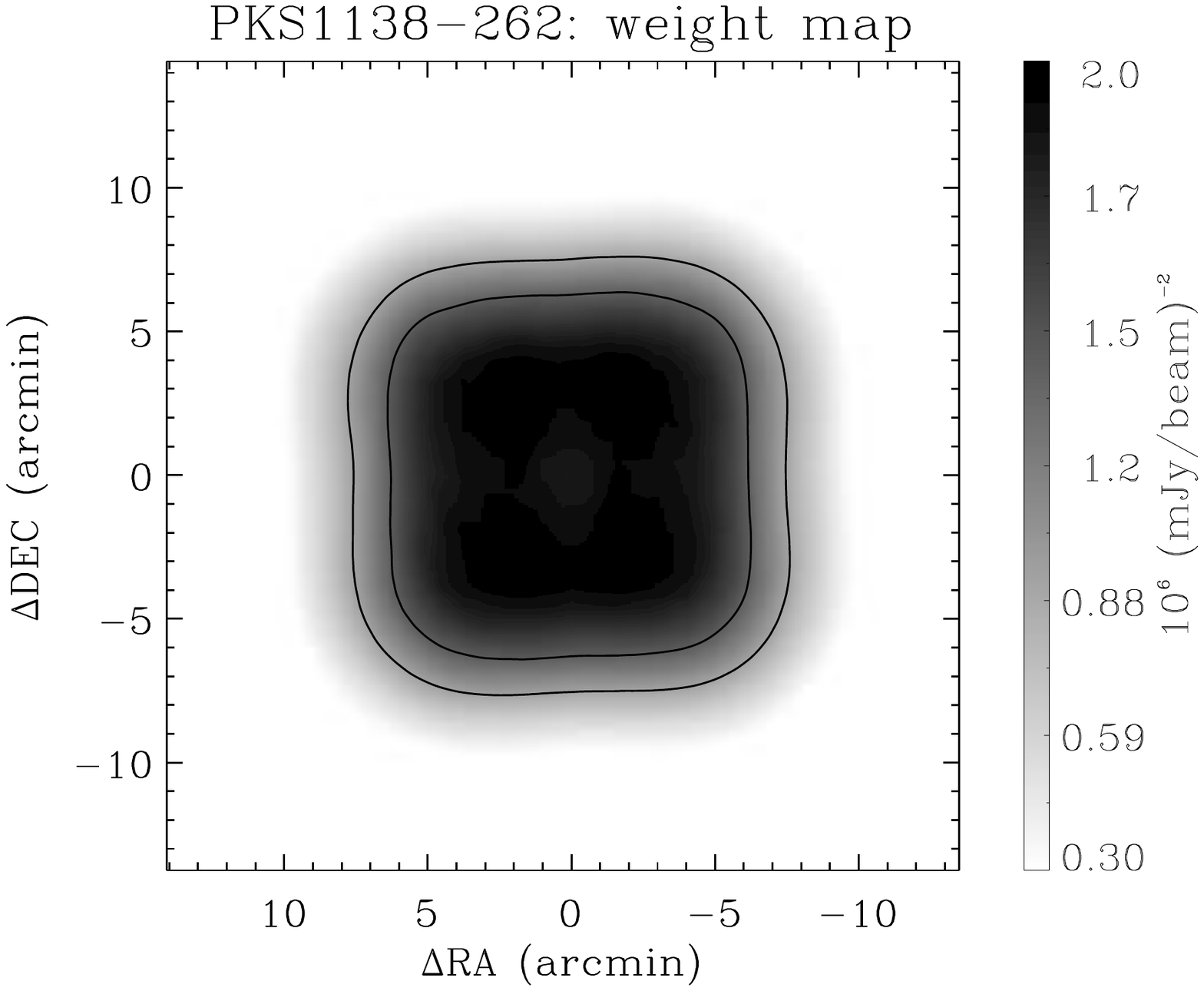}
\caption{AzTEC signal and weight maps for the ACES protocluster candidate towards PKS1138-262. Source candidates with S/N $> 3$ are marked by 30-arcsec diameter circles (FWHM at the ASTE). Contours represent curves of constant noise of 0.84 and 1.04 mJy/beam (75\% and 50\% coverage cuts).}
\label{Map0802}
\end{figure}

Figure~\ref{Map0802} shows the final signal and weight maps towards the field of the radio galaxy PKS1138-262. The rest of the maps can be seen in appendix \ref{appendixA}. Contours represent curves of constant noise and have values of 0.84 and 1.04 mJy/beam. The central part of the map is slightly deeper than the edges, and the noise increases rapidly towards them. Therefore, all the analysis is restricted to the central area where pixel weights are larger than 50\% of the maximum weight (50\% coverage-cut area).

Table~\ref{ACESRGtable_mm} lists the 50\% coverage-cut areas for all the ACES protocluster fields, 15 of which cover areas of $\sim 8$ arcmin radius. The 4C+23.56 map is slightly smaller, with a $\sim 6$ arcmin radius, but large enough to cover a possible protocluster centred at the radio galaxy position and at the radio galaxy redshift (co-moving area of $\sim 10$ Mpc radius at $z=2.48$). On the other hand, the 4C+41.17 map is the largest, covering an area of $\sim 10$ arcmin radius. Table~\ref{ACESRGtable_mm} also shows noise rms values for each ACES protocluster target. Since integration times per pixel slowly decrease from the centre of the map towards the edges, the quoted noise rms values are intervals.  All noise rms values are $< 1.5$ mJy/beam.

Using the standard rule of thumb (one source per 30 beams) and the AzTEC blank-field 1.1 mm source counts, we estimate the confusion limit given the 30 arcsec (18 arcsec) FWHM ASTE (JCMT) beam to be 2.1 (1.0) mJy. Our ACES/ASTE survey is 2 - 4 times deeper than the formal confusion limit. Therefore, in the next sections we consider its effects on the properties of ours map following the analysis in \cite{ScottKS10}.

\subsection{Source catalogues}
\label{subsec:cat}

Source candidates are identified in the S/N maps as local maxima above a S/N threshold of 3.5 for the shallowest maps (TNJ1338-1942 \& 4C+41.17) and 3.0 for the rest. These S/N thresholds are established so that the number of possible sources is maximized while minimizing the number of false detections (noise peaks). Figure~\ref{FDR0802} and appendix \ref{appendixA} show that a reasonably low percentage ($\sim 8-24\%$) of sources with S/N above these defined thresholds could be false positive peaks. Therefore, we considered these limits appropriate since the number of source candidates decreases significantly if higher (lower) thresholds (false detection rates) are selected.

Table~\ref{cluster0802cat} shows the source catalogue for the PKS1138-262 field. This list is in decreasing order of S/N and includes the measured 1.1 mm flux densities and their deboosted fluxes (section \ref{subsec:fluxdeb}). Catalogues for the other ACES protoclusters can be found in appendix \ref{appendixA}.

\begin{table*}
\begin{center}
\caption{\label{cluster0802cat} AzTEC source catalogue for the field of PKS1138-262. The columns show: 1) source id; 2) source name; 3) S/N of the detection; 4) measured 1.1 mm flux density and error; 5) deboosted 1.1 mm flux density and 68\% confidence interval; and 6) probability for the source to have a negative deboosted flux. The catalogue is limited to sources detected at a S/N $> 3$ within 50\% coverage region of the AzTEC map. There are 47 detections, and according to the false detection rate upper limit estimated in Figure~\ref{FDR0802}, at most 4 of these sources could be false (8\%). Nevertheless, all sources show a probability of having a negative deboosted flux $<$ 0.05. Therefore, all of them are considered for the source count analysis in section \ref{sec:nc}.}
\begin{tabular}{lccccc}
\hline
      &     		&     		& S$_{\rm{1.1mm}}$ & S$_{\rm{1.1mm}}$ 	&         		\\
      &      		&     		& (measured)  		& (deboosted) 			&         		\\
Id  & IAU name & S/N 	& (mJy)       		& (mJy)       			& P($<0$) 	\\
\hline
1 	 & 	 MMJ114059.25-263038.40 	  &  	  16.09 	  &  	 11.4$\pm$0.7 	  &  	 11.1$^{+0.7}_{-0.7}$ 	  &  	 0.000 	 \\
2 	 & 	 MMJ114103.73-263340.90 	  &  	  10.41 	  &  	 7.6$\pm$0.7 	  &  	 7.2$^{+0.7}_{-0.7}$ 	  &  	 0.000 	 \\
3 	 & 	 MMJ114059.72-263155.49 	  &  	  10.10 	  &  	 7.1$\pm$0.7 	  &  	 6.7$^{+0.8}_{-0.7}$ 	  &  	 0.000 	 \\
4 	 & 	 MMJ114043.85-262340.86 	  &  	   9.88 	  &  	 7.7$\pm$0.8 	  &  	 7.2$^{+0.8}_{-0.8}$ 	  &  	 0.000 	 \\
5 	 & 	 MMJ114046.73-262538.01 	  &  	   9.60 	  &  	 6.8$\pm$0.7 	  &  	 6.5$^{+0.7}_{-0.8}$ 	  &  	 0.000 	 \\
6 	 & 	 MMJ114052.55-262917.09 	  &  	   9.45 	  &  	 6.9$\pm$0.7 	  &  	 6.5$^{+0.7}_{-0.8}$ 	  &  	 0.000 	 \\
7 	 & 	 MMJ114051.88-263516.75 	  &  	   9.24 	  &  	 7.6$\pm$0.8 	  &  	 7.1$^{+0.8}_{-0.8}$ 	  &  	 0.000 	 \\
8 	 & 	 MMJ114047.42-262910.76 	  &  	   8.41 	  &  	 6.2$\pm$0.7 	  &  	 5.8$^{+0.7}_{-0.8}$ 	  &  	 0.000 	 \\
9 	 & 	 MMJ114044.52-262207.98 	  &  	   7.97 	  &  	 7.5$\pm$0.9 	  &  	 6.9$^{+0.9}_{-1.0}$ 	  &  	 0.000 	 \\
10 	 & 	 MMJ114033.54-263120.28 	  &  	   7.64 	  &  	 5.5$\pm$0.7 	  &  	 5.1$^{+0.7}_{-0.7}$ 	  &  	 0.000 	 \\
11 	 & 	 MMJ114108.87-263422.79 	  &  	   7.47 	  &  	 5.9$\pm$0.8 	  &  	 5.5$^{+0.8}_{-0.8}$ 	  &  	 0.000 	 \\
12 	 & 	 MMJ114101.65-262347.14 	  &  	   7.16 	  &  	 5.7$\pm$0.8 	  &  	 5.2$^{+0.8}_{-0.8}$ 	  &  	 0.000 	 \\
13 	 & 	 MMJ114113.33-263259.16 	  &  	   6.68 	  &  	 5.3$\pm$0.8 	  &  	 4.8$^{+0.8}_{-0.8}$ 	  &  	 0.000 	 \\
14 	 & 	 MMJ114102.20-262732.21 	  &  	   5.60 	  &  	 4.0$\pm$0.7 	  &  	 3.6$^{+0.7}_{-0.8}$ 	  &  	 0.000 	 \\
15 	 & 	 MMJ114031.13-262240.44 	  &  	   5.58 	  &  	 5.0$\pm$0.9 	  &  	 4.3$^{+0.9}_{-0.9}$ 	  &  	 0.000 	 \\
16 	 & 	 MMJ114019.41-263146.52 	  &  	   5.58 	  &  	 4.9$\pm$0.9 	  &  	 4.2$^{+0.9}_{-0.9}$ 	  &  	 0.000 	 \\
17 	 & 	 MMJ114056.79-262644.48 	  &  	   5.12 	  &  	 3.6$\pm$0.7 	  &  	 3.2$^{+0.7}_{-0.8}$ 	  &  	 0.000 	 \\
18 	 & 	 MMJ114038.43-263152.74 	  &  	   4.92 	  &  	 3.5$\pm$0.7 	  &  	 3.0$^{+0.7}_{-0.7}$ 	  &  	 0.000 	 \\
19 	 & 	 MMJ114034.01-263505.27 	  &  	   4.91 	  &  	 4.0$\pm$0.8 	  &  	 3.4$^{+0.8}_{-0.8}$ 	  &  	 0.000 	 \\
20 	 & 	 MMJ114023.25-263022.29 	  &  	   4.62 	  &  	 3.6$\pm$0.8 	  &  	 3.1$^{+0.8}_{-0.8}$ 	  &  	 0.000 	 \\
21 	 & 	 MMJ114058.35-263450.33 	  &  	   4.60 	  &  	 3.6$\pm$0.8 	  &  	 3.0$^{+0.8}_{-0.8}$ 	  &  	 0.000 	 \\
22 	 & 	 MMJ114041.40-262544.02 	  &  	   4.60 	  &  	 3.2$\pm$0.7 	  &  	 2.7$^{+0.7}_{-0.7}$ 	  &  	 0.000 	 \\
23 	 & 	 MMJ114111.53-263610.73 	  &  	   4.52 	  &  	 4.6$\pm$1.0 	  &  	 3.6$^{+1.0}_{-1.0}$ 	  &  	 0.000 	 \\
24 	 & 	 MMJ114108.41-262410.87 	  &  	   4.46 	  &  	 3.5$\pm$0.8 	  &  	 2.9$^{+0.8}_{-0.8}$ 	  &  	 0.000 	 \\
25 	 & 	 MMJ114020.39-262825.87 	  &  	   4.45 	  &  	 3.8$\pm$0.9 	  &  	 3.1$^{+0.9}_{-0.8}$ 	  &  	 0.000 	 \\
26 	 & 	 MMJ114032.18-263243.43 	  &  	   4.32 	  &  	 3.1$\pm$0.7 	  &  	 2.6$^{+0.7}_{-0.7}$ 	  &  	 0.000 	 \\
27 	 & 	 MMJ114120.90-262622.62 	  &  	   4.24 	  &  	 4.1$\pm$1.0 	  &  	 3.2$^{+0.9}_{-1.0}$ 	  &  	 0.001 	 \\
28 	 & 	 MMJ114111.07-263452.90 	  &  	   4.20 	  &  	 3.6$\pm$0.9 	  &  	 2.8$^{+0.9}_{-0.9}$ 	  &  	 0.001 	 \\
29 	 & 	 MMJ114016.13-263105.22 	  &  	   4.20 	  &  	 4.1$\pm$1.0 	  &  	 3.2$^{+1.0}_{-1.0}$ 	  &  	 0.001 	\\
30 	 & 	 MMJ114057.47-262940.84 	  &  	   4.08 	  &  	 2.9$\pm$0.7 	  &  	 2.4$^{+0.7}_{-0.7}$ 	  &  	 0.001 	 \\
31 	 & 	 MMJ114042.27-262710.52 	  &  	   4.05 	  &  	 2.9$\pm$0.7 	  &  	 2.3$^{+0.7}_{-0.7}$ 	  &  	 0.001 	 \\
32 	 & 	 MMJ114024.41-262807.92 	  &  	   4.04 	  &  	 3.1$\pm$0.8 	  &  	 2.5$^{+0.8}_{-0.8}$ 	  &  	 0.001 	 \\
33 	 & 	 MMJ114109.50-262237.90 	  &  	   3.82 	  &  	 3.6$\pm$0.9 	  &  	 2.7$^{+0.9}_{-1.0}$ 	  &  	 0.003 	 \\
34 	 & 	 MMJ114118.70-263349.80 	  &  	   3.66 	  &  	 3.5$\pm$1.0 	  &  	 2.5$^{+1.0}_{-1.0}$ 	  &  	 0.005 	 \\
35 	 & 	 MMJ114048.08-263314.27 	  &  	   3.61 	  &  	 2.6$\pm$0.7 	  &  	 2.0$^{+0.8}_{-0.7}$ 	  &  	 0.003 	 \\
36 	 & 	 MMJ114021.69-263523.07 	  &  	   3.59 	  &  	 3.5$\pm$1.0 	  &  	 2.5$^{+1.0}_{-1.0}$ 	  &  	 0.006 	 \\
37 	 & 	 MMJ114112.46-263023.03 	  &  	   3.58 	  &  	 2.8$\pm$0.8 	  &  	 2.1$^{+0.8}_{-0.8}$ 	  &  	 0.004 	 \\
38 	 & 	 MMJ114046.48-263420.19 	  &  	   3.57 	  &  	 2.7$\pm$0.8 	  &  	 2.1$^{+0.8}_{-0.8}$ 	  &  	 0.004 	 \\
39 	 & 	 MMJ114024.87-262519.63 	  &  	   3.56 	  &  	 2.8$\pm$0.8 	  &  	 2.2$^{+0.8}_{-0.8}$ 	  &  	 0.004 	 \\
40 	 & 	 MMJ114112.65-262508.13 	  &  	   3.53 	  &  	 2.8$\pm$0.8 	  &  	 2.1$^{+0.8}_{-0.8}$ 	  &  	 0.005 	 \\
41 	 & 	 MMJ114026.19-262910.95 	  &  	   3.49 	  &  	 2.6$\pm$0.8 	  &  	 2.0$^{+0.8}_{-0.8}$ 	  &  	 0.005 	 \\
42 	 & 	 MMJ114026.84-263535.05 	  &  	   3.44 	  &  	 3.2$\pm$0.9 	  &  	 2.2$^{+1.0}_{-0.9}$ 	  &  	 0.008 	 \\
43 	 & 	 MMJ114018.60-262422.91 	  &  	   3.32 	  &  	 3.2$\pm$1.0 	  &  	 2.2$^{+1.0}_{-1.0}$ 	  &  	 0.013 	 \\
44 	 & 	 MMJ114026.15-263343.76 	  &  	   3.32 	  &  	 2.6$\pm$0.8 	  &  	 1.9$^{+0.8}_{-0.8}$ 	  &  	 0.009 	 \\
45 	 & 	 MMJ114048.26-262740.46 	  &  	   3.19 	  &  	 2.3$\pm$0.7 	  &  	 1.7$^{+0.7}_{-0.8}$ 	  &  	 0.011 	 \\
46 	 & 	 MMJ114108.01-263532.23 	  &  	   3.08 	  &  	 2.7$\pm$0.9 	  &  	 1.8$^{+0.9}_{-0.9}$ 	  &  	 0.020 	 \\
47 	 & 	 MMJ114106.83-262519.65 	  &  	   3.02 	  &  	 2.2$\pm$0.7 	  &  	 1.5$^{+0.8}_{-0.7}$ 	  &  	 0.018 	\\
\hline
\end{tabular}
\end{center}
\end{table*}

\subsection{Number of false detections}
\label{subsec:fdr}

\begin{figure}
\centering
\includegraphics[width=0.5\textwidth, trim=1cm 13cm 1cm 2cm, clip]{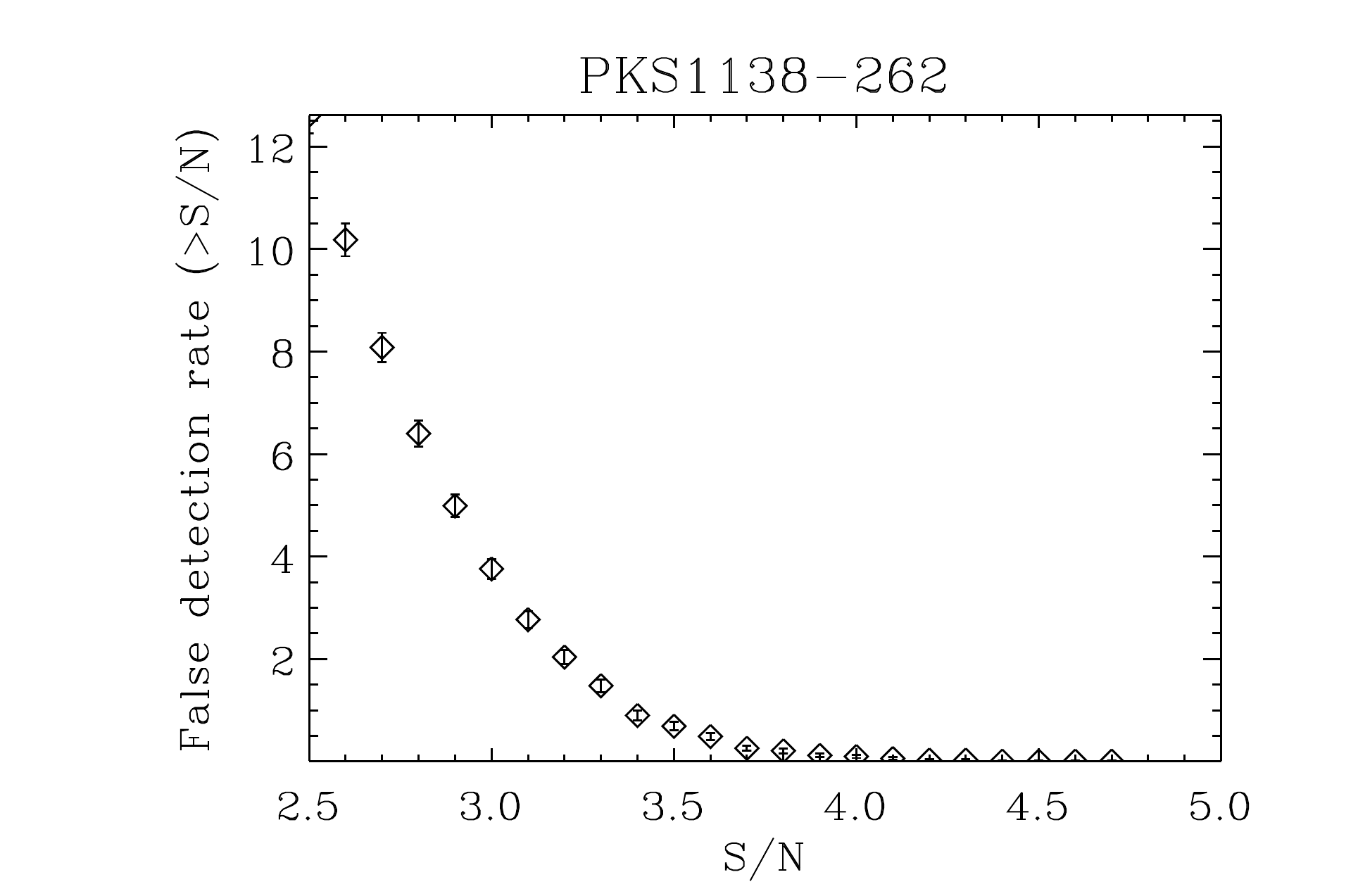}
\caption{Expected number of false detections as a function of limiting S/N produced by jackknifing the timestream data towards the field of PKS1138-262. Error bars denote 68\% intervals for a Poisson distribution. There are 47 detections in this field with S/N $> 3$, and according to the plot, at most 4 of these sources are false detections.}
\label{FDR0802}
\end{figure}

Given the modest S/N of the source candidates, some fraction of the AzTEC sources are expected to be spurious. We identify the number of source detections extracted from the set of noise-only realisations, produced by jackknifing the timestream data of each protocluster field, in order to estimate the number of positive noise peaks that the source detection algorithm would pick up as source candidates. The expected number of false detections as a function of limiting S/N is shown in Figure~\ref{FDR0802} for the field towards PKS1138-262. Appendix \ref{appendixA} shows similar plots for the rest of the sample.

These false detection rates (FDRs) are only upper limits since the number of high-significance positive noise peaks in the signal map decreases because of the existence of real sources, which causes a negative bias in the pixel flux-density distribution of the signal map due to the negative side-lobes of the point source kernel. This effect was first demonstrated for the AzTEC/GOODS-N survey (\citealp{Perera08}) and determined to be particularly strong for maps like ours, with depths below the confusion limit (\citealp{ScottKS10}).

\subsection{Completeness}
\label{subsec:compl}

\begin{figure}
\centering
\includegraphics[width=0.5\textwidth, trim=1cm 13cm 1cm 2cm, clip]{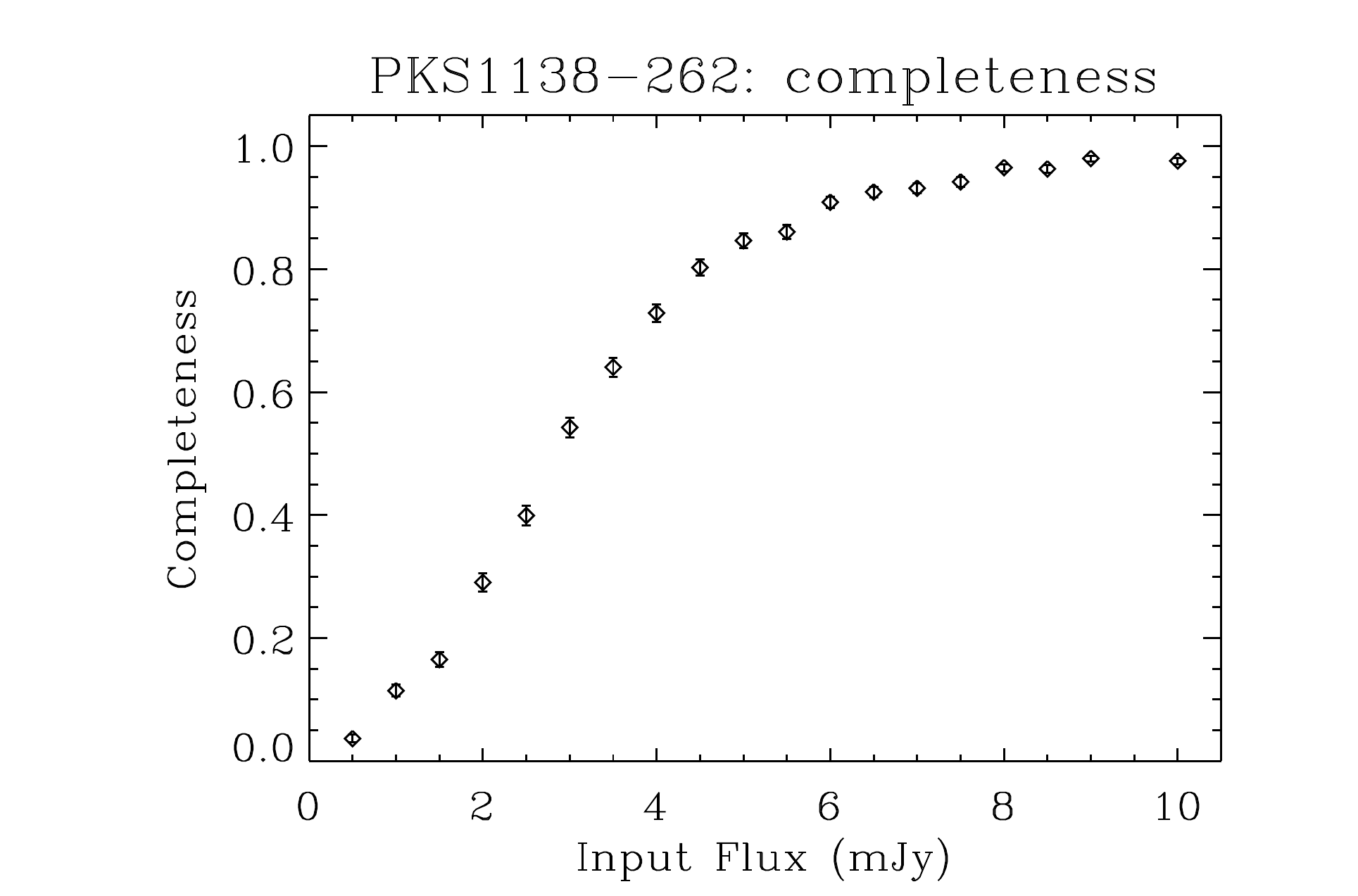}
\caption{Completeness estimation for AzTEC sources in the ACES protocluster field towards PKS1138-262. The data points and 68\% confidence binomial error bars show the completeness estimated by inserting sources of known flux density one at a time into the real signal map. From the image it can be seen that for sources with flux densities $>4$ mJy the completeness is $ >60\%$. According to noise properties of the PKS1138-262 map, sources with S/N $>$ 3 have flux densities $>$ 2.1 mJy.}
\label{Comp0802}
\end{figure}

The detection probability for a given source is affected by both Gaussian random noise and confusion noise from the underlying faint sources. To account for both effects, map completeness is estimated by injecting a total of 1,000 fake sources per flux bin (ranging from 0.5 - 20 mJy), one at a time, into the signal map at random positions, and then checking if they are retrieved by the source identification algorithm of section \ref{subsec:cat}. Adding one source at a time to the signal map provides a valid estimate of the completeness because it accounts for the effects of both random and confusion noise present in the signal map \citep{ScottKS10}. The input positions are restricted to be farther than 17 (10) arcsec from any real source in the ASTE (JCMT) maps, and real sources are defined as having S/N $>4$ in the shallowest maps (TNJ1338-1942 \& 4C+41.17) and S/N $>3.5$ in the rest. Otherwise, the result could be biased because the detection algorithm cannot distinguish between two sources that close. Figure~\ref{Comp0802} shows the map completeness for the field towards PKS1138-262. The plots for the rest of the sample are shown in appendix \ref{appendixA}.

\subsection{Astrometry}
\label{subsec:astro}

Although ACES observations have been corrected for small pointing errors by periodically targeting well-known bright point sources, there is a possibility of a remaining systematic offset in the maps. Without very bright objects on each of the ACES protocluster fields to use as pointing references, or a large catalogue of fainter objects (with high positional certainty such as radio sources) to do individual stacking analysis on each of the signal maps, an overall pointing offset is estimated applying a similar technique as in \cite{Wilson08b}. For the ASTE maps, we averaged the positional offsets of the 9 radio galaxies detected at 1.1 mm; and for the single JCMT map, we averaged positional offsets of 3 AzTEC sources previously observed with submm interferometry (SMA follow-up program, P.I.~D.~H.~Hughes) and the central radio galaxy 4C+41.17.

\begin{figure}
\includegraphics[width=0.5\textwidth, trim=1cm 6cm 1cm 6cm, clip]{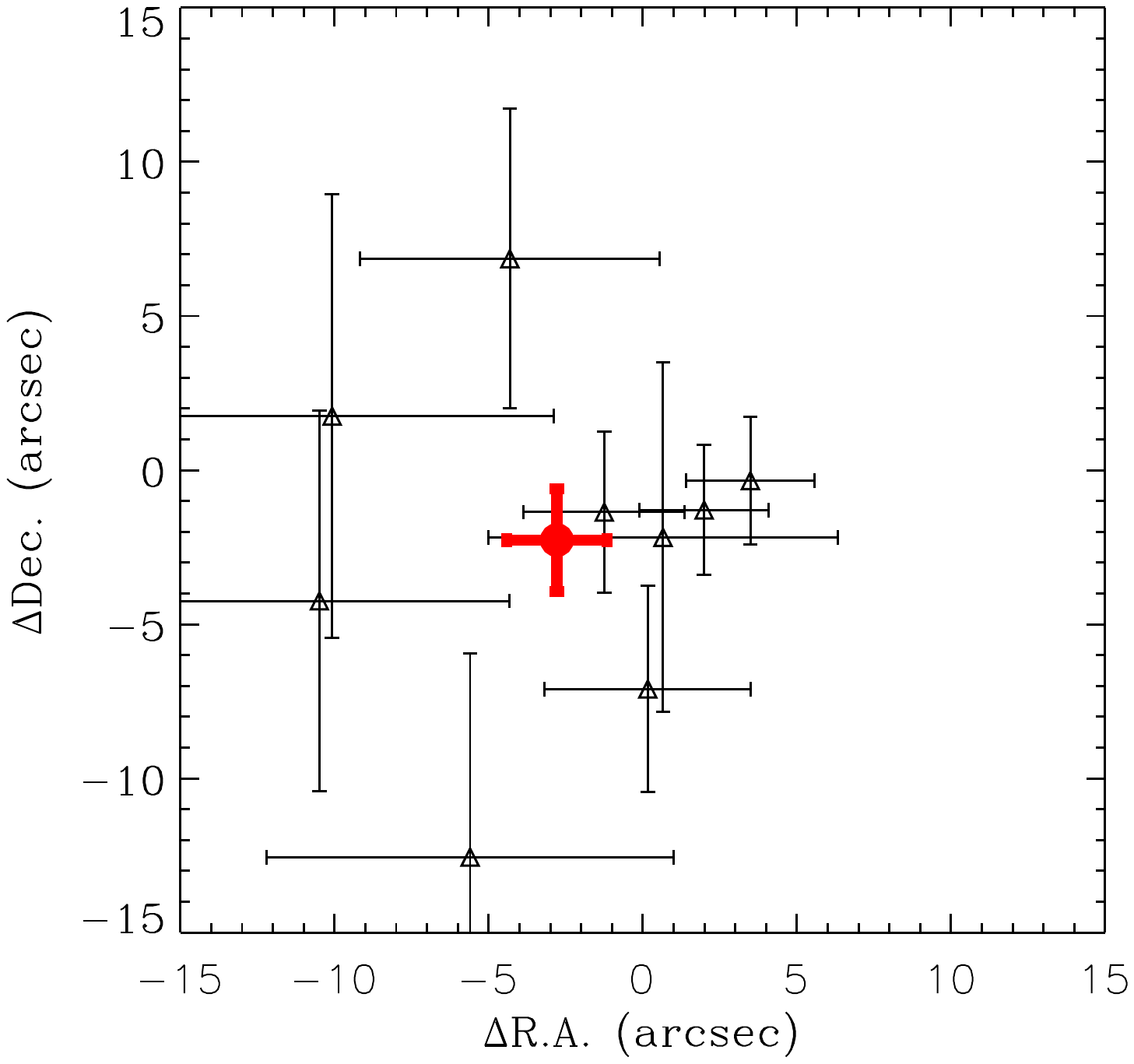}
\includegraphics[width=0.5\textwidth, trim=1cm 6cm 1cm 6cm, clip]{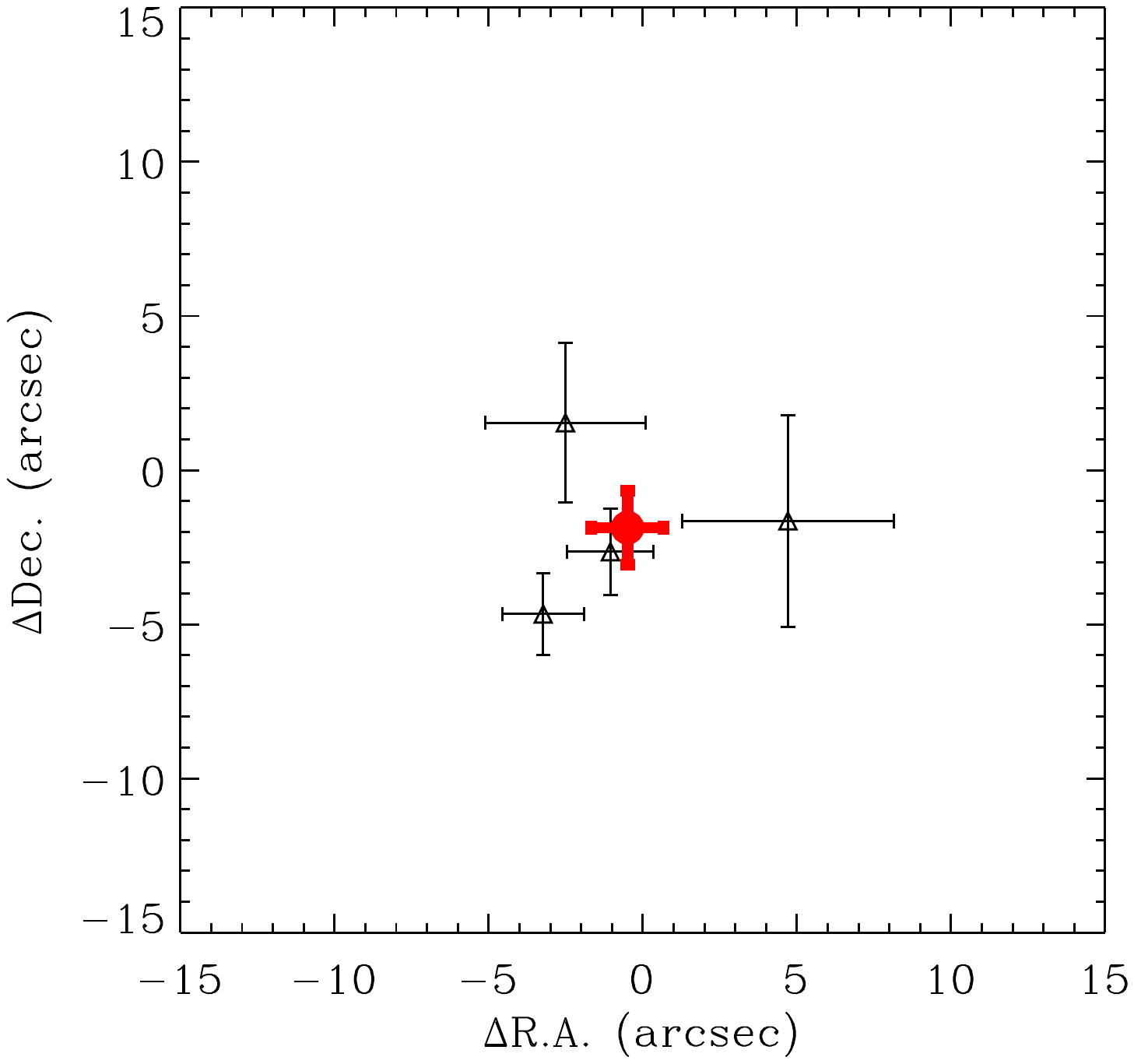}
\caption{Top: Measured pointing offsets in right ascension and declination for 9 radio galaxies detected at 1.1 mm with ASTE. Errors bars represent their positional uncertainties. The thick circle marks the pointing offset and its positional uncertainty estimated by averaging the radio galaxies offsets. No systematic pointing error was found and therefore no correction was applied. Bottom: Measured pointing offsets for 3 SMGs detected with SMA and the central radio galaxy in the 4C+41.17 map taken with the JCMT. The thick circle represents the mean pointing offset and its positional uncertainty. Again, no systematic pointing error was found and therefore no correction was applied.}
\label{astrometryplot}
\end{figure}

The top panel of Figure~\ref{astrometryplot} shows the measured offsets in right ascension and declination for the 9 ASTE radio galaxies. The thick circle represents the mean pointing offset ($\Delta$R.A. = -2.8 arcsec , $\Delta$Dec. = -2.3 arcsec) and 1$\sigma$ error bars (1.6 arcsec in R.A. and Dec.). This is consistent with no systematic pointing error and therefore no correction was applied. The bottom panel of Figure~\ref{astrometryplot} shows the measured offsets for the 3 SMGs detected with SMA and the central radio galaxy in the JCMT map. Again, the thick circle represents the mean pointing offset ($\Delta$ R.A. = -0.5 arcsec, $\Delta$ Dec. = -1.9 arcsec) and 1$\sigma$ error bars (1.2 arcsec in R.A. and Dec.), and no correction was applied.

\begin{figure}
\centering
\includegraphics[width=0.5\textwidth, trim=1cm 7cm 1cm 5cm, clip]{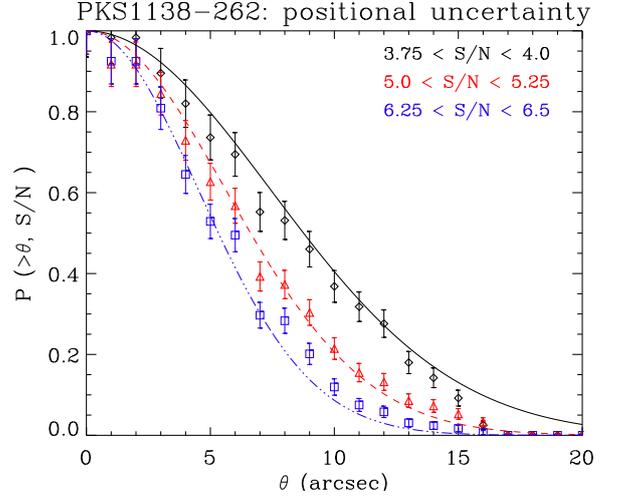}
\caption[]{Positional uncertainty distribution for PKS1138-262 source candidates. The data-points and error bars show the probability P($>\theta$; S/N) that a source detected with a given S/N ratio will be found outside a radial distance $\theta$ from its true location. The curves show the corresponding analytical expression derived in \cite{Ivison07}.}
\label{posuncer0802}
\end{figure}

Random and confusion noise is another source of positional uncertainty. It can cause the peak of a detection to move away from its original location. This effect is considerably notorious for large beam surveys and depends on the S/N ratio of the detection (e.g.~\citealp{Ivison07}). We perform simulations for each ACES protocluster field to determine the probability for a certain source to be displaced some arcseconds away from its original location. The simulations consist of inserting sources with known flux densities in the signal map (one at a time) and determining how many arcseconds away they are recovered by the source detection algorithm. We repeat this procedure 1,000 times for different flux-density bins in the range of 1-20 mJy to obtain a distribution of input-to-output source distances as a function of detected S/N. The probability P ($>\theta$; S/N) that a source will be detected outside of a radial distance $\theta$ of its true position in our map towards PKS1138-262 is shown in Figure~\ref{posuncer0802}. The different symbols show results for three different S/N bins.

\cite{Ivison07} derived an analytical expression for the positional uncertainty of low-S/N sources and found good agreement with the distribution of positional errors measured for the population of submm sources in the SCU\-BA/\-SHA\-DES survey that had radio counterparts. They reported Gaussian profiles for the R.A. and Dec. uncertainties, with $\sigma$ equal to $\Delta \alpha = \Delta \delta = 0.6 \times FWHM \times S/N^{-1}$, where $\Delta \alpha$ and $\Delta \delta$ represent R.A. and Dec. uncertainties respectively, FWHM the full width half maximum of the beam, and S/N the signal-to-noise ratio of the submm source. Thus, the radial offsets follow a profile of the form $\theta e^{-\theta^2/2\sigma^2}$, for which 68\% (95.6\%) of offsets are expected to lie within $1.51\sigma$ ($2.5\sigma$). To check whether our ACES positional uncertainties are consistent with this theoretical expectation, the cumulative probability of the radial offsets is over-plotted in Figure~\ref{posuncer0802} for the same S/N bins given the size of the ASTE beam. The curves show that the analytical expression and the empirical distributions follow roughly the same trend. The positional uncertainty distributions for the other ACES protocluster maps can be found in appendix \ref{appendixA}.

\section{Source counts}
\label{sec:nc}

\subsection{Flux deboosting}
\label{subsec:fluxdeb}

Our AzTEC maps are characterised by low S/N detections and an underlying flux density distribution of sources whose estimated counts have a steep shape \citep{ScottKS12}. Therefore, sources blindly detected in these maps have measured flux densities biased towards higher values (i.e.~flux boosting; \citealp{HoggTurner98}). This can be corrected by constructing the full posterior flux density distribution (PFD) for each source taking as a prior the parameters for a Schechter function that best characterises the underlying distribution of sources. We use N$_{\rm{3 mJy}}$ = 230 deg$^{-2}$ , S' $= 1.7$ mJy and $\alpha = -2.0$, which are the parameters measured for the source counts of SMGs at 1.1 mm towards blank fields \citep{ScottKS12}.

PFDs are derived for all $> 2.5\sigma$ peaks in the ASTE (JCMT) catalogues, and source candidates are selected from this sample if their PFDs show $<5\%$ ($<10\%$) probability of having negative intrinsic flux density. The 5\% threshold has been traditionally used in previous SCUBA \citep{Coppin06} and AzTEC surveys \citep{Perera08,Austermann09} to limit the number of false detections to a near negligible amount. For our JCMT field, however, a 5\% threshold would limit the AzTEC source candidate list to just those with S/N $> 3.7$. \cite{Austermann10} verified through simulations that the use of a higher threshold supplies additional data without introducing any significant biases in the number counts analysis. Therefore, we increase the threshold to 10\% for this particular field.

Since we are using a blank-field prior in fields where overdensities are expected, we examine how dependent the PFD estimation is on the prior. We found that PKS1138-262 is the ACES protocluster field that, on average, has the most deviant source counts with respect to the AzTEC blank-field data we discuss this point further in section \ref{subsec:indnc}). We estimated its overdensity to be as high as 2 by determining the difference, per flux bin, between the differential source counts in the protocluster field and the reference field, and calculating the mean difference for all bins. Then, we used twice the source counts estimated for the AzTEC blank-field data as a prior to determine new PFDs for all the ACES-protocluster sources, and found that the variation in the estimated deboosted fluxes is, on average, $7 \pm 3$ per cent for sources with $3<$ S/N $< 4.0$. These flux-density changes are smaller than the deboosted flux errors, therefore the flux densities estimated with the initial prior are considered good estimates of the intrinsic flux densities of ACES sources.

The deboosted flux densities are listed in column 5 of the ACES protocluster catalogues of Table~\ref{cluster0802cat} and appendix \ref{appendixA}.

\subsection{AGN counterparts}
\label{subsec:agn}

Since we are interested in the number density of SMGs around the ACES AGN, we need to remove possible 1.1 mm counterparts to the active galaxies from our catalogues. We follow a similar approach as in \cite{Humphrey11} and use the AGN coordinates that are most accurate as determined from radio, optical or mid(near)-infrared data. These coordinates are the most likely to mark the position of the radio core. For AGN with no detected radio cores, the preference is to use the centroid of the emission averaged across the 3.6 and 4.5 $\mu$m bands, taken by the Infrared Array Camera (IRAC) on board the \emph{Spitzer} space telescope \citep{Fazio04}. If these data are not available either, the longest wavelength detection in the optical or near infrared is used instead (Table~\ref{ACESRGtable_radio}). Then, we select the closest AzTEC source that has a S/N $> 2.5$ and whose PFD has a probability of having negative intrinsic flux less than the threshold established in the previous section. \cite{Humphrey11}, on the other hand, adopt a selection criteria based only on a S/N $\geq 3.0$. Both approaches minimize the possibility of assigning false detections as AGN counterparts with similar results, and both find possible 1.1 mm detections to 10 ACES AGN. In the case of TNJ2007-1316, however, the closest AzTEC source is a marginal detection with a S/N = 3.0 but a 5.9\% probability of having negative flux. Therefore, it is considered a false positive by us but regarded as counterpart by \cite{Humphrey11}.

Once the possible mm counterpart is identified at a certain distance $\theta$, we use the theoretical approach of \cite{Ivison07}, which is in good agreement with the positional uncertainty distributions found in section \ref{subsec:astro}, to estimate the probability for this counterpart to have been moved a distance $> \theta$ due to confusion and random noise (i.e.~P($> \theta$; S/N)). If that probability is $\leq 5\%$, then the mm detection is discarded as the possible AGN counterpart. As can be seen from the Table~\ref{ACESRG_counterparts}, 10 AGN have possible mm counterparts in the deboosted catalogues.

\begin{table*}
\begin{center}
\caption{\label{ACESRG_counterparts} Possible AGN counterparts at 1.1 mm: 1) AGN name; 2), 3) R.A. and Dec. coordinates of the closest mm source; 4) S/N of the mm source, 5) distance between the AGN and their closest mm source; 6) probability for the mm source to have been moved a distance $> \theta$ due to random and confusion noise; and 7) probability for the mm source to be randomly associated to the AGN (P-statistics). Rows highlighted in bold mark AGN with likely mm counterparts.}
\begin{tabular}{l c c c c c c}
\hline
AGN 		&\multicolumn {6}{c} {CLOSEST 1.1-MM SOURCE}\\
\hline
AGN Name   	&	R.A. (J2000) 	&  Dec. (J2000)	& S/N 	& $\theta$   & $P(> \theta;S/N)$	&	P-stat \\
         			&	    (deg)  		&     (deg)	  	&		& (arcsec)   & ($\%$)              		&     ($\%$) \\
\hline
SDSSJ1030+0524 	 	 &10:30:26.68 	 	 	&+05:25:15.98 	  	  	&6.2  		 &21.9 	  	  & 00.0 	  	   & 3.1\\
\textbf{TNJ0924-2201} 	 &\textbf{09:24:20.31} 	&\textbf{-22:01:28.95}	&\textbf{4.0} 	 &\textbf{13.8}	  &\textbf{11.4} 	   & \textbf{1.6}\\
TNJ1338-1942 		 &13:38:27.35 		 	&-19:42:25.66 		 	&6.3  		 &17.7 	  	  & 00.0 		   & 0.3\\
TNJ2007-1316 		 &20:07:51.03 		 	&-13:15:29.15 		  	&4.0  		 &80.9 	  	  & 00.0 		   &42.2\\
\textbf{4C+41.17} 	 	 &\textbf{06:50:51.93} 	&\textbf{+41:30:33.04}   	&\textbf{3.5} 	 &\textbf{ 5.0}	  &\textbf{35.7} 	   &\textbf{0.4}\\
\textbf{TNJ2009-3040} 	 &\textbf{20:09:48.89} 	&\textbf{-30:40:03.16}   	&\textbf{3.6} 	 &\textbf{11.3}	  &\textbf{18.1} 	   &\textbf{1.5}\\
MRC0316-257 		 	 &03:18:13.14 		 	&-25:35:08.26 		  	&5.5  		 &13.6 		  & 00.3 		   & 1.1\\
\textbf{PKS0529-549} 	 &\textbf{05:30:25.20} 	&\textbf{-54:54:22.02}   	&\textbf{9.7} 	 &\textbf{ 2.4}	  &\textbf{53.4} 	   &\textbf{0.0}\\
\textbf{MRC2104-242} 	 &\textbf{21:06:58.32} 	&\textbf{-24:05:15.96}   	&\textbf{4.3} 	 &\textbf{ 8.2}	  &\textbf{23.7} 	   &\textbf{0.5}\\
4C+23.56 		 	 &21:07:15.63 		 	&+23:31:33.16 		  	&7.7	 		 &16.3 		  & 00.0 		   & 0.9\\
PKS1138-262 	 	  	 &11:40:47.42 		 	&-26:29:10.75 		  	&8.4  		 &12.6 		  & 00.0 		   & 0.2\\
\textbf{MRC0355-037} 	 &\textbf{03:57:48.05} 	&\textbf{-03:34:02.40}   	&\textbf{6.1} 	 &\textbf{ 7.1}	  &\textbf{10.8} 	   &\textbf{0.1}\\
\textbf{MRC2048-272} 	 &\textbf{20:51:04.24} 	&\textbf{-27:03:05.45}   	&\textbf{3.0} 	 &\textbf{10.2}	  &\textbf{36.3} 	   &\textbf{2.6}\\
\textbf{TXS2322-040} 	 &\textbf{23:25:10.19} 	&\textbf{-03:44:44.52}   	&\textbf{3.6} 	 &\textbf{ 2.3}	  &\textbf{92.5} 	   &\textbf{0.1}\\
MRC2322-052 	 	 	 &23:25:18.38 		 	&-04:57:27.48 		  	&3.34 		 &20.5 		  & 00.4 		   & 8.6\\
\textbf{MRC2008-068} 	 &\textbf{20:11:13.99} 	&\textbf{-06:44:03.26}   	&\textbf{9.8} 	 &\textbf{ 3.4}	  &\textbf{27.1} 	   &\textbf{0.0}\\
\textbf{MRC2201-555} 	 &\textbf{22:05:04.97} 	&\textbf{-55:17:42.64}   	&\textbf{7.8} 	 &\textbf{ 1.8}	  &\textbf{79.8} 	   &\textbf{0.0}\\
\hline
\end{tabular}
\end{center}
\end{table*}

In addition, we use the \emph{P-statistic}, $P(r) = 1 - e^{-n \pi r^2}$, to determine whether the possible mm counterpart is associated with the AGN only by chance \citep{Downes86}.  For \emph{n} we use the surface density of mm sources, as determined from the AzTEC blank-field number counts, that have 1.1 mm flux densities greater than or equal to that of the possible counterpart. We assume a blank field population, despite the fact that the environments of our AGN sample possibly contain source clustering or overdensity, because our goal is to find the probability of random association between the AGN and a foreground/background SMG, whose population is better described by a blank field distribution of sources. The AGN-mm association is considered to be random when P $> 5\%$. As can be seen from Table~\ref{ACESRG_counterparts}, each of the 10 previously selected mm counterparts for the AGN are not likely to be random associations.

\subsection{Source counts for individual fields}
\label{subsec:indnc}

Once the possible counterparts to the ACES AGN are removed from the catalogues, we derive estimates for the number density of SMGs as a function of flux density using the Bayesian technique originally outlined in \cite{Coppin05,Coppin06} and used extensively in previous AzTEC publications (e.g.~\citealp{Austermann09,Austermann10,Aretxaga11,ScottKS12}). The PFDs of the source candidates are randomly sampled (with replacement) to determine their intrinsic flux densities. They are sampled 20,000 times and the mock-sources are binned by their flux density (binsize $= 1$ mJy). Each bin in each iteration is then corrected for incompleteness. The source counts are calculated as the mean number of sources in each bin over the 20,000 iterations, and the uncertainties represent the 68\% confidence intervals calculated from the distribution in the counts across those iterations. For the differential counts, the flux densities are the effective bin centres weighted by the assumed prior, and for the integrated counts, the flux densities are the bin edges with the lowest flux-density.

\begin{figure}
\includegraphics[width=0.5\textwidth, trim=3.5cm 9cm 3.5cm 9cm, clip]{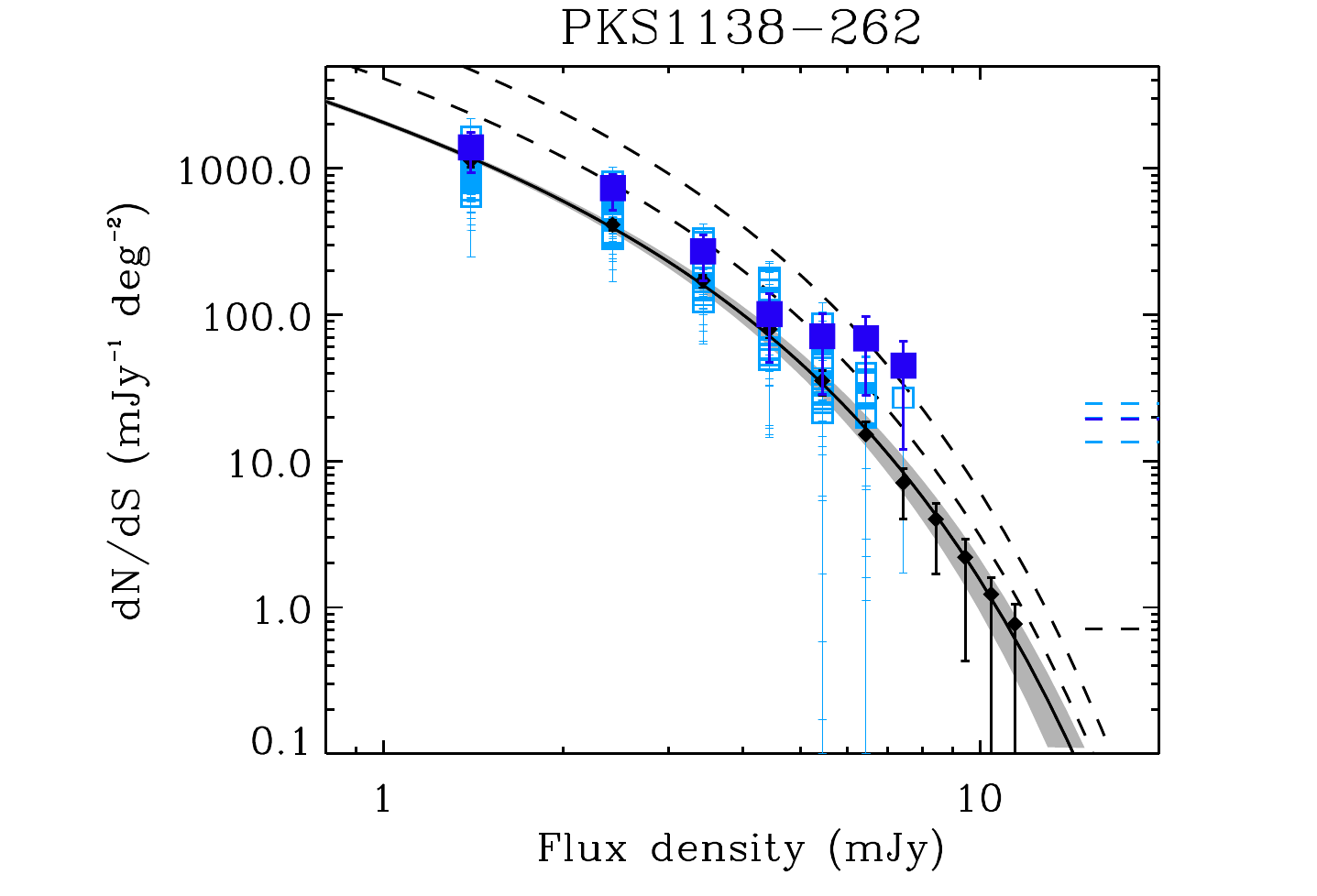}
\includegraphics[width=0.5\textwidth, trim=3.5cm 9cm 3.5cm 9cm, clip]{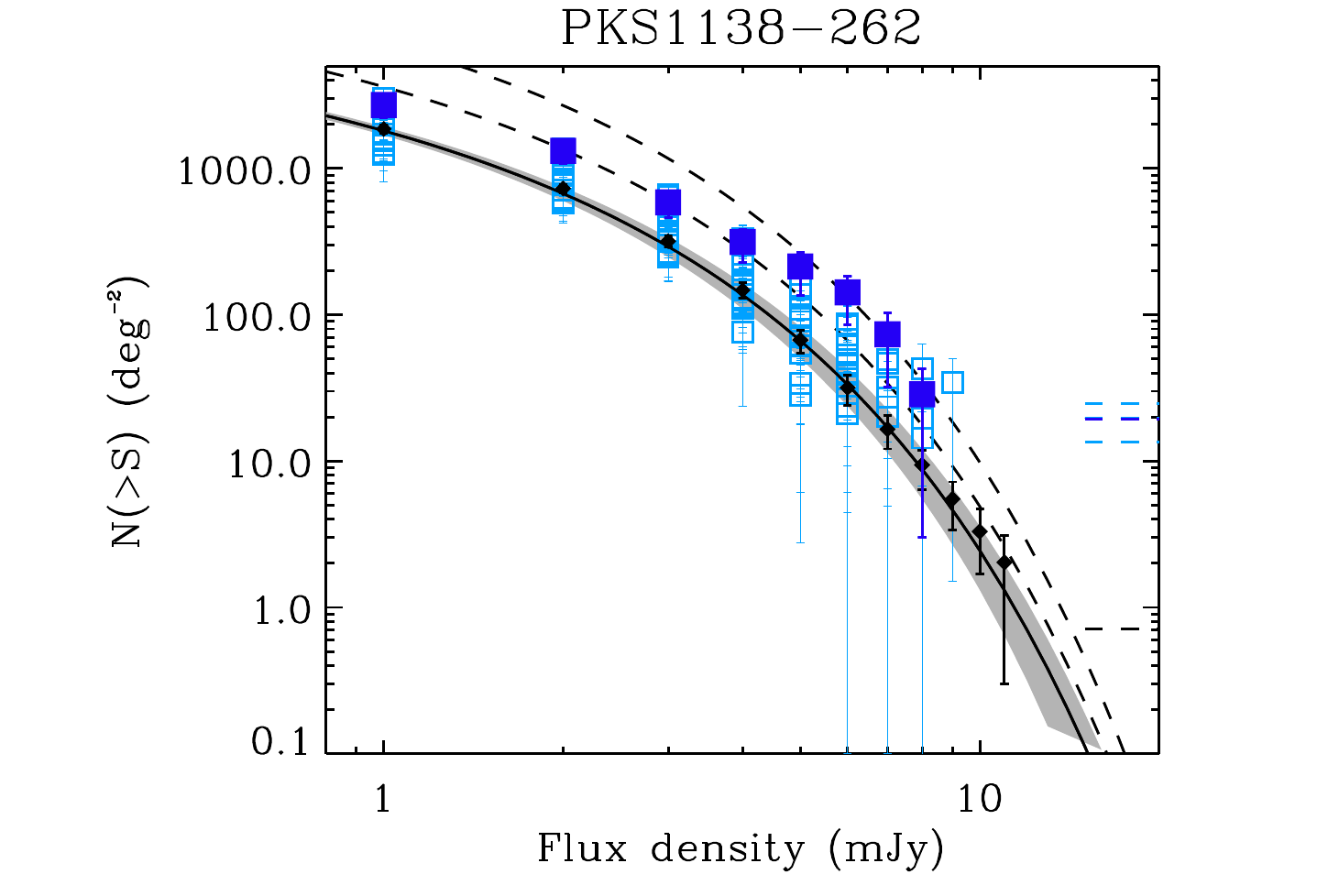}
\caption{AzTEC 1.1 mm differential and integrated source counts for the field of PKS1138-262 (solid squares). Source counts for the AzTEC blank fields used as reference are also shown (diamonds; \citealp{ScottKS12}). The solid line and grey shading represent the best fit of a Schechter function to the reference counts and its 68\% confidence interval. Overplotted in light colour are the counts for the other 16 ACES protocluster fields. This is to show the scattering in the data that is mainly due to sample variance. Dashed lines show twice and four times the reference-field density of sources. Horizontal dashed lines represent the survey-limit, defined as the source density (inside de map area) that will Poisson deviate to zero sources 32\% of the time.}
\label{indNC0802}
\end{figure}

\begin{figure}
\centering
\includegraphics[width=0.5\textwidth, trim=3.5cm 8.5cm 3.5cm 9cm, clip]{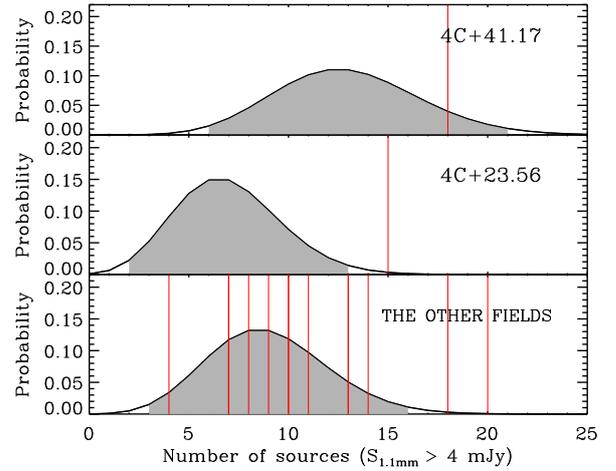}
\caption[]{Poisson probability distributions of the number of sources with S$_{\rm{1.1mm}} > 4$ mJy expected for blank-field maps with sizes similar to the ACES protocluster maps. The 95\% confidence intervals are represented by gray shaded areas. Since not all the ACES maps have similar sizes, the probability distributions that correspond to the 4C+41.17 (area = 303 sq arcmin) and the 4C+23.56 (area = 166 sq arcmin) maps are plotted separately. The probability distribution for the rest of the protocluster maps, which have individual areas of $\sim 200$ sq arcmin, is plotted at the bottom of the image. The numbers of sources found in each protocluster map are plotted as vertical lines. 14 of the 17 ACES fields show number of sources falling within the 95\% confidence interval of their corresponding blank-field data. Therefore, sample variance cannot be ruled out as being the dominant effect on the large scattering observed in the individual number counts of Figure~\ref{indNC0802}. Fields whose numbers of sources with S$_{\rm{1.1mm}} > 4$ mJy that exceed their 95\% confidence intervals are 4C+23.56, PKS1138-262 and MRC0355-037.}
\label{SampleVariance}
\end{figure}

Table~\ref{indNC0802_table} and Figure~\ref{indNC0802} show the 1.1 mm differential and integrated source counts derived for the field towards PKS1138-262. Appendix \ref{appendixA} shows similar plots and tables for the rest of the sample. As can be seen from the graphs, the counts from different fields show a lot of variance, especially when extreme cases such as PKS1138-262 and TNJ1338-1942 are compared. A look at the 4-mJy flux-density bin, which is above the average $3\sigma$ detection level for all ACES maps, allow a fair comparison among fields with different noise levels. The significant scattering observed in this flux-density bin (and in higher flux-density bins) could be explained by a combination of sample variance, which largely affects small-size maps like these ones, and intrinsic clustering variations. According to the AzTEC blank-field source counts \citep{ScottKS12}, $\sim 9$ sources with S$_{\rm{1.1mm}} > 4$ mJy are expected to populate maps as large as $\sim 200$ sq arcmin ($\sim 7$ in the case of the smaller 166-sq-arcmin map of 4C+23.56 and $\sim 13$ in the case of the larger 303-sq-arcmin map of 4C+41.17). In 14 of the 17 ACES maps, the number of sources with S$_{\rm{1.1mm}} > 4$ mJy fall within the 95\% confidence interval of a Poisson probability distribution of the blank-field counts (Figure~\ref{SampleVariance}). Therefore, sample variance cannot be ruled out as being the dominant effect on the large scattering observed in the individual counts.

In the fields towards 4C+23.56, PKS1138-262 and MRC0355-037, however, the numbers of sources with S$_{\rm{1.1mm}} > 4$ mJy exceeds the number of expected sources in a blank field by a factor of $\sim 2$. The significance of these overdensities is $> 3\sigma$, which means that the probability of finding  overdensities like these ones by chance is $< 0.3\%$.

In order to reduce field-to-field scattering and improve the statistical significance of any overdensity result, the source counts analysis is repeated with all the ACES protocluster fields combined.

\begin{table}
\begin{center}
\caption{\label{indNC0802_table} Differential and integrated source counts calculated for the ACES field towards PKS1138-262.}
\begin{tabular}{cccc}
\hline
Flux density      	&   dN/dS          				&  Flux density		&	N($>$S)\\
(mJy)  		&   (mJy$^{-1}$ deg$^{-2}$) 	& (mJy) 			&       (deg$^{-2}$)\\
\hline 
	  1.4 	 & $ 	 1400 	  ^{+  	 370 	  }_{-  	 450 	 }$ &  	  1.0 	  & $ 	 2700 	  ^{+  	 410 	  }_{-  	 500 	 }$\\[3pt]
	  2.4 	 & $ 	 730 	  ^{+  	 170 	  }_{-  	 210 	 }$ &  	  2.0 	  & $ 	 1320 	  ^{+  	 220 	  }_{-  	 230 	 }$\\[3pt]
	  3.4 	 & $ 	 270 	  ^{+  	 77 	  }_{-  	 99 	 }$ &  	  3.0 	  & $ 	 590 	  ^{+  	 110 	  }_{-  	 120 	 }$\\[3pt]
	  4.4 	 & $ 	 100 	  ^{+  	 38 	  }_{-  	 53 	 }$ &  	  4.0 	  & $ 	 314 	  ^{+  	 68 	  }_{-  	 87 	 }$\\[3pt]
	  5.4 	 & $ 	 71 	  ^{+  	 29 	  }_{-  	 42 	 }$ &  	  5.0 	  & $ 	 213 	  ^{+  	 51 	  }_{-  	 77 	 }$\\[3pt]
	  6.4 	 & $ 	 68 	  ^{+  	 28 	  }_{-  	 40 	 }$ &  	  6.0 	  & $ 	 142 	  ^{+  	 40 	  }_{-  	 57 	 }$\\[3pt]
	  7.4 	 & $ 	 44 	  ^{+  	 20 	  }_{-  	 32 	 }$ &  	  7.0 	  & $ 	 73 	  ^{+  	 29 	  }_{-  	 41 	 }$\\[3pt]
	  8.4 	 & $ 	 10 	  ^{+  	 9 	  }_{-  	 10 	 }$ &  	  8.0 	  & $ 	 28 	  ^{+  	 13 	  }_{-  	 25 	 }$\\[3pt]
\hline
\end{tabular}
\end{center}
\end{table}

\subsection{Source counts for the combined fields}
\label{subsec:cnc}

All 17 catalogues and completeness estimations are combined to produce a single study of the ACES protocluster source counts based on a total surveyed area of 1.01 square degrees. The differential and integrated counts are determined applying the same technique explained in section \ref{subsec:indnc} but taking into account that completeness depends on noise level and it is different for each field. The differential and cumulative counts from the combined biased fields are listed in Table~\ref{combinedNC_table}, together with their 68\% confidence interval uncertainties.

\begin{figure}
\centering
\includegraphics[width=0.5\textwidth, trim=4.8cm 9cm 4.5cm 9cm, clip]{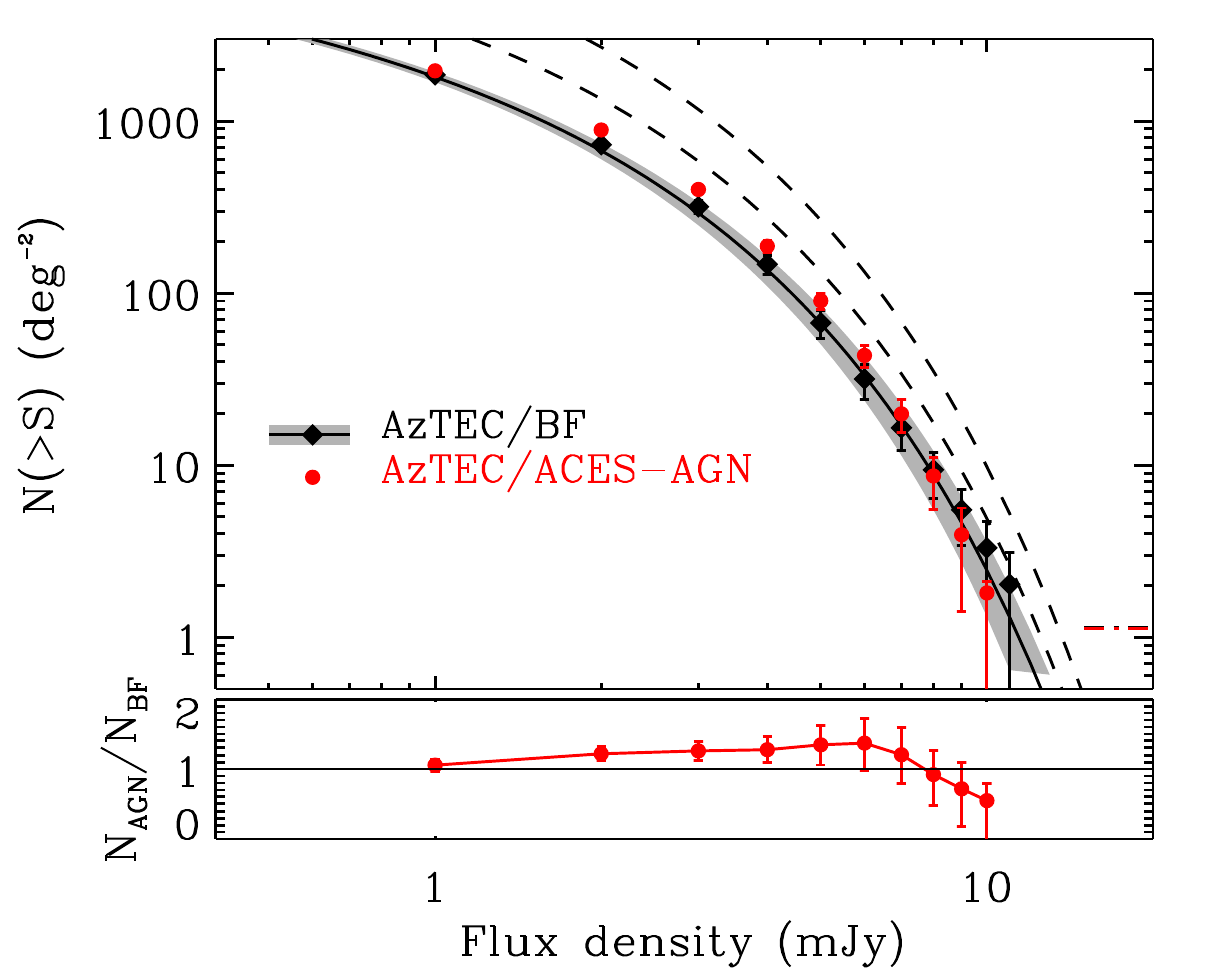} 
\caption{Top: AzTEC 1.1 mm integrated source counts for the combined 17 ACES protocluster fields (circles) along with the AzTEC blank-field data used as reference (diamonds). Dashed lines represent overdensities of 2 and 4 compared to the AzTEC blank-field data. The horizontal dashed line represents the survey limit. Bottom: Ratio between the source counts of the ACES protocluster maps and the source counts that are predicted from the combined AzTEC blank-field data. The ACES protocluster counts lie very close to the blank field data, which may not contradict previous studies around AGN since an overdensity in small areas close to the centre of the protoclusters might have been diluted while estimating an average source density in larger areas.}
\label{combinedNC}
\end{figure}

In order to find out if there is an overdensity of mm sources around the AGN, we compare the ACES protocluster source counts to the ones of the AzTEC blank fields (Figure~\ref{combinedNC}). Overall, the combined ACES protocluster counts (circles) lie very close to the AzTEC blank-field estimation (diamonds). Since the surveyed area for the ACES protocluster project was nearly one square degree, it is possible that if there was an overdensity spanning a few Mpc (a few arcmin) from the centre of the protocluster maps, it was diluted while estimating an average source density in a larger area (see section \ref{subsec:sim_small_overd}). For instance, previous submm surveys around AGN (e.g.~\citealp{Stevens03,Priddey08,Stevens10}) have claimed overdensity detections of $\sim 2-4$ in areas of $\sim 1.5$-arcmin radius (SCUBA field of view).

\begin{table}
\begin{center}
\caption{\label{combinedNC_table} Combined differential and integrated source counts for the 17 ACES protocluster fields.}
\begin{tabular}{cccc}
\hline
Flux density      	&   dN/dS          				&  Flux density		&	N($>$S)\\
(mJy)  		&   (mJy$^{-1}$ deg$^{-2}$) 	& (mJy) 			&       (deg$^{-2}$)\\
\hline 
1.4	&	1071	$^{+	79	}_{-	87	}$ &	1.0	&	1960	$^{+	88	}_{-	96	}$\\ [+3pt]
2.4	&	487	$^{+	34	}_{-	35	}$ &	2.0	&	887	$^{+	40	}_{-	42	}$\\ [+3pt]
3.4	&	212	$^{+	18	}_{-	17	}$ &	3.0	&	400	$^{+	22	}_{-	23	}$\\ [+3pt]
4.4	&	97	$^{+	10	}_{-	11	}$ &	4.0	&	187	$^{+	14	}_{-	15	}$\\ [+3pt]
5.4	&	46	$^{+	7	}_{-	8	}$ &	5.0	&	90	$^{+	9	}_{-	10	}$\\ [+3pt]
6.4	&	23	$^{+	5	}_{-	5	}$ &	6.0	&	43	$^{+	6	}_{-	7	}$\\ [+3pt]
7.4	&	11	$^{+	3	}_{-	4	}$ &	7.0	&	20	$^{+	4	}_{-	5	}$\\ [+3pt]
8.4	&	4.7	$^{+	1.8	}_{-	2.5	}$ &	8.0	&	8.7	$^{+	2.2	}_{-	3.4	}$\\ [+3pt]
9.4	&	2.1	$^{+	1.0	}_{-	1.7	}$ &	9.0	&	3.9	$^{+	1.3	}_{-	2.1	}$\\ [+3pt]
10.4	&	1.2	$^{+	0.6	}_{-	1.2	}$ &	10.0	&	1.8	$^{+	0.8	}_{-	1.3	}$\\ [+3pt]
\hline
\end{tabular}
\end{center}
\end{table}

\subsection{Source counts at different radii from the AGN}
\label{subsec:ncradii}

\begin{figure}
\centering
\includegraphics[width=0.5\textwidth, trim=4.8cm 9cm 4.5cm 9cm, clip]{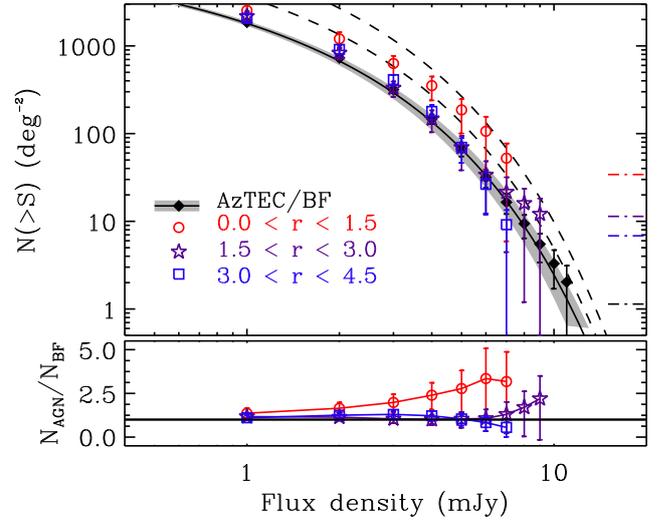} 
\caption{Top: AzTEC 1.1 mm integrated source counts estimated for the combined 17 ACES protocluster fields inside 3 different areas centred at the AGN: an inner circle of 1.5-arcmin radius (circles), and 2 annuli between radii of 1.5 \& 3 arcmin (stars) and between 3 \& 4.5 arcmin (squares). The AzTEC blank-field data used as reference is also plotted as solid diamonds. Beyond a radius of 1.5 arcmin, the source density looks like that of a blank field, and could be also used as a control field. Bottom: ratio between the source counts estimated for the protocluster fields inside the 3 chosen areas and those from the AzTEC blank-field data.}
\label{combinedNC_annuli}
\end{figure}

We test the number density of SMGs in small areas around our AGN, first extracting a circle of 1.5-arcmin radius to let us conduct a direct comparison with previous studies, and then over 2 subsequent annuli between radii of 1.5 and 3 arcmin, and between 3 and 4.5 arcmin (Table~\ref{combinedNC_radii_table}). As can be seen in Figure~\ref{combinedNC_annuli}, source counts for the inner circle ($r < 1'.5$) are indeed larger than those of a typical blank-field, even though the overdensity does not include the AGN 1.1-mm counterparts (section \ref{subsec:agn}). The overdensity reaches a factor $\gtrsim 3$ at S$_{\rm{1.1mm}} \ge 4$ mJy. If we exclude the 3 fields that show individual overdensities (4C+23.56, PKS1138-262 and MRC0355-037), we still detect a global overdensity of a factor of $\sim2$ at these flux densities. Beyond a radius of 1.5 arcmin, the source density looks like that of a blank field, and could be also used as a control field.

\begin{table}
\begin{center}
\caption{\label{combinedNC_radii_table} Combined differential and integrated source counts for the 17 ACES protocluster fields calculated inside 3 different areas centred at the AGN.}
\begin{tabular}{cccc}
\hline
Flux density      	&   dN/dS          				&  Flux density		&	N($>$S)\\
(mJy)  		&   (mJy$^{-1}$ deg$^{-2}$) 	& (mJy) 			&       (deg$^{-2}$)\\
\hline 
\multicolumn{4}{c}{Circular area of 1.5-arcmin radius} \\ [3pt]
1.4 	  & $ 	         1335^{+         441}_{         -540} 	 $ &  	 1.0 	  & $ 	         2538^{+492}_{-608} 	 $ \\ [3pt]
2.4 	  & $ 	          572^{+         179}_{         -221} 	 $ &  	 2.0 	  & $ 	         1203^{+218}_{-279} 	 $ \\ [3pt]
3.4 	  & $ 	          279^{+          96}_{         -125} 	 $ &  	 3.0 	  & $ 	          631^{+125}_{-170} 	 $ \\ [3pt]
4.4 	  & $ 	          166^{+          65}_{         -85} 	 $ &  	 4.0 	  & $ 	          352^{+ 81}_{-116} 	 $ \\ [3pt]
5.4 	  & $ 	           80^{+          36}_{          -59} 	 $ &  	 5.0 	  & $ 	          186^{+ 48}_{-78} 	 $\\ [3pt]
6.4 	  & $ 	           54^{+          26}_{          -47} 	 $ &  	 6.0 	  & $ 	          106^{+31}_{-51} 	 $\\ [3pt]
\multicolumn{4}{c}{Annulus between radii of 1.5 and 3.0 arcmin} \\ [3pt]
1.4 	  & $ 	         1344^{+         262}_{         -301} 	 $ &  	 1.0 	  & $ 	         2176^{+         290}_{         -320} 	 $ \\ [3pt]
2.4 	  & $ 	          501^{+         103}_{         -116} 	 $ &  	 2.0 	  & $ 	          832^{+         120}_{         -130} 	 $ \\ [3pt]
3.4 	  & $ 	          184^{+          46}_{          -58} 	 $ &  	 3.0 	  & $ 	          331^{+          61}_{          -70} 	 $ \\ [3pt]
4.4 	  & $ 	           78^{+          26}_{          -34} 	 $ &  	 4.0 	  & $ 	          148^{+          36}_{          -44} 	 $ \\ [3pt]
5.4 	  & $ 	           36^{+          15}_{          -22} 	 $ &  	 5.0 	  & $ 	           70^{+          24}_{          -30} 	 $ \\ [3pt]
6.4 	  & $ 	           12^{+           6}_{          -12} 	 $ &  	 6.0 	  & $ 	           34^{+           15}_{          -21} 	 $  \\ [3pt]
7.4	  &$		    6^{+	    6}_{	     -6}	 $ &	7.0	&$		22^{+           10}_{          -17} 	 $  \\ [3pt]
\multicolumn{4}{c}{Annulus between radii of 3.0 and 4.5 arcmin} \\ [3pt]
1.4 	  & $ 	         1144^{+         204}_{         -206} 	 $ &  	 1.0 	  & $ 	         2058^{+         223}_{        -233} 	 $\\ [3pt]
2.4 	  & $ 	          500^{+          75}_{          -93} 	 $ &  	 2.0 	  & $ 	          913^{+          98}_{         -107} 	 $\\ [3pt]
3.4 	  & $ 	          234^{+          43}_{          -48} 	 $ &  	 3.0 	  & $ 	          413^{+          53}_{          -61} 	 $\\ [3pt]
4.4 	  & $ 	          109^{+          25}_{          -30} 	 $ &  	 4.0 	  & $ 	          179^{+          33}_{          -35} 	 $ \\ [3pt]
5.4 	  & $ 	           43^{+          14}_{          -19} 	 $ &  	 5.0 	  & $ 	           70^{+          19}_{          -23} 	 $\\ [3pt]
6.4 	  & $ 	           17^{+           8}_{          -12} 	 $ &  	 6.0 	  & $ 	           26^{+          11}_{          -14} 	 $\\ [3pt]
7.4 	  & $ 	            8^{+           4}_{           -8} 	 $ 	    &  	 7.0 	  & $ 	            9^{+           4}_{           9} 	 $\\ [3pt]
\hline
\end{tabular}
\end{center}
\end{table}

\begin{figure}
\centering
\includegraphics[width=0.45\textwidth, trim=2cm 6cm 2cm 5cm, clip]{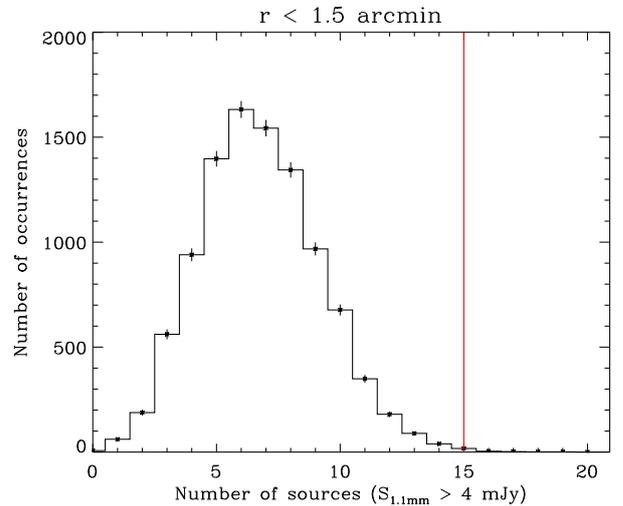}
\caption{Number of sources with S$_{\rm{1.1mm}} > 4$ mJy extracted from the ACES protocluster maps inside an area centred on the AGN and of 1.5 arcmin radius (vertical line) compared to the distribution of the number of sources extracted from 10,000 synthetic maps with similar noise properties to the ACES maps but populated as AzTEC blank fields (histogram). According to this distribution, the probability of finding 15 or more sources is 0.25\%.}
\label{simulatedNC_1p5}
\end{figure}

In order to estimate the significance of the overdensity found in the inner area, we construct synthetic maps with a blank-field source population for each protocluster field. This is performed by populating the noise maps with the number of sources described by the best fit of a Schechter function to the AzTEC blank-field data, properly scaled to the size of the maps, and Poisson deviated to introduce sample variance. We iteratively construct 10,000 maps for each protocluster and extract sources with flux densities $> 4$ mJy inside circular areas of 1.5-arcmin radius placed at the centre of the maps. In each iteration we generate a set of 17 maps corresponding to the 17 protocluster fields. We add up their number of extracted sources, and at the end we can construct the distribution of the number of sources with flux densities $> 4$ mJy that is characteristic of a blank field the size of 17 circles of 1.5-arcmin radius. The source extraction is performed with the same algorithm used in the observed maps, and with a flux density threshold rather than a S/N one because our protocluster maps have different noise levels. Since we compare the number of sources extracted from the observed maps directly to the one extracted from the synthetic maps, no deboosting or completeness correction needs to be applied. The result is shown in Figure~\ref{simulatedNC_1p5}, where a vertical line denotes the 15 sources that were found in the protocluster fields. As can be seen, the probability of finding $\geq 15$ sources with S$_{\rm{1.1mm}} > 4$ mJy in a blank-field area of 120 sq arcmin is 0.25\%, i.e.~the significance of the ACES protocluster overdensity is 99.75\% ($\gtrsim$ 3 $\sigma$).

\subsection{Probability of missing a compact overdensity in large maps}
\label{subsec:sim_small_overd}

\begin{figure}
\centering
\includegraphics[width=0.45\textwidth, trim=2cm 6cm 2cm 5cm, clip]{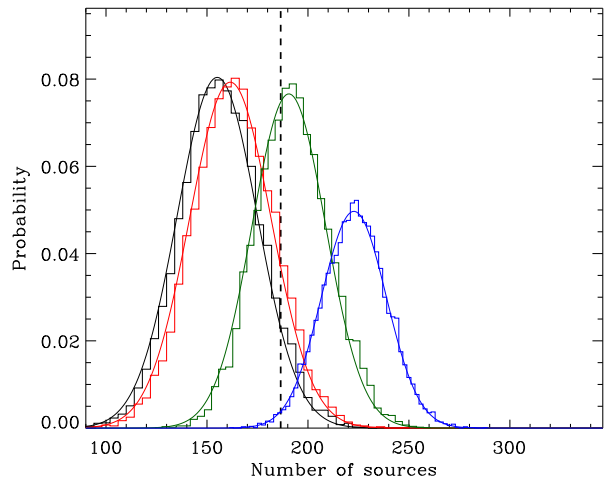}
\caption[]{Distribution of sources with flux densities S$_{\rm{1.1mm}} > 4$ mJy for the 17 ACES-like synthetic maps with only a blank-field population (in black). This distribution peaks close to the 150 sources expected from the AzTEC blank-field source counts. The other three distributions are obtained from the same synthetic maps but with an inserted overdensity of 2 spatially distributed following King density profiles with $r_{c} = 0.3$ (in red), $0.7$ (in green), and $1$ (in blue) arcmin. These $r_{c}$ values are selected to account for overdensities with 95\% of the sources distributed inside radii of 1.5, 3 and 4.5 arcmin respectively. The plot shows that 95\% of the time, a blank-field population is characterised by $< 186$ sources (vertical dashed line). Therefore, the overdensity of 2 is confused with a blank field 91\% of the time if it is distributed following a King profile with $r_{c} = 0.3$ arcmin. This confusion rate decreases to 69\% if $r_{c} = 0.7$ arcmin, and continues to fall to only 4\% if  $r_{c} = 1$ arcmin.}
\label{simprob}
\end{figure}

We also use these simulations to estimate the probability for a central overdensity of 2 to be diluted in the analysis of large maps as ours. We obtain the probability distribution of the number of sources with flux densities $> 4$ mJy found in 17 blank fields the size of our ACES protocluster maps. Then, we populate the central areas of these maps with twice the number of sources estimated for circular areas of 1.5 arcmin radius (following a King density profile with $r_{c} = 0.3$ arcmin, which is characterised by having 95\% of the sources inside a radius of 1.5 arcmin) and obtain the new probability distribution of the number of sources. Comparing these two distributions shows that 95\% of the time a blank-field population is characterised by $< 186$ sources, and that the overdensity of 2 is confused with a blank field 91\% of the time (Figure~\ref{simprob}).

These simulations explain why it is not surprising that an overdensity of 2 inside a 1.5 arcmin radius of the AGN is missed when we analyze the whole 1.01 square degrees of the ACES AGN survey. In addition, it offers an explanation for the non-detection of overdensities in 14 of our ACES individual maps. The overdensities can get confused with the expected sample variance of our individual fields if they only cover a small central area.

\section{Alignment of SMGs and radio jets}
\label{sec:alignment}

Numerical simulations suggest that alignments occur naturally on various scales in hierarchical models of structure formation such as the $\Lambda$CDM model (e.g.~\citealp{Basilakos06,Faltenbacher08,Velliscig15}). Under this paradigm, primordial alignments along the large-scale filamentary structures can originate from a combination of different mechanisms such as tidal sheering, produced by the matter distribution around galaxies, and galaxy-galaxy interactions in the direction of the filaments.

Since SMGs are a high-redshift population, undergoing violent episodes of star formation and, as shown in the previous section, associated with protocluster candidates, they are likely to trace large-scale structure (e.g.~\citealp{Tamura09,Umehata15}). Therefore, we may expect to find them preferentially forming inside the filaments feeding our cluster progenitors. Since our ACES maps cover areas with radii of $\sim$6-8 arcmin (Table~\ref{ACESRGtable_mm}), which are equivalent to co-moving distances of 4-20 Mpc, we may be capable of identifying filamentary structure in them.
 
Although we have no redshift information for our sources, we can test whether they are randomly distributed around the AGN or aligned in a preferred direction. Sixteen of our ACES AGN are high-redshift radio galaxies with known jet position angles (PAs; Table~\ref{ACESRGtable_radio}). The direction of their radio jets is thought to be the direction of the angular momentum axis of the super massive black hole that drives their nuclear activity. Since the black hole spin axis may be strongly coupled to the surrounding large-scale structure (e.g.~\citealp{Rees78,Taylor16}), identifying a preferred direction for the distribution of mm sources with respect to the radio jets could be an indication of filamentary structure. 

\subsection{Principal axes}
\label{subsec:tensor}

\begin{figure}
\centering
\includegraphics[width=0.5\textwidth, trim=2cm 6cm 2cm 5cm, clip]{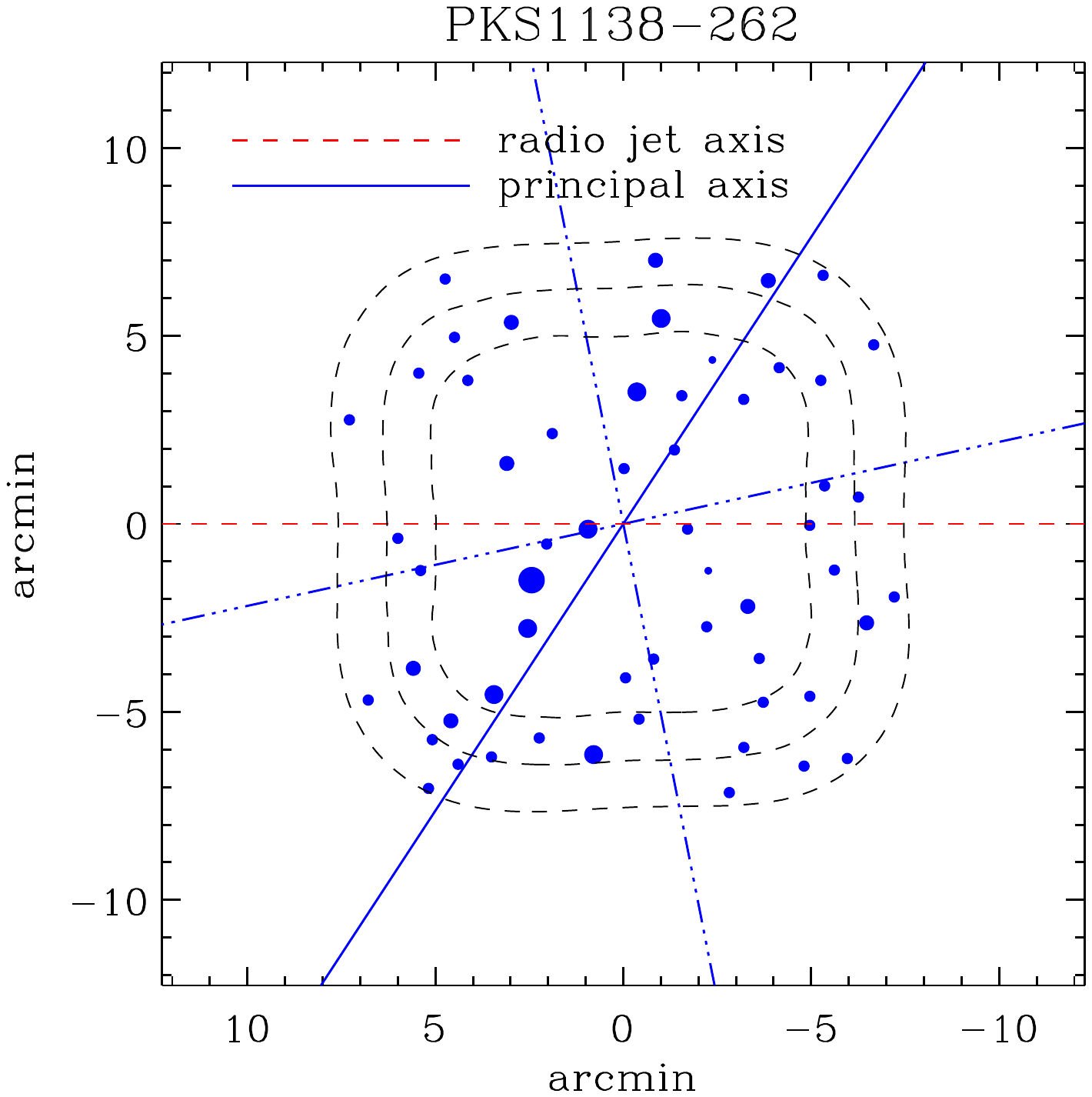}
\includegraphics[width=0.45\textwidth, trim=2cm 6cm 2cm 5cm, clip]{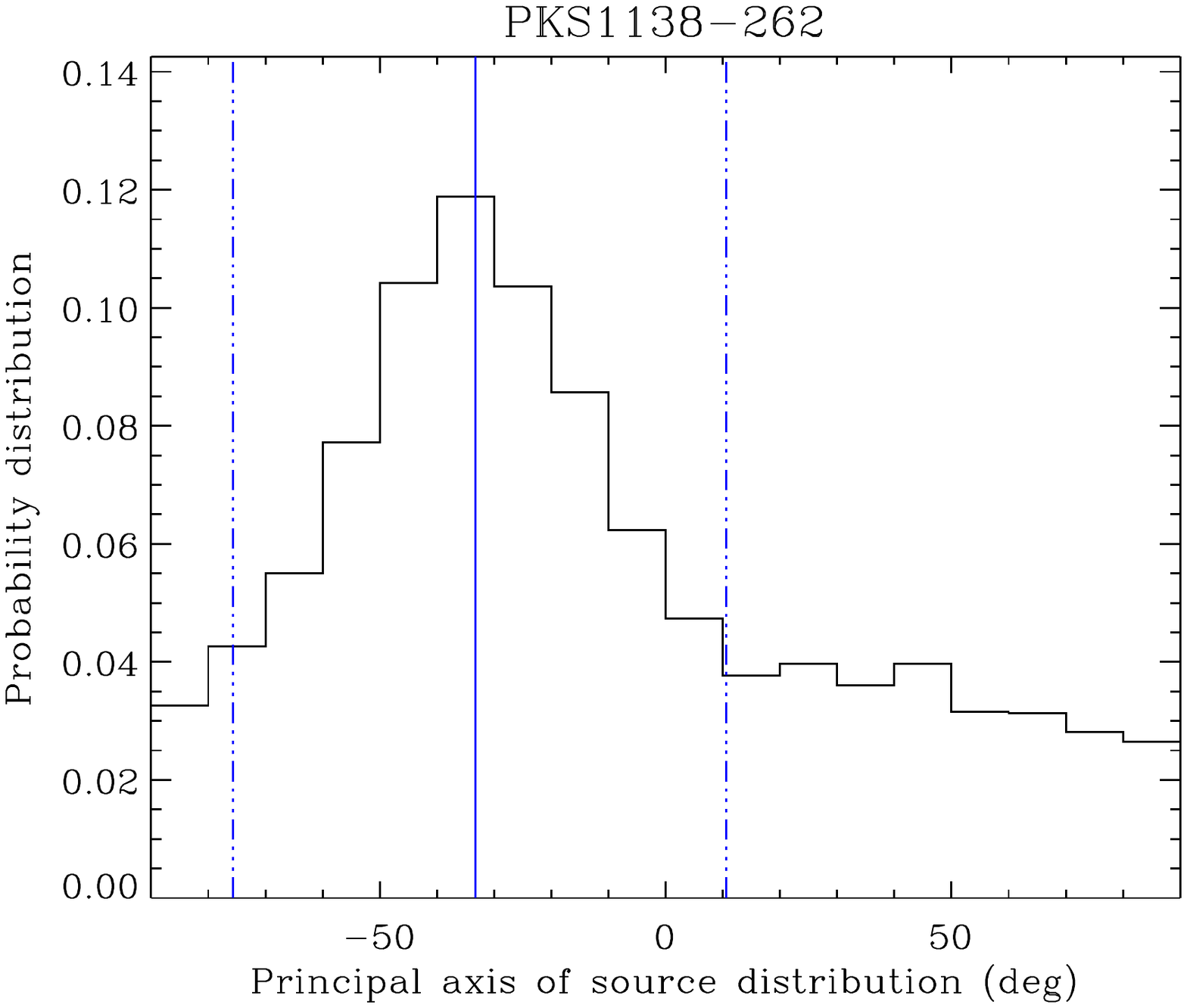}
\caption[]{Top: Position map of the AzTEC sources around the radio galaxy PKS1138-262. The circle diameter is proportional to the source S/N. The radio jet position angle is obtained from the literature (Table~\ref{ACESRGtable_radio}) and is represented by the horizontal dashed line. The principal axis of the source distribution is determined using equation 1, motivated by the form of the inertia tensor, and is marked by the solid line. Its $68\%$ confidence interval is estimated from the bootstrapped probability distribution and is represented by the two dot-dashed lines. Both, the radio jet and the principal axis directions are measured North to East. Dashed contours show different noise levels (50\%, 75\% and 95\% coverage cuts) which show no correlation with the principal axis. Bottom: Principal axis probability distribution determined by bootstrapping. The vertical line marks the principal axis calculated from the source distribution and the two dot-dashed lines the $68\%$ confidence interval.}
\label{IndAlign0802}
\end{figure}

To determine if SMGs around our radio galaxies align in a preferred direction, we find the major axis of their spatial distribution using a tool motivated by the form of the inertia tensor. Since we have no redshift information, we calculate the major axis in two dimensions. In addition, since moments of inertia are heavily dependent on the distance between the sources and the central radio galaxy, we calculate a form of reduced version \citep{Gerhard83}, which uses only directional information to estimate the shape of the distribution and weights sources equally regardless of their distance to the centre,

\begin{equation}
I = \left[ \begin{array}{cc}
\sum\limits_{i=1}^{n}{\frac{x^2}{r^2}} & -\sum\limits_{i=1}^{n}{\frac{xy}{r^2}}  \\
 & \\
-\sum\limits_{i=1}^{n}{\frac{xy}{r^2}} & \sum\limits_{i=1}^{n}{\frac{y^2}{r^2}}  \end{array} \right],
\end{equation}

\noindent where $n$ is the number of sources, $r$ is the distance from the radio galaxy to the source, and $x,y$ are the horizontal and vertical components of that distance. The major axis of the distribution is defined by the major eigenvector of equation 1. We call this the \emph{principal axis}. In order to estimate the error on this measurement we sample the source distribution 6,000 times with replacement (i.~e.~it is possible to sample a source more than once in each iteration) and calculate the principal axis in each iteration to produce a probability distribution of angles (bootstrapping). For all 16 fields, the peak of the distribution lies close to the \emph{principal axis}, and it is used as starting point for the $68\%$ confidence interval estimation. The position map of SMGs around the radio galaxy PKS1138-262, together with the probability distribution of its principal axis, is plotted in Figure~\ref{IndAlign0802}. Appendix \ref{appendixA} shows similar plots for the rest of the sample. Angles are measured North-East.

\subsection{Alignment in individual maps}
\label{subsec:pas}

\begin{figure}
\centering
\includegraphics[width=0.43\textwidth, trim=2cm 7cm 2cm 7cm, clip]{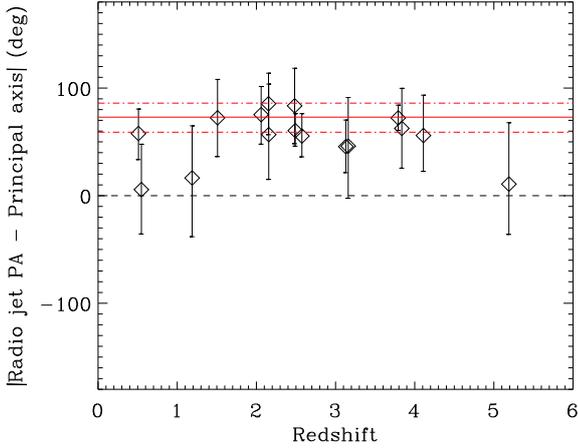}
\caption[]{Angular difference (absolute value) between the radio galaxy jet and the principal axis of the surrounding source distribution for the 16 ACES radio galaxy fields, plotted against redshift. In most cases the radio jet direction and the principal axis are separated more than 45 deg, although the error bars are quite large. A similar analysis using all fields is performed in section \ref{subsec:calign}. The result is plotted as solid and dot-dashed horizontal lines.}
\label{IndAlignments}
\end{figure}

Position angles (PAs) for the ACES radio galaxy jets were obtained from the literature and are listed in Table~\ref{ACESRGtable_radio}. For all radio galaxies except two, radio images show at least two components or radio lobes. In the cases of MRC2201-555 and TNJ2009-3040, their jet PAs denote the orientation of the radio emission since it is not symmetric. Using these angles, it is possible to estimate the angular separation between radio jets and the \emph{principal axes} of the surrounding source distributions. Figure~\ref{IndAlignments} shows absolute values for these separations against redshift. In 13 cases the radio jet direction and the \emph{principal axis} are separated more than 45 degrees, indicating a possible signal of misalignment, but the error bars are large. Combining the information for all the 16 fields will enhance this signal.

\subsection{Combined analysis}
\label{subsec:calign}

\begin{figure}
\centering
\includegraphics[width=0.43\textwidth, trim=2cm 6cm 2cm 7cm, clip]{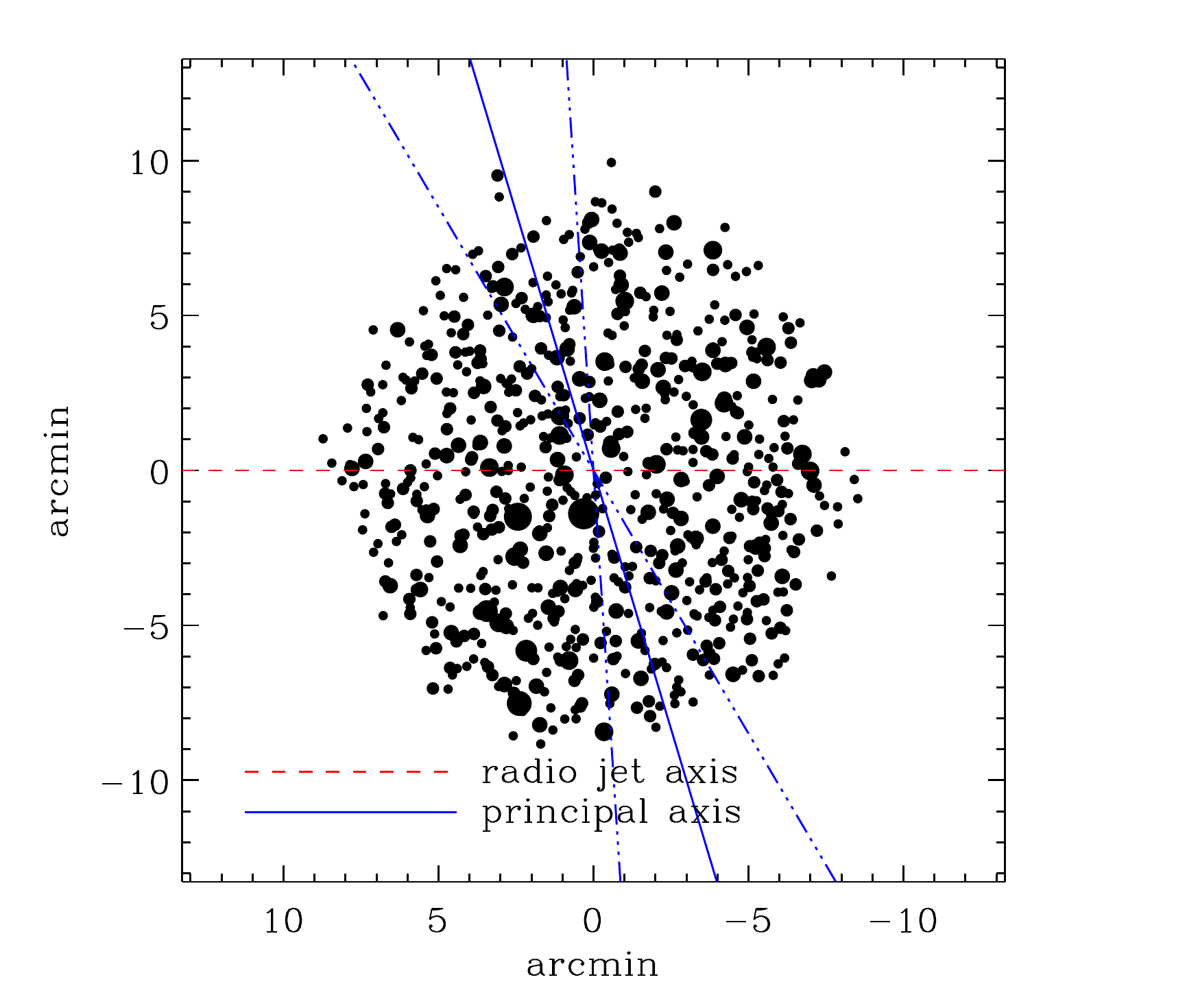}
\includegraphics[width=0.43\textwidth, trim=1cm 6cm 1cm 6.5cm, clip]{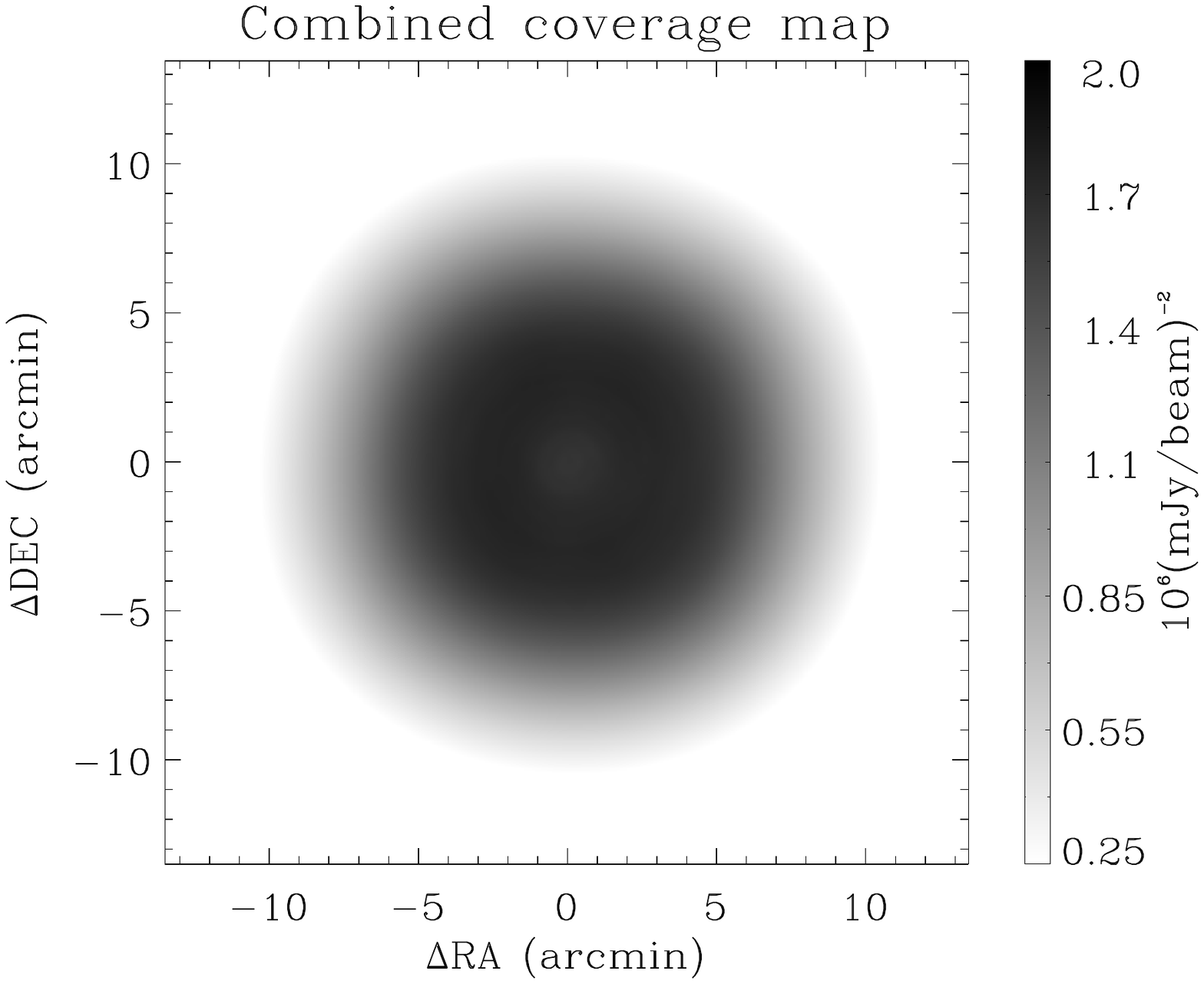}
\caption[]{Top: Principal axis for the source distribution of the stacked 16 ACES radio galaxy fields. Circles represent sources and their diameters are proportional to the S/N ratio. The dashed line represents the horizontally aligned radio jet directions, and the solid and dot-dashed lines mark the preferred direction of the source distribution and its $68\%$ confidence interval (see Figure~\ref{ComAlignProb}). As can be seen, the principal axis is $73^{+13}_{-14}$ degrees away from the radio jets, which makes this a $\sim 5\sigma$ detection of a misalignment. Bottom: Combined coverage map for the 16 ACES radio galaxy fields that shows an isotropic coverage; and therefore, that the principle axis determination is not bias towards deeper parts of the map.}
\label{ComAlign}
\end{figure}

\begin{figure}
\centering
\includegraphics[width=0.45\textwidth, trim=2cm 6cm 2cm 6cm, clip]{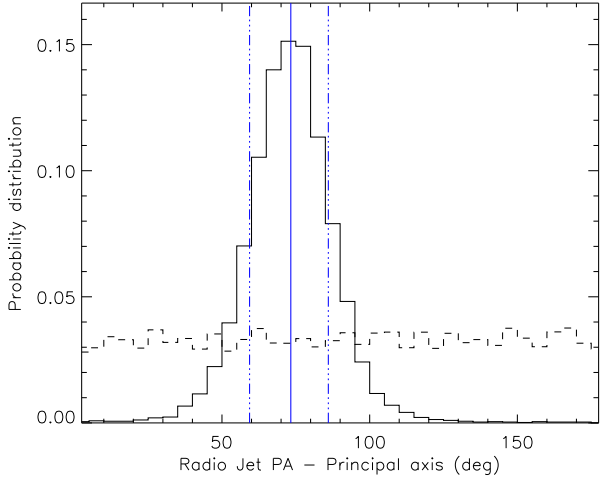}
\caption[]{Estimated probability distribution for the difference between the principal axis of the stacked distribution of sources around the 16 ACES radio galaxy fields and the aligned radio jets (see Figure~\ref{ComAlign}). The probability distribution is represented as the solid-line histogram and was obtained by sampling the stacked source distribution with replacement. The vertical solid line represents the value found directly from the map and the dot-dashed lines mark its 68\% confidence interval. The dashed-line histogram shows the corresponding probability distribution for 10,000 simulations of a 1.1mm blank-field population randomly distributed on the same area.}
\label{ComAlignProb}
\end{figure}

The 16 radio galaxy maps are rotated so that their radio jets lie horizontally and aligned to each other. Then, all the maps are stacked at the position of the targeted radio galaxies, and the \emph{principal axis} and its 68\% confidence interval are calculated in the same manner as with the individual maps. The top part of Figure~\ref{ComAlign} shows a position map of the stacked distribution of sources together with the radio jets (dashed line) and the principal axis (solid line) directions. As can be seen from the bottom part, the combined coverage map for the 16 ACES radio galaxy fields shows no obvious bias in the principle axis determination towards deeper map areas. The probability distribution for the \emph{principal axis} is plotted in Figure~\ref{ComAlignProb}. As can be seen from the images, the principal axis is $73^{+13}_{-14}$ degrees away from the radio jets direction, which makes this a $\sim 5 \sigma$ detection of a misalignment. We caution the reader that all projections are measured on the plane of the sky, since we do not have redshift information for the SMGs, and that we do not expect a strong dilution of the alignment due to this projection since the modified inertia tensor only takes into account the direction and not the distance to the AGN. However, the chances for source blending could be severe if the intrinsic alignment were perpendicular to the plane of the sky.

In order to test if there are additional biases in the way the \emph{principal axis} is estimated, the modified tensor of inertia technique is applied to each of the 10,000 blank-field simulated maps used in section \ref{subsec:ncradii}. Because the sources are randomly distributed, the probability distribution of their principal axes should be characteristic of fields without any alignment. The result can be seen in Figure~\ref{ComAlignProb} as the dashed-line histogram. As expected, there is no preferred direction or no principal axis and the distribution is quite flat. Although a principal axis can be defined for any individual simulation, and the angle can be at any given value between 0 and 180~degrees, the resulting probability of the 68 per cent confidence intervals is quite wide, with a median value of 60~degrees. The distribution of widths has a tail towards high values, and widths as small as 27~degrees happen by chance only 2 per cent of the time. However, these distributions are significantly different when a Kolmogorov-Smirnov test is used, and in no case we find a shape coincidence with a significance greater than 95 per cent.

\begin{figure}
\centering
\includegraphics[width=0.45\textwidth, trim=2cm 6cm 2cm 6cm, clip]{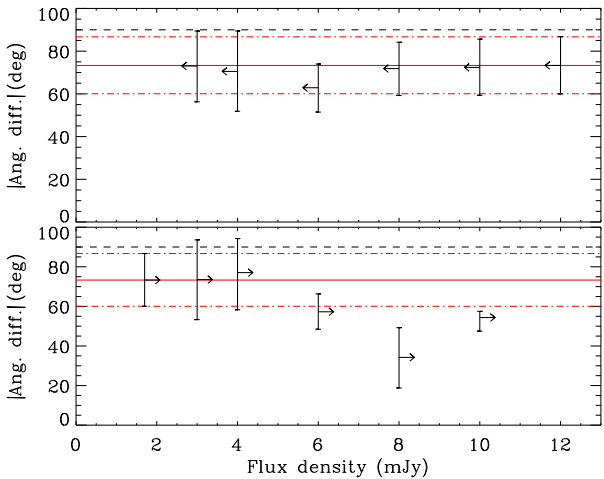}
\caption[]{Angular differences between the aligned radio jets and the principal axes for the stacked distribution of sources around the 16 ACES radio galaxies as a function of limiting flux density (arrows). The vertical lines are their associated error. The red horizontal lines (solid and dot-dashed lines) represent the 73-deg principal axis found for the whole distribution of sources and its 68\% confidence interval respectively. The dashed line marks 90 deg. Top: angular differences for sources with flux densities less than a certain value. When the brightest sources are gradually removed, no significant change in the misalignment is found. Bottom: angular differences for sources with flux densities greater than a certain value.}
\label{ComAlignbyFlux}
\end{figure}

At submm/mm wavelengths higher flux densities usually mean higher luminosities (due to a flat K correction for objects with 1 $<$ z $<$ 10). Therefore, in order to test if the misalignment is traced by the most luminous SMGs, we determine principal axes of our stacked distribution of sources after applying different flux density cuts. Top panel of Figure~\ref{ComAlignbyFlux} shows the angular differences between the radio jets and the principal axes of distributions of sources with flux densities less than a certain value. The bottom panel shows a similar plot but for sources with flux densities greater than those values. The red horizontal lines (solid and dot-dashed lines) represent the 73-degree principal axis found for the whole distribution of sources and its 68\% confidence interval. As can be seen from the plots, while gradually removing the brightest galaxies causes no significant change in the alignment, gradually removing the faintest galaxies shows some variation, mainly because brighter sources are much less common. Nevertheless, most estimated axes are above 45 degrees, which points to a consistent trend for the principal axes.

\section{Discussion}
\label{sec:discussion}

\subsection{Overdensity of SMGs towards AGN}
\label{subsec:overddis}

Our source counts analysis performed individually on the 17 fields show that only in the surroundings of 3 radio galaxies (4C+23.56, PKS1138-262, MRC0355-037) the number density of sources with S$_{\rm{1.1mm}} > 4$ mJy exceeds that of a blank field. The overdensity factors are  $\sim2$ and have a significance $\gtrsim$ 3$\sigma$, which means that the probability of finding overdensities like these ones by chance is $< 0.3\%$. This finding is in line with previous studies, which described 4C+23.56 and PKS1138-262 fields as overdense environments via narrow line emission surveys \citep{Tanaka11,Kurk00} and, more recently, via IRAC studies at 3.6 and 4.5 $\mu$m \citep{Galametz12}, and MIPS studies at 24 $\mu$m \citep{Mayo12}. PKS1138-262 was also reported as SMG overdense by \cite{Dannerbauer14}. In addition, these three fields belong to the CARLA sample \citep{Wylezalek13a}, which found overdensities in IRAC 3.6 and 4.5 $\mu$m data deeper than that used by \cite{Galametz12}. One point worth noticing is that these three fields have redshifts between $2.0 < z < 2.5$. Arguably, this may suggest that the field density of dust-enshrouded star forming galaxies in protoclusters has a dependence on redshift, and that the epoch of protocluster peak activity is z $\sim 2$, which is consistent with the epoch of peak activity for blank-field SMGs (e.g.~\citealp{Chapman05}) and other populations like luminous quasars and X-ray selected AGN (e.g~\citealp{Schmidt91,Boyle98}). Nevertheless, our sample is too small to fully support such a statement, but future surveys with tens of protocluster candidates per redshift bin will help clarify this issue.

The rest of the ACES fields show source counts in agreement with those of a blank field. We showed, however, that in maps as large as ours an overdensity of 2 covering small areas (such as a circle of 1.5 arcmin radius) is confused with a blank field 91\% of the time (section \ref{subsec:sim_small_overd}). Therefore, we cannot discard the presence of overdensities diluted by the expected sample variance of these individual maps. In the case of 4C+41.17, the field was found tentatively overdense by \cite{Ivison00} using SCUBA observations with a field of view of 2.5 arcmin. However, \citealp{Wylezalek13b} used Herschel/SPIRE observations to estimate that the excess of galaxies is at z $\sim$ 2.5 and not at the redshift of the radio galaxy (z = 3.792).

In order to reduce sample variance, we performed the same source count analysis in the combined fields. We detected an overdensity $\gtrsim 3$ at S$_{\rm{1.1mm}} \ge 4$ mJy occurring only inside areas of 1.5-arcmin radius centred at the AGN. The number density of sources falls to reach a blank-field density for successive and concentric annuli with 1.5-arcmin widths ($\Delta r = 1.5$ arcmin). The size of our ACES sample enabled us to detect this overdensity with a statistical significance $> 99.75\%$ (for sources with flux densities $> 4$ mJy), which corresponds to a $3.5\sigma$ significance when the distribution of number of sources is approximated by a Gaussian. Another key advantage of the ACES survey is that both the sample and the blank-field reference data were observed with the same instrument (AzTEC), under very similar conditions, and reduced and analyzed using the same techniques. Therefore, any bias or systematic error that could not have been identified during the observation and reduction stages affects both sets of data equally, allowing a fair comparison between them. In addition, the source counts used as reference come from the analysis of the combined data from six blank-field surveys carried out with AzTEC, which is the best estimation to-date of the 1.1-mm source counts at flux densities between $S_{\rm{1.1mm}} = 1 - 12$ mJy \citep{ScottKS12}.

Since the redshift range of our sample of radio galaxies and quasar spans from $z= 0.5 - 6.3$, a 1.5-arcmin radius corresponds to co-moving diameters between 1.7 and 7.5 Mpc. This implies that at all redshifts the extension of the overdensity covers at least an area equivalent to the core of an average galaxy protocluster, whose diameter ($2 \times R_{\rm{200}}$) can range from $\sim$ 0.4 to 2 co-moving Mpc according to simulations (e.g.~\citealp{Chiang17}) and recent observations \cite{Miller18}. This result is in agreement with that of previous works that claim that high-redshift radio galaxies reside in rich environments and may indicate the presence of protoclusters. It also supports the idea that SMGs trace dense environments at high redshifts (e.g.~\citealp{Umehata15}), although it is important to mention that there are observational and simulated data suggesting that SMGs do not always trace the most massive ones (e.g.~\citealp{Chapman09,Miller15}). 

Our measured overdensity shows a tendency to increase from a factor of $\sim$2 to 3.3 for sources with higher flux densities. This could be interpreted as the environment of protoclusters enhancing star-formation rates, probably through an increment in the merging rate, as expected under the standard model of structure formation ($\Lambda$CDM model). But it is important to remember that due to the AzTEC large beam size, very bright sources can also be the result of blending multiple fainter sources. Were this the case, we could still argue that around our AGN sample, the source density is higher than that in the blank-field, implying that our fields could indeed be clusters in the process of formation. Nevertheless, it should be noted that recovery rates for sources with low signal-to-noise ratios are difficult to estimate accurately and could affect the completeness calculations for low flux-density sources. Therefore, the decrement in the overdensity signal for lower flux-density sources could be artificial.

The SMGs contributing to our detected overdensity have 1.1-mm flux densities ranging from about 1-10 mJy with their SFRs ranging between 200-1800 M$_{\odot}$ yr$^{-1}$ \citep{Kennicutt98}. If Arp 220 is indeed a good analogue \citep{Stevens10,Lapi11,Magnelli12,Contini13}, and if the starburst can be sustained for a few hundred million years or so, then a stellar mass equivalent to that of the bulge of a large galaxy will be assembled ($10^{11} - 10^{12}$ M$_{\odot}$). It is, therefore, logical to think that these overdense regions at high redshift contain galaxies that are capable of building large populations of stars that will evolve over cosmic time to become the massive elliptical galaxies that dominate the population in the cluster cores of the local Universe.

\subsubsection{Comparison with previous submm surveys}
\label{subsec:comp_other}

\begin{figure}
\centering
\includegraphics[width=0.45\textwidth, trim=0cm 12cm 1cm 1cm, clip]{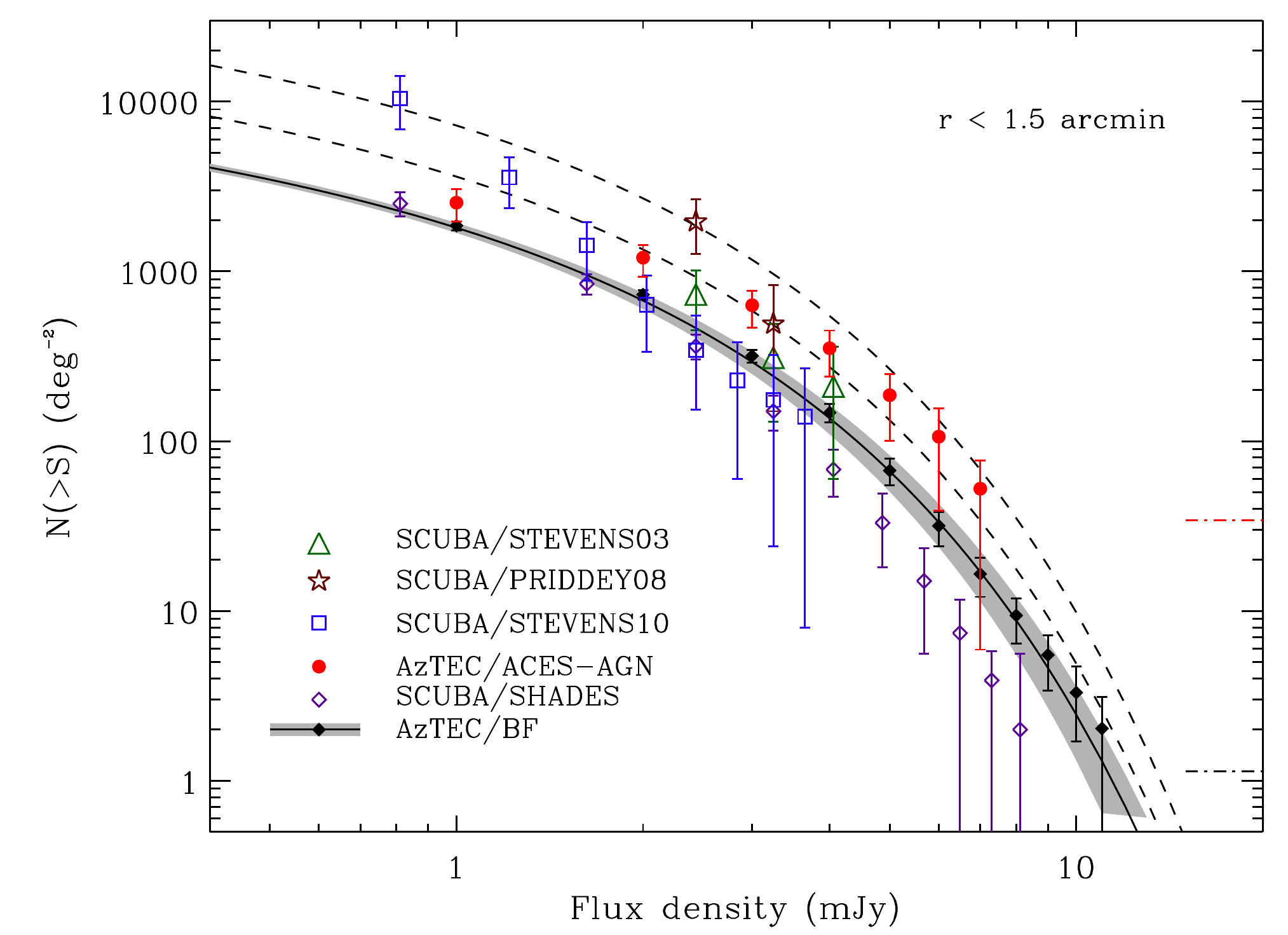} 
\caption[]{AzTEC 1.1 mm integrated source counts for the combined 17 ACES protocluster fields inside an area of 1.5-arcmin radius centred at the ACES AGN (solid circles). Source counts from previous studies with SCUBA at 850 $\mu$m by \cite{Stevens03} (triangles), \cite{Priddey08} (stars), and \cite{Stevens10} (squares) are shown for comparison. The AzTEC combined blank-field (AzTEC/BF; solid diamonds) and SCUBA/SHADES (diamonds) number counts are plotted for reference. SCUBA data is scaled to 1.1 mm using a dust emissivity index of 1.5 \citep{Casey14}. Dashed lines represent overdensities of 2 and 4 compared to the AzTEC/BF, and horizontal dashed lines mark survey limits.}
\label{combinedNC_1p5}
\end{figure}

Figure~\ref{combinedNC_1p5} compares our results to previous submm studies acquired with SCUBA at 850 $\mu$m. Their observations were centred at AGN and covered areas of $\sim$1.5-arcmin radius. Their source counts are scaled to 1.1 mm using a dust emissivity index of 1.5, which is a common assumption since several works find its value in the range 1 - 2 \citep{Casey14}. Figure~\ref{combinedNC_1p5} shows that the magnitude and extension of the overdensity we found is mostly in agreement with the SCUBA studies, but taking into account some caveats.

\cite{Stevens03} surveyed the environments of 7 radio galaxies with redshifts between $z = 2.2 - 4.3$, and found an overdensity of $\sim 2$ although not statistically significant ($< 2\sigma$) due to their small sample. Meanwhile, \cite{Priddey08} surveyed the environments of 3 optically selected quasars with redshifts between $z = 5 - 6.3$, and found an overdensity of $\sim 4$, although the magnitude of this overdensity must be interpreted carefully because no corrections for incompleteness or flux boosting were applied on the data, in contrast to the corrections applied on SCUBA/SHADES \citep{Coppin06}, their control field. Figure~\ref{combinedNC_1p5} also shows that SHADES data points fall below AzTEC blank-field data, indicating that the assumed dust emissivity index may be $<$ 1.5. Recently determined 850-$\mu$m source counts shows that these SHADES data points are in agreement with source counts analysis from the $\sim$5-sq deg.~SCUBA-2 Cosmology Legacy Survey \citep{Geach17}.

\cite{Stevens10}, on the other hand, found an opposite trend. They surveyed the environment of 5 X-ray-selected quasars with redshifts between $z= 1.7- 2.8$ and reported an overdensity of $\sim 4$ at the lower end of their source counts (S$_{\rm{1.1mm}}< 2$ mJy). This overdensity decreases at higher flux densities reaching their reference blank-field source density. A reason for such different trend could be that the lower flux-density bins are not well-corrected for incompleteness and therefore the trend is artificial. But there is also the fact that \cite{Stevens10} surveyed the environments of radio-quiet quasars in contrast to our study that concentrates on the environments of radio galaxies. Following the work of \cite{Falder10}, who found evidence for larger overdensities of galaxies in the environments of radio-loud objects compared to the environments of radio-quiet ones (based on observations at 3.6 $\mu$m), \cite{Stevens10} suggested that if their AGN are indeed located in less dense regions of the Universe, the detection of an overdensity of sources only in the lowest flux-density bins (and therefore with the lowest SFRs) is feasible. Another possible explanation, specially regarding the fact that \cite{Priddey08} detected an overdensity of SMGs towards $z>5$ radio-quiet quasars at S$_{\rm{1.1mm}}> 2$ mJy, is that obscured star formation activity in protoclusters is incremental with redshift. This is in agreement with studies at submm wavelengths on high redshift radio galaxies showing that their detection rates increases at $z>2.5$ (e.g.~\citealp{Archibald01,Reuland03}).

In addition to SCUBA, Hershel/SPIRE (250, 350 and 500 $\mu$m) also performed surveys towards high redshift biased regions. \cite{Rigby14} studied the environment of 19 high-redshift radio galaxies and reported a marginal excess of 500 $\mu$m sources within 6 co-moving Mpc of the radio galaxy. When the analysis was restricted to potential protocluster members only (identified using a far-infrared colour selection), it revealed that two fields have significant overdensities, one of 1.5 ($3.9\sigma$) and the other of 1.9 ($4.3\sigma$). Both the extension and the magnitude of these overdensities are in good agreement with our ACES results.

More recently, and based on studies suggesting that obscured AGN could be more strongly clustered and inhabit denser environments than unobscured AGN (e.g.~\citealp{Donoso14}), \cite{Jones15,Jones17} used 850 $\mu$m SCUBA-2 observations to study ovedensities of SMGs around obscured active galaxies. They studied the surroundings of 10 hot dust-obscured galaxies (Hot DOGs) and 30 WISE (Wide-field Infrared Survey Explorer, \cite{Wright10})/radio-selected AGN, finding overdensities inside 1.5 arcmin scale maps of a factor of $\sim$ 2.4 and $\sim$ 5.6, respectively. In addition, \cite{Silva15} used 870 $\mu$m high-resolution (0.45 - 1.24 arcsec) ALMA observations to study the surroundings of 49 WISE/radio-selected dusty, hyper-luminous quasars to find that the number of detected sources is 10 times greater than what is expected for unbiased regions. These results surely add to the idea that dusty powerful AGN are signposts of dense regions in the early Universe. The overdensity factors, however, are not straightforward to compare. While the SCUBA-2 results appear to be in agreement with ours, the level of overdensity measured in the ALMA study appears 5 times higher. \cite{Silva15} suggested that it could be an effect of source blending due to the large beam sizes of submm/mm single dish telescopes.

\subsubsection{Comparison with \emph{Spitzer} studies}
\label{subsec:comp_spitzer}

IRAC 3.6 and 4.5 $\mu$m studies were carried out towards 48 high-redshift radio galaxies by \cite{Galametz12}, and more extensively towards 387 radio loud AGN by \cite{Wylezalek13a}, both within a redshift range of $1.2 < z < 3.2$. They restricted the analysis to sources with colours [3.6] - [4.5] $>$ -0.1 (AB) in order to select z $> 1.3$ galaxies, and found a clear rise in surface density of sources towards the position of the AGN. This rise sharpens at distances $< 1$ arcmin, coinciding with the extension of the overdensity measured in our combined fields.

The 24-$\mu$m study by \cite{Mayo12} of the environments of 63 radio galaxies between redshifts $1 \leq z \leq 5.2$ also found an average overdensity of $2.2 \pm 1.2$ in 1.75-arcmin-radius circular cells centred on the radio galaxies. In this case both the extension and magnitude of the overdensity are in agreement with our results. Previous blank-field studies have shown that very red dust-obscured galaxies (DOGs) observed at 24 $\mu$m (R-[24] colour$ > 14$) have similar properties to bright SMGs (S$_{\rm{850\mu m}}> 6$ mJy), which could suggest that these two populations are associated, at least in an evolutionary sequence (e.g.~\citealp{Dey08,Pope08,Wu12}). Consequently, a coincident $\sim2$ times overdensity of 24 $\mu$m and mm sources around radio galaxies adds to the idea that, first, radio galaxies are indeed very good protoclusters candidates, and second, these populations could very well form part of an evolutionary succession in which SMGs represent the early phase of the formation of a massive galaxy (the starburst-dominated phase, possibly $< 1$ Gyr long) while DOGs represent the transition from a starburst-dominated phase to an AGN-dominated phase (possibly $\sim$ 3-5 times shorter than the starburst-dominated phase - e.g.~\cite{Coppin10}).

Higher resolution submm/mm experiments like ALMA or the 50-m LMT will help clarify this picture allowing more accurate multi-wavelength counterpart identifications, better SED determinations, and therefore, a more complete comparison of these two populations.

\subsection{Misalignment of mm sources with respect to radio jets}
\label{subsec:aligndis}

We found that there is a trend for SMGs to align closer to a perpendicular direction to the ACES radio galaxy jets. This result is measured on the plane of the sky, since we do not have redshift information for the SMGs. The misalignment is found over a projected co-moving scale of 4-20 Mpc, departs from perfect alignment (0 deg) by $\sim 5\sigma$ and is independent of source luminosity. We propose, under the assumption that the SMGs are at similar redshifts to the radio galaxies, that this misalignment could be the result of SMGs preferentially inhabiting dominant filaments feeding the protocluster structures that contain the radio galaxies. 

Using dark matter simulations, several studies have investigated primordial alignments between dark matter halos and their large-scale filamentary structures (e.g.~\citealp{Aragon-Calvo07,Zhang09,Codis12}). They found that the orientation of the spin axes of the dark matter halos is mass-dependent. Low-mass halos ($\sim$ M$< 10^{12}$ M$_{\odot}$) have a tendency to have their spin axes oriented along the parent structure, while high-mass halos have their spin axes perpendicular to it. Regarding baryonic matter, however, fewer studies have investigated the alignment between galaxy spin axes and their embedding cosmic web, mainly because of the high computational cost of the simulations. Recently, \cite{Dubois14} used a large-scale hydrodynamical cosmological simulation to investigate the alignment between galaxy spins and their surrounding cosmic filaments at $1.2 < z < 1.8$. They showed that the spin of low-mass blue galaxies is preferentially aligned with their neighbouring filaments, while high-mass red galaxies tend to have a perpendicular spin as a result of mergers.

Observationally, \cite{Tempel13} and \cite{Zhang13} also found tentative evidence of such alignments in the Sloan Digital Sky Survey. In addition, one of the best known examples of preferential alignments is the orientation of the major axes of bright cluster galaxies (BCGs) along the distribution of cluster members and how they point towards other nearby clusters on scales of $\sim 10-20$ Mpc \citep{Carter80,Binggeli82,Struble90,Plionis94,Hashimoto08}. This alignment could be explained as clusters forming at the intersection of different filaments but with a dominant filamentary feature that is likely to exert the most profound influence on the final cluster-BCG orientation \citep{West94}.

Considering all this evidence and the fact that high-redshift radio galaxies are very massive sources, we may expect their spin axes, both of their dark matter halo and the galaxy itself, to be aligned perpendicularly to the direction of their embedding dominant filaments. Currently, there is no way to observationally corroborate how dark matter spins in galaxies. In some cases, however, radio jets position angles were found to act as a proxy for the direction of the major axis of the baryonic matter in their host galaxies. For instance, extended radio jets were found preferentially aligned with the optical minor axes of their galaxies, particularly for elliptical galaxies (e.g.~\citealp{Condon91,Battye09}). If we assume that the radio jets in our sample are aligned with their galaxies' minor axes, the fact that we find their position angles to be closer to a perpendicular direction with respect to the principal axis of the SMGs spatial distribution could be interpreted as SMGs tracing large-scale structure and inhabiting dominant filaments feeding the protocluster structures that contain the radio galaxies.

\cite{Stevens03} also studied alignments of SMGs around their high-redshift radio galaxy sample. They compared the radio jet position angle with the location of the brightest submm source in the map, apart from the AGN. The location of this bright submm companion was expected to give an idea about the orientation of the large scale structure around the radio galaxy. They found a possible alignment between the radio jet and the large-scale structure. We propose that this opposite outcome was the result of having small maps (2.5-arcmin diameter) and a limited data set (7 fields). If we artificially reduce our field of views to what \cite{Stevens03} would have observed, we find that 13 of the 16 fields have at least one mm companion. From these 13, 11 have their brightest companion more than 45 deg away from the radio jet, i.e.~in agreement with our misalignment result.

\section{Conclusions}
\label{sec:conclusions}

We explored the spatial distribution of SMGs towards the environments of 16 powerful high-redshift radio galaxies and a quasar using continuum observations at 1.1 mm taken with the AzTEC camera. We targeted the environments of powerful high-redshift AGN in order to pinpoint the location of the progenitors of the richest galaxy clusters we see today in the local Universe. After removing possible mm counterparts to the AGN, we estimated source counts for individual fields, but in the majority of cases the density of sources with S$_{\rm{1.1mm}} > 4$ mJy fell within the 95\% confidence interval of the density of sources in a comparison sample of unbiased blank fields. Only in the surroundings of 4C+23.56, PKS1138-262 and MRC0355-037 did we detect individual overdensity signals of $\sim2$ with a significance of $\sim 3\sigma$. Performing simulations, however, we found that 91\% of the time an overdensity of a factor 2 covering a small area of 1.5 arcmin radius is lost in a number density analysis of sources with S$_{\rm{1.1mm}} > 4$ mJy populating maps as large as our ACES maps. Therefore, we cannot discard the presence of overdensities confused with the expected sample variance of our individual maps.

When we performed a combined analysis on the complete sample, we found an overdensity $\gtrsim 3$ at S$_{\rm{1.1mm}} \ge 4$ mJy with greater statistical significance, covering an area of 1.5-arcmin radius centred on the AGN (corresponding to a co-moving diameter of 1.7-7.5 Mpc over the redshift range $0.5<z<6.3$ of the sample, and a co-moving diameter of 4.6-7.5 Mpc over $2<z<6.3$, where most of our targets lie). The large size of our maps allowed us to establish that beyond a radius of 1.5 arcmin, the radial surface density of SMGs falls to reach a typical value for a blank field distribution of SMGs. The measured angular extent of this overdensity is in agreement with protocluster core simulations and observations. In addition, we found that the overdensity shows a tendency to increase with higher flux densities.

We interpreted this as an enhancement in the dust-obscured star-formation activity towards protocluster environments, detected either as an increment in the star formation rates of individual galaxies or as an increment in the number of sources that get blended due to the large size of the AzTEC beam.

The data used as reference to measure the magnitude and extent of the overdensity is composed of six blank-field surveys carried out also with AzTEC. These observations provide the best estimation to-date of the 1.1-mm source counts towards blank fields at $S_{\rm{1.1mm}} = 1 - 12$ mJy. In addition, our protocluster targets and the blank-field data were observed, reduced, and analyzed under very similar conditions and using the same techniques. Therefore, a fair comparison between them is possible. 

We also examined if there was a preferred direction, on the plane of the sky, in which the SMGs align around our sample of high-redshift radio galaxies. Using a tool motivated by the form of the tensor of inertia, we found that there is a trend for SMGs to align along an orientation that is closer to a perpendicular direction with respect to the radio jets (73$_{+13}^{-14}$ degrees) than to the parallel direction. This misalignment was found over projected co-moving scales of 4-20 Mpc, departs from perfect alignment (0 deg) by $\sim 5\sigma$ and apparently has no dependence on the source luminosity, although the dynamical range of our flux-limited sample is probably not large enough to draw a definite conclusion.

Since our radio galaxies are thought to be massive sources, and simulations predict that their dark matter halo spin axes align perpendicularly to the direction of the dominant filament feeding them, we suggest that this misalignment could be the result of SMGs preferentially inhabiting the mass-dominant filaments funneling material towards the protoclusters that contain our radio galaxies. This suggestion is based on the assumptions that the distribution of baryonic matter roughly follows the distribution of dark matter, and that the radio jets in our sample are a proxy for their galaxies' minor axes.

The properties of the SMG distribution described above are consistent with the idea that powerful AGN reside in massive dark matter haloes, and that these protocluster regions are sites of enhanced dust-obscured star formation. In the local Universe the centres of rich clusters are inhabited preferentially by massive elliptical galaxies. Assuming their progenitors go through a submm phase, we speculate that SMGs in these environments are forming a large fraction of the stellar population of the massive ellipticals. Moreover, these properties also show that SMGs are probably contributing to the formation of the stellar population in the filamentary structure around them, since they appear to be tracing at least the most dominant structures.

\section*{Acknowledgments}

We would like to thank everyone who supported the AzTEC/ASTE and AzTEC/JCMT observations of the ACES fields. This work has been supported by CONACYT projects CB-2011-01-167291, FDC-2016-1848 and CONACYT studentship 12812. AH acknowledges Funda\c c\^ao para a Ci\^encia e a Tecnologia (FCT) support through UID/FIS/04434/2013, and through project FCOMP-01-0124-FEDER-029170 (Reference FCT PTDC/FIS-AST/3214/2012) funded by FCT-MEC (PIDDAC) and FEDER (COMPETE), in addition to FP7 project PIRSES-GA-2013-612701, and FCT grant SFRH/BPD/107919/2015. The ASTE project is driven by the Nobeyama Radio Observatory (NRO), a branch of the National Astronomical Observatory of Japan (NAOJ), in collaboration with the University of Chile and Japanese institutions including the University of Tokyo, Nagoya University, Osaka Prefecture University, Ibaraki University and Hokkaido University. The James Clerk Maxwell Telescope is operated by the Joint Astronomy Centre on behalf of the Science and Technology Facilities Council of the United Kingdom, the Netherlands Organisation for Scientific Research, and the National Research Council of Canada.








\appendix
\section{Data reduction and analysis products}
\label{appendixA}

Set of products for each ACES protocluster field. The sets are listed in decreasing order according to the redshift of the central AGN (as shown in Table~\ref{ACESRGtable_radio}). Each set contains the following plots and tables:\\

{\bf 1)} AzTEC signal and weight maps. Source candidates have S/N $> 3.5$ for the fields of TNJ1338-1942 and 4C+41.17, and S/N $> 3.0$ for the rest of the maps. ASTE maps have their source candidates marked by 30-arcsec diameter circles, while the JCMT map has 18-arcsec diameter circles. Contours represent the 75\% and 50\% coverage cuts.\\

{\bf 2)} Positional uncertainty distributions for sources within 3 different S/N bins: $3.75<S/N<4.0$ (diamonds), $5.0 <S/N< 5.25$ (triangles), and $6.25<S/N< 6.5$ (squares). The curves show the analytical expression derived in \cite{Ivison07} for the corresponding S/N bins.\\

{\bf 3)} False detection rate, estimated using jackknife maps, for sources with S/N greater than a certain value. The error bars represent the 68\% confidence interval from a Poisson distribution.\\

{\bf 4)} Completeness estimation as a function of flux density. The error bars represent the 68\% confidence interval from a binomial distribution.\\

{\bf 5)} AzTEC 1.1-mm differential source counts (solid squares) compared to those of the reference field (diamonds). Differential source counts for the other 16 fields are also shown as comparison (open squares). The solid line and grey shading represent the best fit of a Schechter function to the density of sources of the reference field and its 68\% confidence interval. Dashed lines show twice and four times the source counts described by the fit. Horizontal dashed lines represent the survey-limit, defined as the source density (inside de map area) that will Poisson deviate to zero sources 32\% of the time.\\

{\bf 6)} AzTEC 1.1-mm integrated source counts. Symbols and lines are coded as with the differential source counts.\\

{\bf 7)} Position maps for the AzTEC sources around the central radio galaxy, together with the principal axis of the source distribution (solid line), its 68\% confidence interval (two dot-dashed lines), and the radio jet direction (dashed line). Dashed contours show different noise levels (50\%, 75\% and 95\% coverage cuts).\\

{\bf 8)} Probability distribution of the principal axis described in item 7 and determined by bootstrapping. The vertical line marks the principal axis calculated from the source distribution and the two dot-dashed lines the $68\%$ confidence interval.\\

{\bf 9)} Differential and integrated source counts table.\\

{\bf 10)} Catalogue of AzTEC sources with S/N $> 3.5$ for the shallowest maps (TNJ1338-1942 \& 4C+41.17) and S/N $> 3.0$ for the rest. The columns show: 1) source id; 2) source name; 3) S/N of the detection; 4) measured 1.1 mm flux density and error; 5) deboosted 1.1 mm flux density and 68\% confidence interval; and 6) probability for the source to have a negative deboosted flux. The catalogue is limited to sources detected within the 50\% coverage region of the map.\\

\FloatBarrier
\begin{figure*}
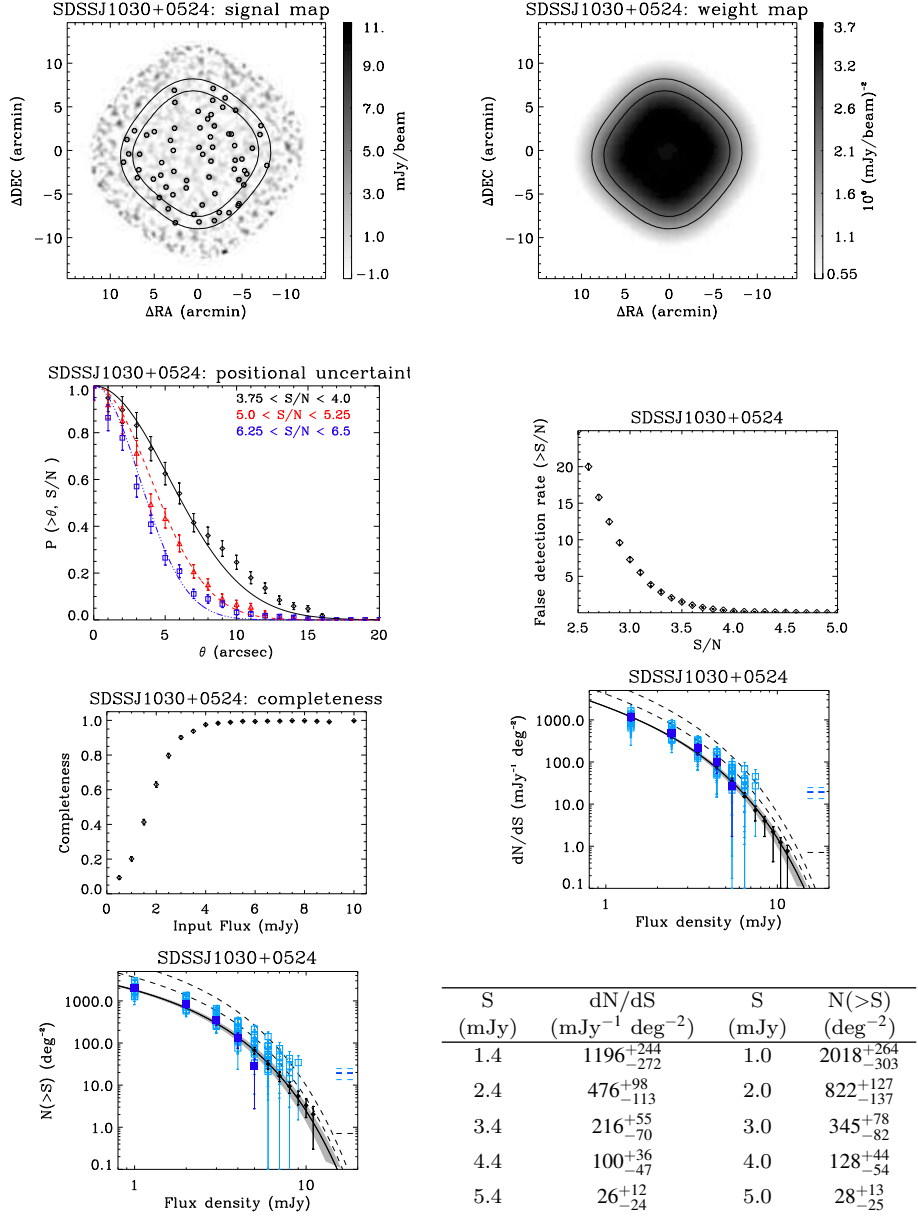

\centering
\caption[]{SDSSJ1030+0524 products}

\end{figure*}

\FloatBarrier
\begin{table*}
\begin{center}
\scriptsize
\caption{SDSSJ1030+0524 source catalogue.}
\begin{tabular}{lccccc}
\hline
      &     		&     		& S$_{\rm{1.1mm}}$ & S$_{\rm{1.1mm}}$ 	&         		\\
      &      		&     		& (measured)  		& (deboosted) 			&         		\\
Id  & IAU name & S/N 	& (mJy)       		& (mJy)       			& P($<0$) 	\\
\hline
1 	 & 	 MMJ103019.72+052743.31 	  &  	   9.37 	  &  	 5.0$\pm$0.5 	  &  	 4.7$^{+0.6}_{-0.5}$ 	  &  	 0.000 	 \\
2 	 & 	 MMJ103008.45+052133.48 	  &  	   8.59 	  &  	 5.0$\pm$0.6 	  &  	 4.7$^{+0.6}_{-0.6}$ 	  &  	 0.000 	 \\
3 	 & 	 MMJ103044.81+051930.54 	  &  	   8.30 	  &  	 5.3$\pm$0.6 	  &  	 5.0$^{+0.6}_{-0.7}$ 	  &  	 0.000 	 \\
4 	 & 	 MMJ103025.24+051839.75 	  &  	   8.14 	  &  	 4.6$\pm$0.6 	  &  	 4.3$^{+0.6}_{-0.6}$ 	  &  	 0.000 	 \\
5 	 & 	 MMJ103032.40+052128.05 	  &  	   8.12 	  &  	 4.3$\pm$0.5 	  &  	 4.0$^{+0.6}_{-0.5}$ 	  &  	 0.000 	 \\
\hline
\end{tabular}
\end{center}
\end{table*}

\FloatBarrier
\begin{figure*}
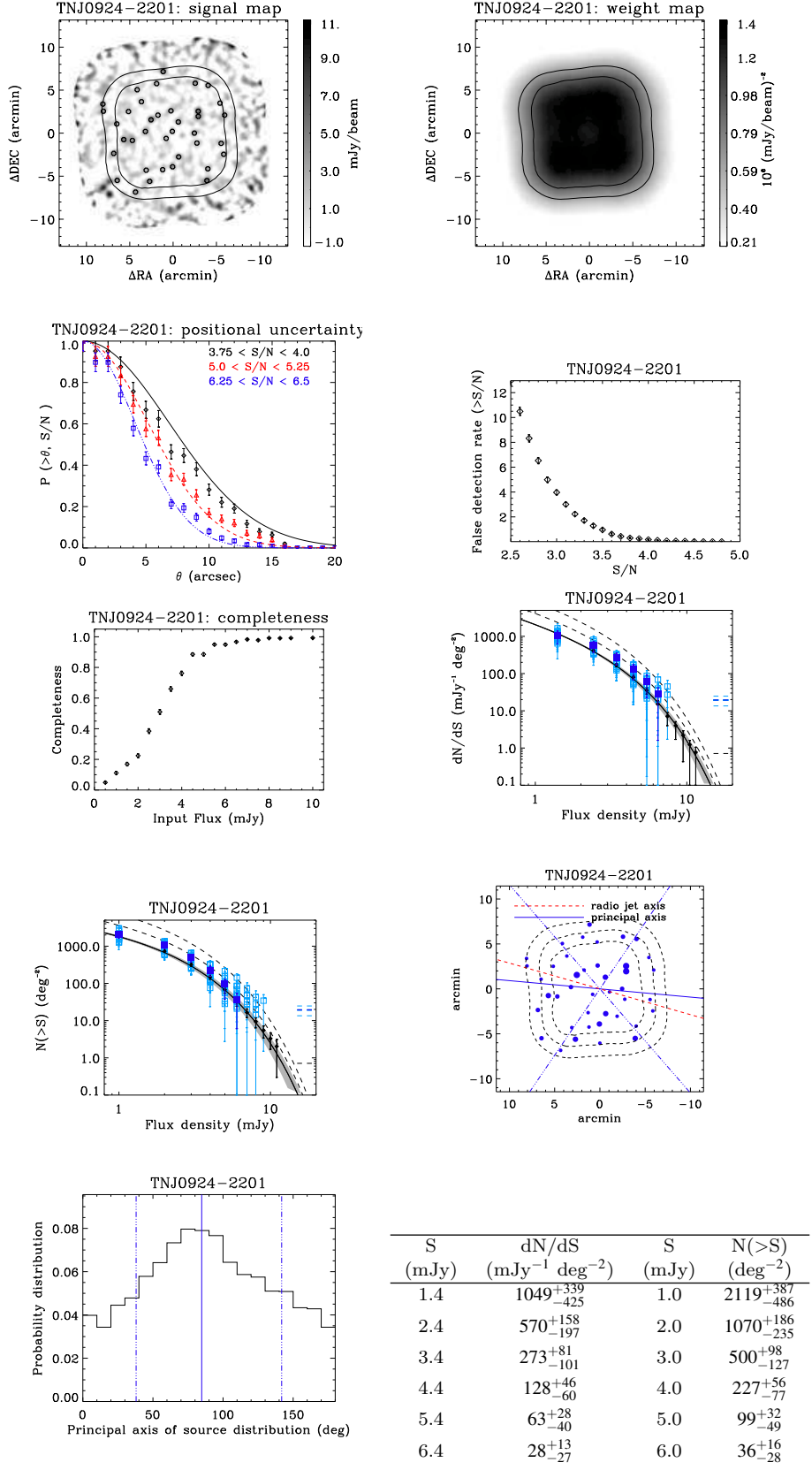

\centering
\caption[]{TNJ0924-2201 products}
%
\end{figure*}

\FloatBarrier
\begin{table*}
\begin{center}
\scriptsize
\caption{TNJ0924-2201 source catalogue.}
\begin{tabular}{lccccc}
\hline
      &     		&     		& S$_{\rm{1.1mm}}$ & S$_{\rm{1.1mm}}$ 	&         		\\
      &      		&     		& (measured)  		& (deboosted) 			&         		\\
Id  & IAU name & S/N 	& (mJy)       		& (mJy)       			& P($<0$) 	\\
\hline
1 	 & 	 MMJ092430.92-220008.37 	  &  	   7.73 	  &  	 6.8$\pm$0.9 	  &  	 6.1$^{+1.0}_{-0.8}$ 	  &  	 0.000 	 \\
2 	 & 	 MMJ092407.41-215908.67 	  &  	   7.61 	  &  	 6.7$\pm$0.9 	  &  	 6.1$^{+0.9}_{-0.9}$ 	  &  	 0.000 	 \\
3 	 & 	 MMJ092407.39-215944.35 	  &  	   6.86 	  &  	 6.0$\pm$0.9 	  &  	 5.4$^{+0.9}_{-0.9}$ 	  &  	 0.000 	 \\
4 	 & 	 MMJ092403.07-220712.00 	  &  	   6.06 	  &  	 6.1$\pm$1.0 	  &  	 5.2$^{+1.1}_{-0.9}$ 	  &  	 0.000 	 \\
5 	 & 	 MMJ092444.49-220226.45 	  &  	   5.82 	  &  	 5.6$\pm$1.0 	  &  	 4.8$^{+1.0}_{-0.9}$ 	  &  	 0.000 	 \\
\hline
\end{tabular}
\end{center}
\end{table*}

\FloatBarrier
\begin{figure*}
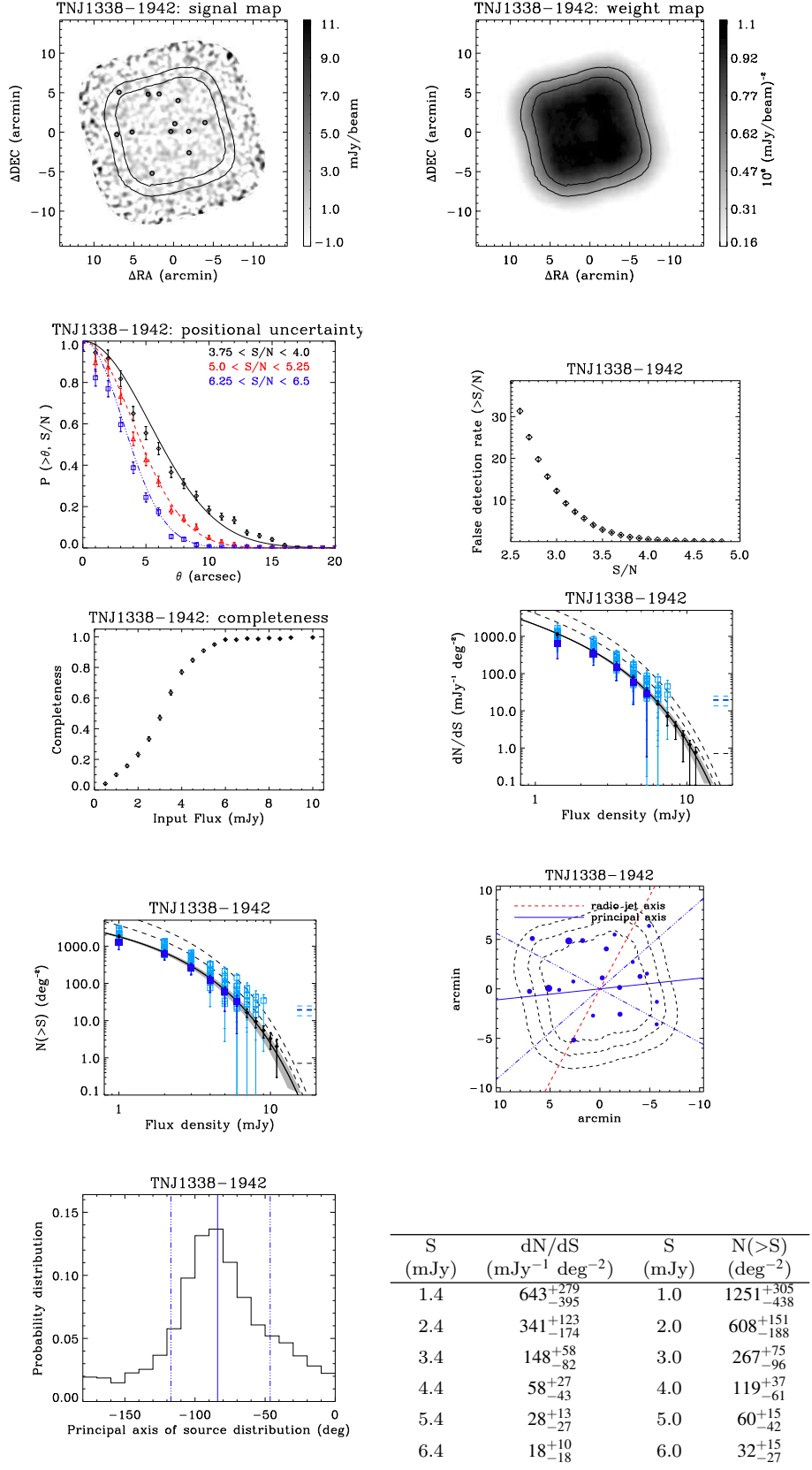

\centering
\caption[]{TNJ1338-1942 products}
%
\end{figure*}

\FloatBarrier
\begin{table*}
\begin{center}
\scriptsize
\caption{TNJ1338-1942 source catalogue.}
\begin{tabular}{lccccc}
\hline
      &     		&     		& S$_{\rm{1.1mm}}$ & S$_{\rm{1.1mm}}$ 	&         		\\
      &      		&     		& (measured)  		& (deboosted) 			&         		\\
Id  & IAU name & S/N 	& (mJy)       		& (mJy)       			& P($<0$) 	\\
\hline
1 	 & 	 MMJ133839.49-193742.74 	  &  	   7.31 	  &  	 7.9$\pm$1.1 	  &  	 6.9$^{+1.1}_{-1.0}$ 	  &  	 0.000 	 \\
2 	 & 	 MMJ133848.07-194230.00 	  &  	   6.65 	  &  	 6.8$\pm$1.0 	  &  	 6.0$^{+1.0}_{-1.1}$ 	  &  	 0.000 	 \\
3 	 & 	 MMJ133827.35-194225.65 	  &  	   6.28 	  &  	 6.5$\pm$1.0 	  &  	 5.6$^{+1.1}_{-1.1}$ 	  &  	 0.000 	 \\
4 	 & 	 MMJ133823.52-193831.12 	  &  	   4.64 	  &  	 4.5$\pm$1.0 	  &  	 3.6$^{+1.0}_{-1.0}$ 	  &  	 0.000 	 \\
5 	 & 	 MMJ133856.25-194248.63 	  &  	   4.54 	  &  	 5.7$\pm$1.3 	  &  	 4.2$^{+1.4}_{-1.3}$ 	  &  	 0.001 	 \\
6 	 & 	 MMJ133833.62-193740.75 	  &  	   4.41 	  &  	 4.6$\pm$1.0 	  &  	 3.6$^{+1.1}_{-1.1}$ 	  &  	 0.001 	 \\
7 	 & 	 MMJ133817.83-194225.07 	  &  	   4.23 	  &  	 4.2$\pm$1.0 	  &  	 3.2$^{+1.0}_{-1.0}$ 	  &  	 0.002 	 \\
8 	 & 	 MMJ133825.23-194127.00 	  &  	   4.16 	  &  	 4.2$\pm$1.0 	  &  	 3.2$^{+1.1}_{-1.1}$ 	  &  	 0.002 	 \\
9 	 & 	 MMJ133855.03-193728.17 	  &  	   3.97 	  &  	 5.1$\pm$1.3 	  &  	 3.5$^{+1.3}_{-1.5}$ 	  &  	 0.009 	 \\
10 	 & 	 MMJ133837.39-194743.56 	  &  	   3.95 	  &  	 4.1$\pm$1.0 	  &  	 3.0$^{+1.0}_{-1.1}$ 	  &  	 0.005 	 \\
11 	 & 	 MMJ133809.13-194118.36 	  &  	   3.71 	  &  	 3.8$\pm$1.0 	  &  	 2.7$^{+1.1}_{-1.2}$ 	  &  	 0.011 	 \\
12 	 & 	 MMJ133817.65-194507.40 	  &  	   3.68 	  &  	 3.6$\pm$1.0 	  &  	 2.5$^{+1.1}_{-1.1}$ 	  &  	 0.010 	 \\
\hline
\end{tabular}
\end{center}
\end{table*}

\FloatBarrier
\begin{figure*}
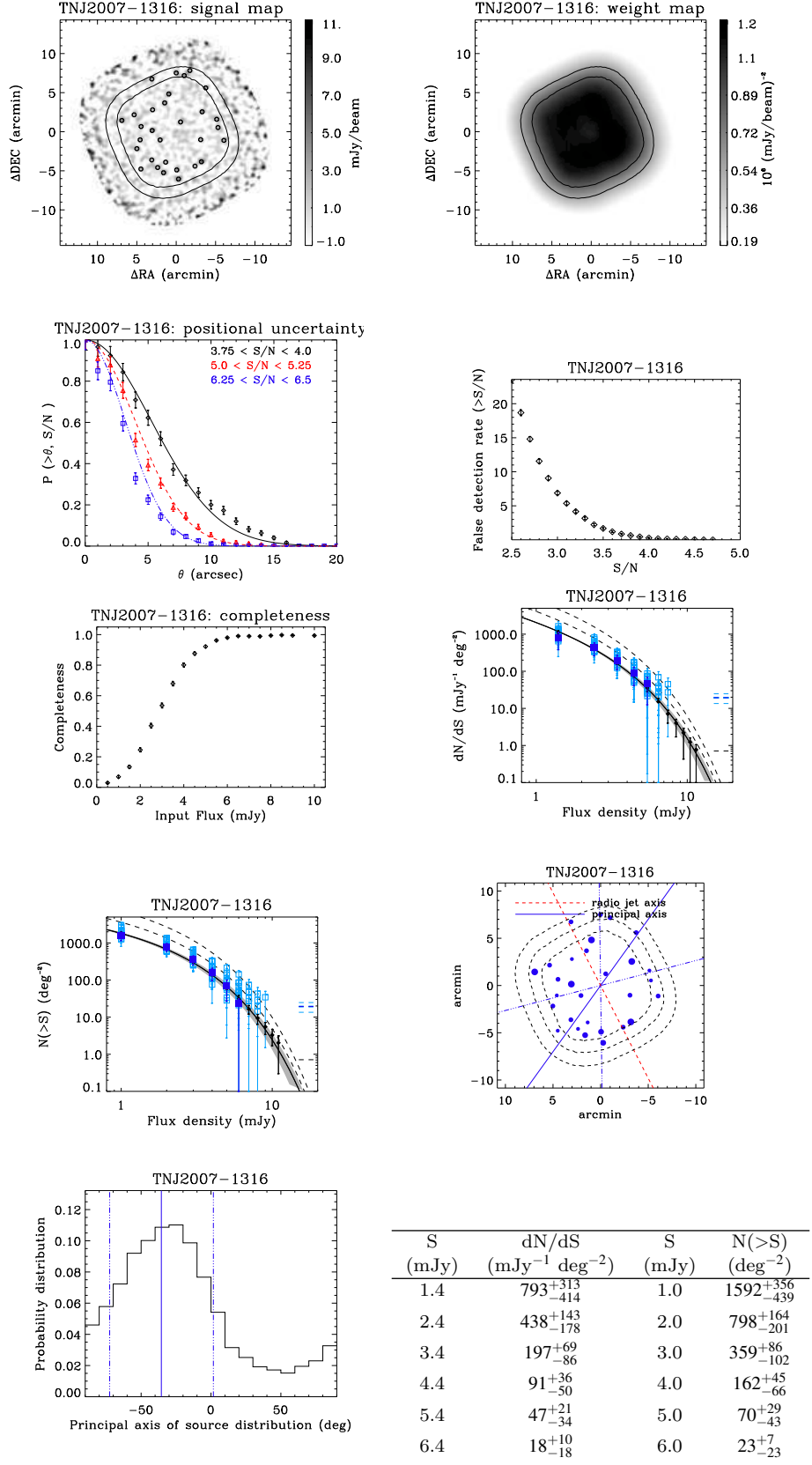

\centering
\caption[]{TNJ2007-1316 products}
%
\end{figure*}

\FloatBarrier
\begin{table*}
\begin{center}
\scriptsize
\caption{TNJ2007-1316 source catalogue.}
\begin{tabular}{lccccc}
\hline
      &     		&     		& S$_{\rm{1.1mm}}$ & S$_{\rm{1.1mm}}$ 	&         		\\
      &      		&     		& (measured)  		& (deboosted) 			&         		\\
Id  & IAU name & S/N 	& (mJy)       		& (mJy)       			& P($<0$) 	\\
\hline
1 	 & 	 MMJ200740.16-132034.59 	  &  	   7.09 	  &  	 6.8$\pm$1.0 	  &  	 6.0$^{+1.0}_{-1.0}$ 	  &  	 0.000 	 \\
2 	 & 	 MMJ200805.85-131634.54 	  &  	   6.19 	  &  	 5.7$\pm$0.9 	  &  	 5.1$^{+0.9}_{-1.0}$ 	  &  	 0.000 	 \\
3 	 & 	 MMJ200821.68-131517.16 	  &  	   6.01 	  &  	 6.2$\pm$1.0 	  &  	 5.3$^{+1.1}_{-1.0}$ 	  &  	 0.000 	 \\
4 	 & 	 MMJ200757.17-131154.25 	  &  	   5.99 	  &  	 5.6$\pm$0.9 	  &  	 4.9$^{+1.0}_{-1.0}$ 	  &  	 0.000 	 \\
5 	 & 	 MMJ200739.78-131410.92 	  &  	   5.96 	  &  	 5.6$\pm$0.9 	  &  	 4.9$^{+0.9}_{-1.0}$ 	  &  	 0.000 	 \\
\hline
\end{tabular}
\end{center}
\end{table*}

\FloatBarrier
\begin{figure*}
\centering
\caption[]{4C+41.17 products}
%
\end{figure*}

\FloatBarrier
\begin{table*}
\begin{center}
\scriptsize
\caption{4C+41.17 source catalogue.}
\begin{tabular}{lccccc}
\hline
      &     		&     		& S$_{\rm{1.1mm}}$ & S$_{\rm{1.1mm}}$ 	&         		\\
      &      		&     		& (measured)  		& (deboosted) 			&         		\\
Id  & IAU name & S/N 	& (mJy)       		& (mJy)       			& P($<0$) 	\\
\hline
1 	 & 	 MMJ065043.59+412918.63 	  &  	   8.93 	  &  	 8.9$\pm$1.0 	  &  	 8.2$^{+1.0}_{-1.0}$ 	  &  	 0.000 	 \\
2 	 & 	 MMJ065015.97+412944.58 	  &  	   8.50 	  &  	 8.7$\pm$1.0 	  &  	 8.0$^{+1.0}_{-1.1}$ 	  &  	 0.000 	 \\
3 	 & 	 MMJ065050.17+412820.22 	  &  	   6.33 	  &  	 6.3$\pm$1.0 	  &  	 5.5$^{+1.0}_{-1.0}$ 	  &  	 0.000 	 \\
4 	 & 	 MMJ065048.98+413130.17 	  &  	   6.00 	  &  	 5.9$\pm$1.0 	  &  	 5.0$^{+1.1}_{-0.9}$ 	  &  	 0.000 	 \\
5 	 & 	 MMJ065030.59+413940.08 	  &  	   5.94 	  &  	 7.9$\pm$1.3 	  &  	 6.5$^{+1.4}_{-1.4}$ 	  &  	 0.000 	 \\
\hline
\end{tabular}
\end{center}
\end{table*}

\FloatBarrier
\begin{figure*}
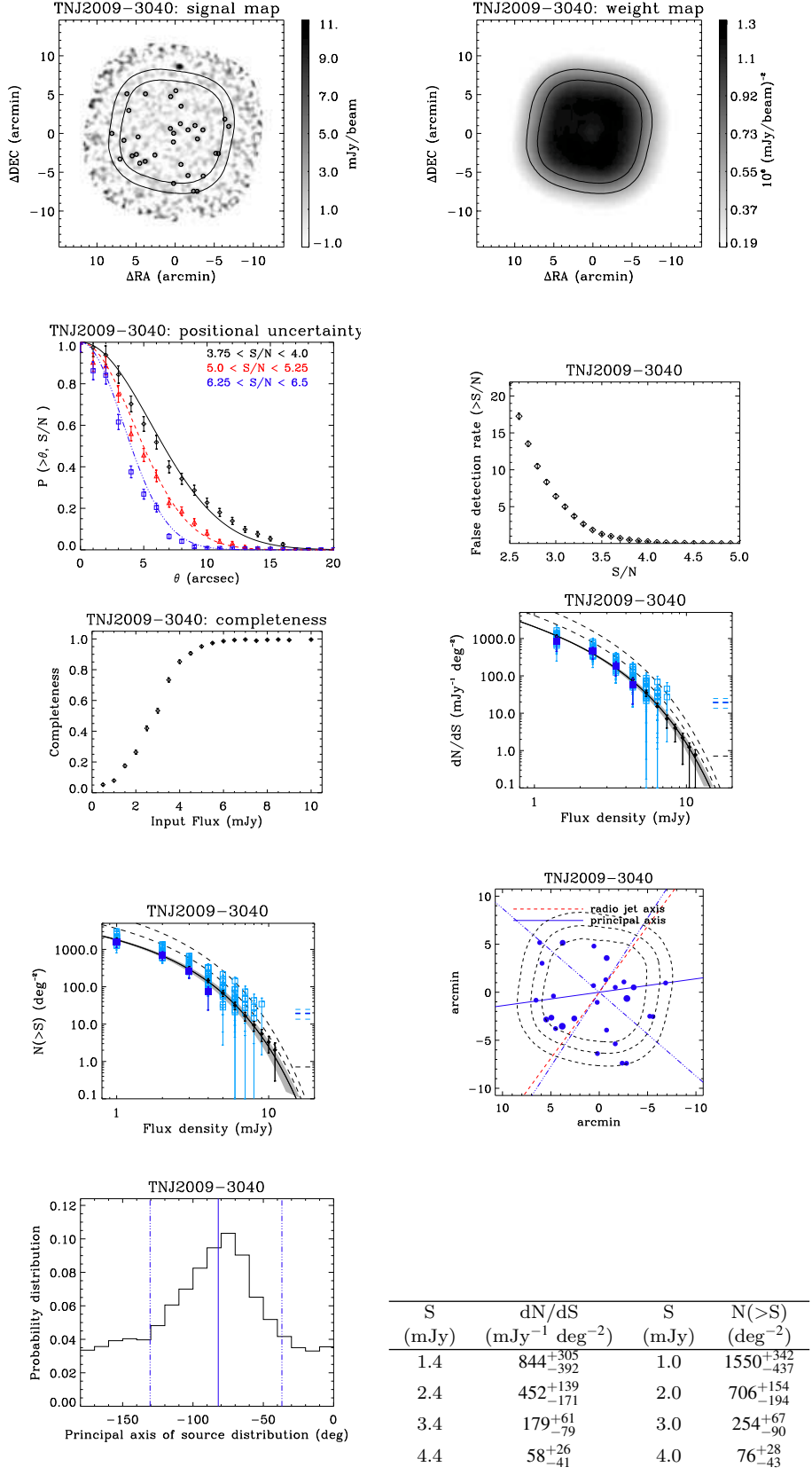

\centering
\caption[]{TNJ2009-3040 products}
%
\end{figure*}

\FloatBarrier
\begin{table*}
\begin{center}
\scriptsize
\caption{TNJ2009-3040 source catalogue.}
\begin{tabular}{lccccc}
\hline
      &     		&     		& S$_{\rm{1.1mm}}$ & S$_{\rm{1.1mm}}$ 	&         		\\
      &      		&     		& (measured)  		& (deboosted) 			&         		\\
Id  & IAU name & S/N 	& (mJy)       		& (mJy)       			& P($<0$) 	\\
\hline
1 	 & 	 MMJ201005.87-304339.86 	  &  	   5.85 	  &  	 5.5$\pm$0.9 	  &  	 4.8$^{+0.9}_{-1.0}$ 	  &  	 0.000 	 \\
2 	 & 	 MMJ200934.77-304046.39 	  &  	   5.24 	  &  	 4.8$\pm$0.9 	  &  	 4.0$^{+1.0}_{-0.9}$ 	  &  	 0.000 	 \\
3 	 & 	 MMJ200931.51-303936.94 	  &  	   4.58 	  &  	 4.2$\pm$0.9 	  &  	 3.4$^{+1.0}_{-0.9}$ 	  &  	 0.000 	 \\
4 	 & 	 MMJ201013.58-304257.78 	  &  	   4.44 	  &  	 4.3$\pm$1.0 	  &  	 3.4$^{+1.0}_{-1.0}$ 	  &  	 0.001 	 \\
5 	 & 	 MMJ201000.08-304251.29 	  &  	   4.37 	  &  	 4.0$\pm$0.9 	  &  	 3.1$^{+0.9}_{-1.0}$ 	  &  	 0.001 	 \\
\hline
\end{tabular}
\end{center}
\end{table*}

\FloatBarrier
\begin{figure*}
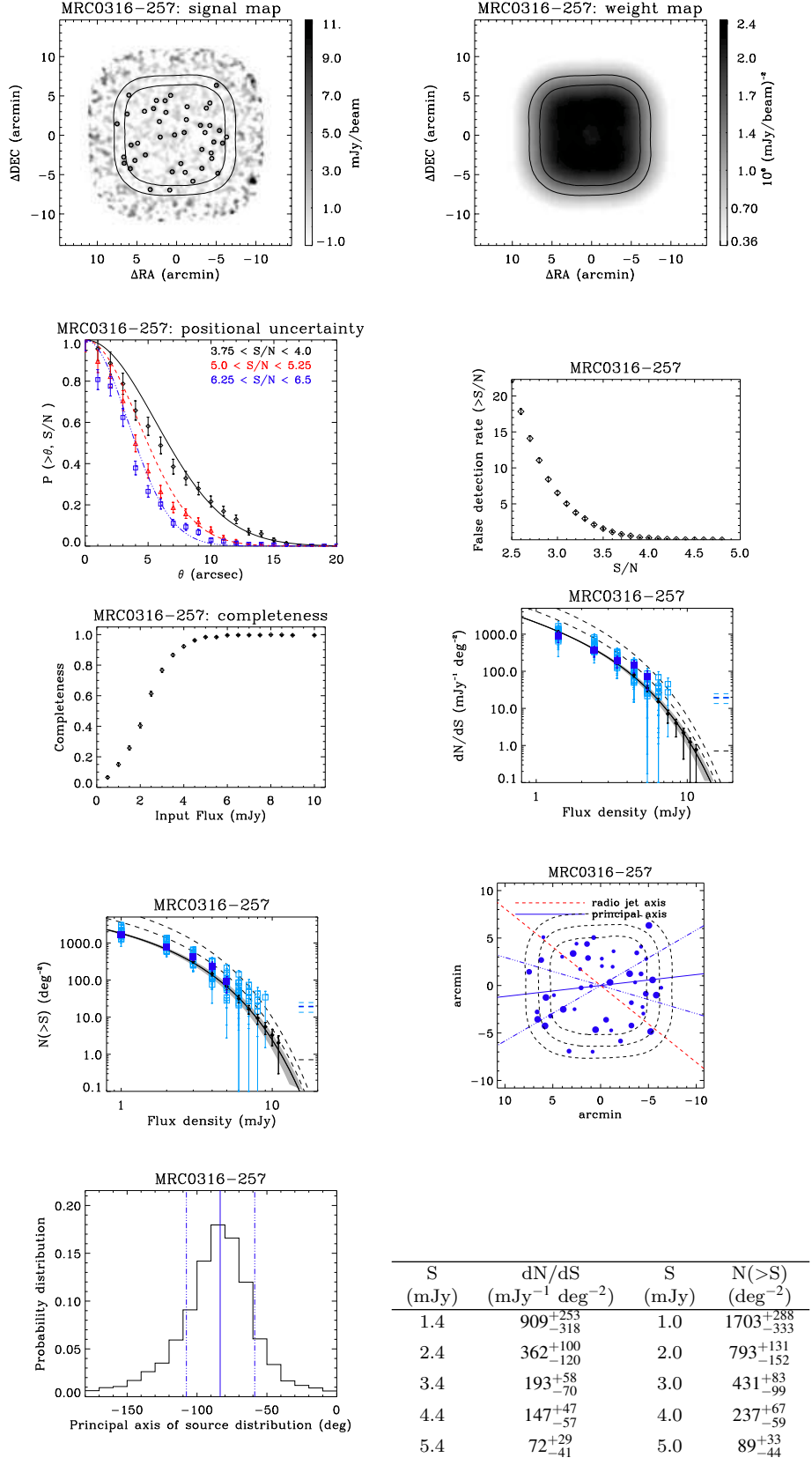

\centering
\caption[]{MRC0316-257 products}
%
\end{figure*}

\FloatBarrier
\begin{table*}
\begin{center}
\scriptsize
\caption{MRC0316-257 source catalogue.}
\begin{tabular}{lccccc}
\hline
      &     		&     		& S$_{\rm{1.1mm}}$ & S$_{\rm{1.1mm}}$ 	&         		\\
      &      		&     		& (measured)  		& (deboosted) 			&         		\\
Id  & IAU name & S/N 	& (mJy)       		& (mJy)       			& P($<0$) 	\\
\hline
1 	 & 	 MMJ031758.69-253355.57 	  &  	   7.88 	  &  	 5.2$\pm$0.7 	  &  	 4.9$^{+0.7}_{-0.7}$ 	  &  	 0.000 	 \\
2 	 & 	 MMJ031837.94-253925.24 	  &  	   7.81 	  &  	 6.0$\pm$0.8 	  &  	 5.6$^{+0.8}_{-0.8}$ 	  &  	 0.000 	 \\
3 	 & 	 MMJ031824.87-253147.88 	  &  	   7.72 	  &  	 5.1$\pm$0.7 	  &  	 4.7$^{+0.7}_{-0.7}$ 	  &  	 0.000 	 \\
4 	 & 	 MMJ031814.48-253949.57 	  &  	   7.37 	  &  	 5.0$\pm$0.7 	  &  	 4.6$^{+0.7}_{-0.7}$ 	  &  	 0.000 	 \\
5 	 & 	 MMJ031746.31-253611.03 	  &  	   7.34 	  &  	 5.5$\pm$0.8 	  &  	 5.1$^{+0.7}_{-0.8}$ 	  &  	 0.000 	 \\
\hline
\end{tabular}
\end{center}
\end{table*}

\FloatBarrier
\begin{figure*}
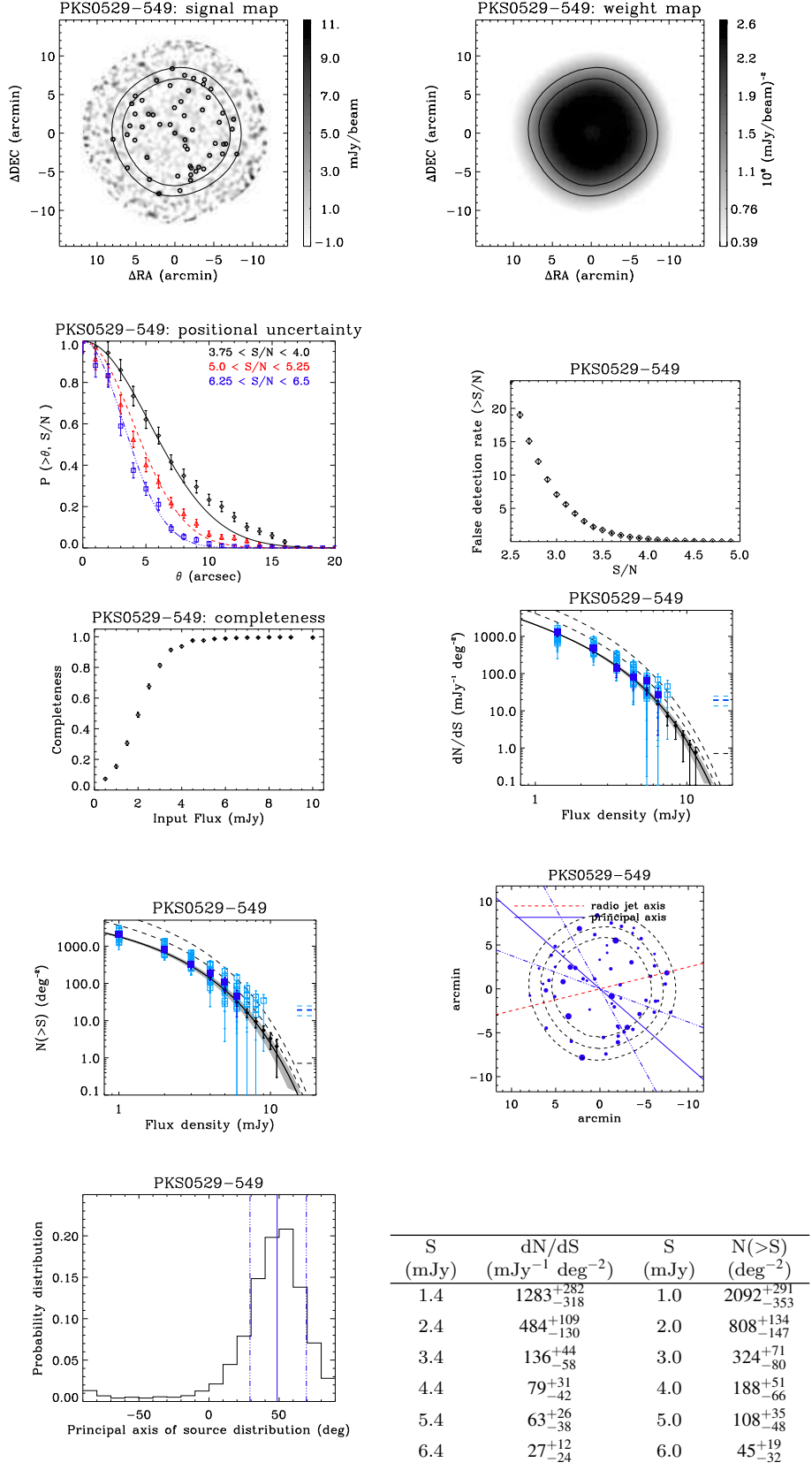

\centering
\caption[]{PKS0529-549 products}
%
\end{figure*}

\FloatBarrier
\begin{table*}
\begin{center}
\scriptsize
\caption{PKS0529-549 source catalogue.}
\begin{tabular}{lccccc}
\hline
      &     		&     		& S$_{\rm{1.1mm}}$ & S$_{\rm{1.1mm}}$ 	&         		\\
      &      		&     		& (measured)  		& (deboosted) 			&         		\\
Id  & IAU name & S/N 	& (mJy)       		& (mJy)       			& P($<0$) 	\\
\hline
1 	 & 	 MMJ053025.20-545422.02 	  &  	   9.72 	  &  	 6.4$\pm$0.7 	  &  	 6.0$^{+0.7}_{-0.6}$ 	  &  	 0.000 	 \\
2 	 & 	 MMJ053039.52-550212.93 	  &  	   9.68 	  &  	 8.9$\pm$0.9 	  &  	 8.3$^{+0.9}_{-0.9}$ 	  &  	 0.000 	 \\
3 	 & 	 MMJ053013.09-544852.52 	  &  	   9.32 	  &  	 6.2$\pm$0.7 	  &  	 5.8$^{+0.7}_{-0.7}$ 	  &  	 0.000 	 \\
4 	 & 	 MMJ053050.23-545730.72 	  &  	   9.16 	  &  	 6.0$\pm$0.7 	  &  	 5.6$^{+0.7}_{-0.6}$ 	  &  	 0.000 	 \\
5 	 & 	 MMJ053016.15-545513.12 	  &  	   8.66 	  &  	 5.6$\pm$0.6 	  &  	 5.2$^{+0.6}_{-0.7}$ 	  &  	 0.000 	 \\
\hline
\end{tabular}
\end{center}
\end{table*}

\FloatBarrier
\begin{figure*}
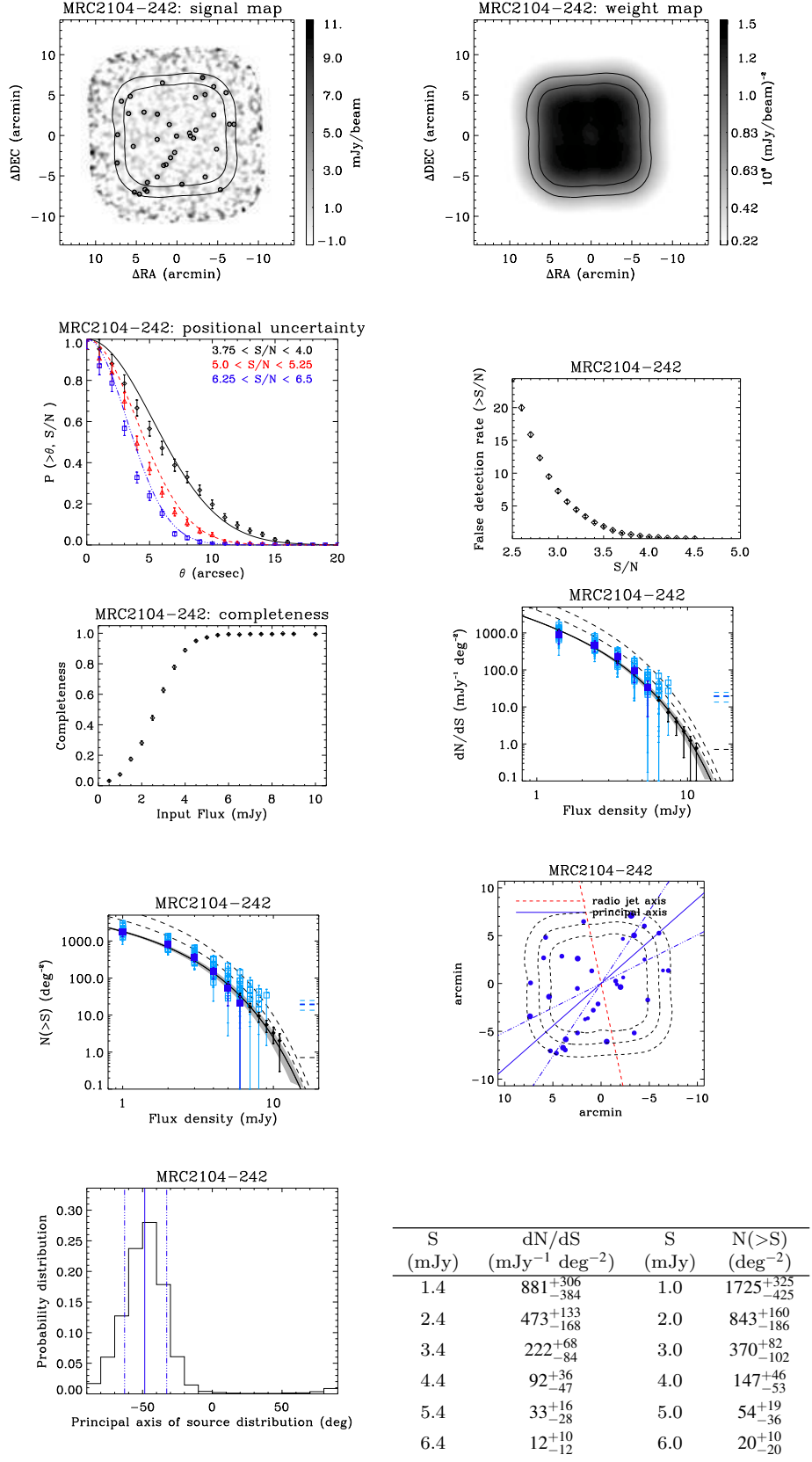

\centering
\caption[]{MRC2104-242 products}
%
\end{figure*}

\FloatBarrier
\begin{table*}
\begin{center}
\scriptsize
\caption{MRC2104-242 source catalogue.}
\begin{tabular}{lccccc}
\hline
      &     		&     		& S$_{\rm{1.1mm}}$ & S$_{\rm{1.1mm}}$ 	&         		\\
      &      		&     		& (measured)  		& (deboosted) 			&         		\\
Id  & IAU name & S/N 	& (mJy)       		& (mJy)       			& P($<0$) 	\\
\hline
1 	 & 	 MMJ210644.32-235801.57 	  &  	   6.90 	  &  	 7.8$\pm$1.1 	  &  	 6.8$^{+1.2}_{-1.1}$ 	  &  	 0.000 	 \\
2 	 & 	 MMJ210642.99-240007.56 	  &  	   5.60 	  &  	 5.1$\pm$0.9 	  &  	 4.4$^{+0.9}_{-0.9}$ 	  &  	 0.000 	 \\
3 	 & 	 MMJ210708.64-240232.83 	  &  	   5.52 	  &  	 4.7$\pm$0.8 	  &  	 4.0$^{+0.9}_{-0.9}$ 	  &  	 0.000 	 \\
4 	 & 	 MMJ210714.07-241059.63 	  &  	   5.41 	  &  	 5.2$\pm$1.0 	  &  	 4.4$^{+1.0}_{-1.0}$ 	  &  	 0.000 	 \\
5 	 & 	 MMJ210655.44-241113.81 	  &  	   5.33 	  &  	 5.3$\pm$1.0 	  &  	 4.4$^{+1.0}_{-1.0}$ 	  &  	 0.000 	 \\
\hline
\end{tabular}
\end{center}
\end{table*}

\FloatBarrier
\begin{figure*}
\centering
\caption[]{4C+23.56 products}
%
\end{figure*}

\FloatBarrier
\begin{table*}
\begin{center}
\scriptsize
\caption{4C+23.56 source catalogue.}
\begin{tabular}{lccccc}
\hline
      &     		&     		& S$_{\rm{1.1mm}}$ & S$_{\rm{1.1mm}}$ 	&         		\\
      &      		&     		& (measured)  		& (deboosted) 			&         		\\
Id  & IAU name & S/N 	& (mJy)       		& (mJy)       			& P($<0$) 	\\
\hline
1 	 & 	 MMJ210719.67+233050.70 	  &  	  18.37 	  &  	 10.3$\pm$0.6 	  &  	 10.1$^{+0.5}_{-0.6}$ 	  &  	 0.000 	 \\
2 	 & 	 MMJ210743.23+232718.06 	  &  	  13.56 	  &  	 9.5$\pm$0.7 	  &  	 9.2$^{+0.7}_{-0.7}$ 	  &  	 0.000 	 \\
3 	 & 	 MMJ210737.89+232830.12 	  &  	  10.24 	  &  	 6.2$\pm$0.6 	  &  	 6.0$^{+0.6}_{-0.6}$ 	  &  	 0.000 	 \\
4 	 & 	 MMJ210713.83+233348.31 	  &  	  10.06 	  &  	 5.8$\pm$0.6 	  &  	 5.5$^{+0.6}_{-0.6}$ 	  &  	 0.000 	 \\
5 	 & 	 MMJ210726.11+233353.38 	  &  	   9.84 	  &  	 5.5$\pm$0.6 	  &  	 5.3$^{+0.6}_{-0.6}$ 	  &  	 0.000 	 \\
\hline
\end{tabular}
\end{center}
\end{table*}

\FloatBarrier
\begin{figure*}
\centering
\caption[]{MRC0355-037 products}
%
\end{figure*}

\FloatBarrier
\begin{table*}
\begin{center}
\scriptsize
\caption{MRC0355-037 source catalogue.}
\begin{tabular}{lccccc}
\hline
      &     		&     		& S$_{\rm{1.1mm}}$ & S$_{\rm{1.1mm}}$ 	&         		\\
      &      		&     		& (measured)  		& (deboosted) 			&         		\\
Id  & IAU name & S/N 	& (mJy)       		& (mJy)       			& P($<0$) 	\\
\hline
1 	 & 	 MMJ035812.35-033738.25 	  &  	   9.75 	  &  	 9.0$\pm$0.9 	  &  	 8.5$^{+0.8}_{-1.0}$ 	  &  	 0.000 	 \\
2 	 & 	 MMJ035742.05-034125.45 	  &  	   7.50 	  &  	 7.1$\pm$0.9 	  &  	 6.5$^{+0.9}_{-1.0}$ 	  &  	 0.000 	 \\
3 	 & 	 MMJ035732.89-034028.04 	  &  	   7.42 	  &  	 7.5$\pm$1.0 	  &  	 6.6$^{+1.1}_{-1.0}$ 	  &  	 0.000 	 \\
4 	 & 	 MMJ035759.69-033035.79 	  &  	   7.28 	  &  	 5.9$\pm$0.8 	  &  	 5.4$^{+0.8}_{-0.8}$ 	  &  	 0.000 	 \\
5 	 & 	 MMJ035811.19-033438.14 	  &  	   7.16 	  &  	 5.9$\pm$0.8 	  &  	 5.4$^{+0.8}_{-0.9}$ 	  &  	 0.000 	 \\
\hline
\end{tabular}
\end{center}
\end{table*}

\FloatBarrier
\begin{figure*}
\centering
\caption[]{MRC2048-272 products}
%
\end{figure*}

\FloatBarrier
\begin{table*}
\begin{center}
\scriptsize
\caption{MRC2048-272 source catalogue.}
\begin{tabular}{lccccc}
\hline
      &     		&     		& S$_{\rm{1.1mm}}$ & S$_{\rm{1.1mm}}$ 	&         		\\
      &      		&     		& (measured)  		& (deboosted) 			&         		\\
Id  & IAU name & S/N 	& (mJy)       		& (mJy)       			& P($<0$) 	\\
\hline
1 	 & 	 MMJ205136.12-270103.24 	  &  	   8.06 	  &  	 7.9$\pm$1.0 	  &  	 7.2$^{+1.0}_{-1.0}$ 	  &  	 0.000 	 \\
2 	 & 	 MMJ205054.99-270947.22 	  &  	   8.00 	  &  	 7.5$\pm$0.9 	  &  	 6.8$^{+1.1}_{-0.8}$ 	  &  	 0.000 	 \\
3 	 & 	 MMJ205040.21-265814.51 	  &  	   7.55 	  &  	 6.6$\pm$0.9 	  &  	 5.9$^{+0.9}_{-0.8}$ 	  &  	 0.000 	 \\
4 	 & 	 MMJ205038.67-265717.40 	  &  	   5.83 	  &  	 5.7$\pm$1.0 	  &  	 5.0$^{+0.9}_{-1.1}$ 	  &  	 0.000 	 \\
5 	 & 	 MMJ205121.30-270200.05 	  &  	   5.73 	  &  	 4.3$\pm$0.8 	  &  	 3.8$^{+0.8}_{-0.7}$ 	  &  	 0.000 	 \\
\hline
\end{tabular}
\end{center}
\end{table*}

\FloatBarrier
\begin{figure*}
\centering
\caption[]{TXS2322-040 products}
%
\end{figure*}

\FloatBarrier
\begin{table*}
\begin{center}
\scriptsize
\caption{TXS2322-040 source catalogue.}
\begin{tabular}{lccccc}
\hline
      &     		&     		& S$_{\rm{1.1mm}}$ & S$_{\rm{1.1mm}}$ 	&         		\\
      &      		&     		& (measured)  		& (deboosted) 			&         		\\
Id  & IAU name & S/N 	& (mJy)       		& (mJy)       			& P($<0$) 	\\
\hline
1 	 & 	 MMJ232445.77-034219.54 	  &  	  10.00 	  &  	 7.6$\pm$0.8 	  &  	 7.2$^{+0.7}_{-0.8}$ 	  &  	 0.000 	 \\
2 	 & 	 MMJ232510.01-035132.31 	  &  	   9.81 	  &  	 6.8$\pm$0.7 	  &  	 6.4$^{+0.7}_{-0.7}$ 	  &  	 0.000 	 \\
3 	 & 	 MMJ232500.57-035213.99 	  &  	   8.10 	  &  	 6.6$\pm$0.8 	  &  	 6.0$^{+0.8}_{-0.8}$ 	  &  	 0.000 	 \\
4 	 & 	 MMJ232453.16-035002.48 	  &  	   8.01 	  &  	 6.1$\pm$0.8 	  &  	 5.6$^{+0.8}_{-0.8}$ 	  &  	 0.000 	 \\
5 	 & 	 MMJ232446.55-034605.21 	  &  	   7.87 	  &  	 5.3$\pm$0.7 	  &  	 4.9$^{+0.7}_{-0.7}$ 	  &  	 0.000 	 \\
\hline
\end{tabular}
\end{center}
\end{table*}

\FloatBarrier
\begin{figure*}
\centering
\caption[]{MRC2322-052 products}
%
\end{figure*}

\FloatBarrier
\begin{table*}
\begin{center}
\scriptsize
\caption{MRC2322-052 source catalogue.}
\begin{tabular}{lccccc}
\hline
      &     		&     		& S$_{\rm{1.1mm}}$ & S$_{\rm{1.1mm}}$ 	&         		\\
      &      		&     		& (measured)  		& (deboosted) 			&         		\\
Id  & IAU name & S/N 	& (mJy)       		& (mJy)       			& P($<0$) 	\\
\hline
1 	 & 	 MMJ232504.54-045724.13 	  &  	   7.09 	  &  	 4.9$\pm$0.7 	  &  	 4.5$^{+0.7}_{-0.7}$ 	  &  	 0.000 	 \\
2 	 & 	 MMJ232514.75-045736.21 	  &  	   6.68 	  &  	 4.7$\pm$0.7 	  &  	 4.2$^{+0.7}_{-0.7}$ 	  &  	 0.000 	 \\
3 	 & 	 MMJ232543.27-050236.04 	  &  	   6.55 	  &  	 6.0$\pm$0.9 	  &  	 5.3$^{+0.9}_{-0.9}$ 	  &  	 0.000 	 \\
4 	 & 	 MMJ232520.22-050600.69 	  &  	   6.55 	  &  	 6.1$\pm$0.9 	  &  	 5.3$^{+1.0}_{-0.9}$ 	  &  	 0.000 	 \\
5 	 & 	 MMJ232456.70-045857.92 	  &  	   5.92 	  &  	 4.4$\pm$0.7 	  &  	 3.9$^{+0.8}_{-0.7}$ 	  &  	 0.000 	 \\
\hline
\end{tabular}
\end{center}
\end{table*}

\FloatBarrier
\begin{figure*}
\centering
\caption[]{MRC2008-068 products}
%
\end{figure*}

\FloatBarrier
\begin{table*}
\begin{center}
\scriptsize
\caption{MRC2008-068 source catalogue.}
\begin{tabular}{lccccc}
\hline
      &     		&     		& S$_{\rm{1.1mm}}$ & S$_{\rm{1.1mm}}$ 	&         		\\
      &      		&     		& (measured)  		& (deboosted) 			&         		\\
Id  & IAU name & S/N 	& (mJy)       		& (mJy)       			& P($<0$) 	\\
\hline
1 	 & 	 MMJ201113.99-064403.26 	  &  	   9.84 	  &  	 8.6$\pm$0.9 	  &  	 8.0$^{+0.9}_{-0.8}$ 	  &  	 0.000 	 \\
2 	 & 	 MMJ201112.78-064453.62 	  &  	   8.20 	  &  	 7.1$\pm$0.9 	  &  	 6.6$^{+0.8}_{-0.9}$ 	  &  	 0.000 	 \\
3 	 & 	 MMJ201110.11-064925.25 	  &  	   8.11 	  &  	 7.2$\pm$0.9 	  &  	 6.6$^{+0.9}_{-0.9}$ 	  &  	 0.000 	 \\
4 	 & 	 MMJ201059.29-063742.32 	  &  	   7.66 	  &  	 8.7$\pm$1.1 	  &  	 7.8$^{+1.1}_{-1.2}$ 	  &  	 0.000 	 \\
5 	 & 	 MMJ201058.04-064841.54 	  &  	   7.38 	  &  	 7.1$\pm$1.0 	  &  	 6.4$^{+1.0}_{-1.0}$ 	  &  	 0.000 	 \\
\hline
\end{tabular}
\end{center}
\end{table*}

\FloatBarrier
\begin{figure*}
\centering
\caption[]{MRC2201-555 products}
%
\end{figure*}

\FloatBarrier
\begin{table*}
\begin{center}
\scriptsize
\caption{MRC2201-555 source catalogue.}
\begin{tabular}{lccccc}
\hline
      &     		&     		& S$_{\rm{1.1mm}}$ & S$_{\rm{1.1mm}}$ 	&         		\\
      &      		&     		& (measured)  		& (deboosted) 			&         		\\
Id  & IAU name & S/N 	& (mJy)       		& (mJy)       			& P($<0$) 	\\
\hline
1 	 & 	 MMJ220528.18-551939.97 	  &  	  10.24 	  &  	 7.7$\pm$0.8 	  &  	 7.3$^{+0.8}_{-0.7}$ 	  &  	 0.000 	 \\
2 	 & 	 MMJ220504.97-551742.63 	  &  	   7.84 	  &  	 6.1$\pm$0.8 	  &  	 5.7$^{+0.8}_{-0.8}$ 	  &  	 0.000 	 \\
3 	 & 	 MMJ220540.28-551134.15 	  &  	   6.93 	  &  	 7.2$\pm$1.0 	  &  	 6.4$^{+1.0}_{-1.1}$ 	  &  	 0.000 	 \\
4 	 & 	 MMJ220459.69-551352.65 	  &  	   6.91 	  &  	 5.3$\pm$0.8 	  &  	 4.8$^{+0.8}_{-0.8}$ 	  &  	 0.000 	 \\
5 	 & 	 MMJ220544.24-551204.60 	  &  	   5.89 	  &  	 6.1$\pm$1.0 	  &  	 5.2$^{+1.1}_{-1.1}$ 	  &  	 0.000 	 \\
\hline
\end{tabular}
\end{center}
\end{table*}

\FloatBarrier


\bsp	
\label{lastpage}
\end{document}